\documentclass[fleqn]{2017SCGE}
\usepackage{color}
\usepackage{mwe,tikz,ulem}
\usepackage{enumitem}                
\setlist{                            
  listparindent=\parindent,          
  parsep=0pt,                        
  itemsep=0pt,                       
  topsep=0pt,                        
  leftmargin=3.5mm                   
}
\usepackage{graphicx}
 \usepackage{natbib}
\def\degr{$^\circ$}                  
\def\arcmin{\hbox{$^\prime$}}        
\def\arcsec{\hbox{$^{\prime\prime}$}} 
\newcommand{\extp}{\textsl{eXTP}\xspace}
\newcommand{\loft}{\textsl{LOFT}\xspace}
\newcommand{\lad}{LAD\xspace}
\newcommand{\wfm}{WFM\xspace}
\newcommand{\xmm}{\textsl{XMM-Newton}\xspace}
\newcommand{\lfa}{SFA\xspace}
\newcommand{\gpd}{PFA\xspace}
\newcommand{\cgsflux}{\ensuremath{\mathrm{erg}\,\mathrm{s}^{-1}\,\mathrm{cm}^{-2}}}
\newcommand{\sci}{Science\xspace}
\newcommand{\apj}{ApJ\xspace}
\newcommand{\apjl}{ApJ\xspace}
\newcommand{\apjs}{ApJ Supp.\xspace}
\newcommand{\mnras}{MNRAS\xspace}
\newcommand{\aap}{A\&A\xspace}
\newcommand{\aj}{AJ\xspace}
\newcommand{\aaps}{A\&A Suppl.\xspace}
\newcommand{\nat}{Nature\xspace}
\newcommand{\araa}{ARAA\xspace}

\newcommand{\ssr}{Space Sci.\ Rev.\xspace}

\newcommand{\pasj}{PASJ\xspace}
\newcommand{\pasp}{PASP\xspace}
\newcommand{\prl}{PRL\xspace}
\newcommand{\prd}{Phys. Rev. D.\xspace}
\newcommand{\ssrv}{Sp. Sc. Rev.\xspace}

\begin{document}

\ensubject{subject}

\ArticleType{Invited Review}
\SpecialTopic{The X-ray Timing and Polarimetry Frontier with eXTP}
\Year{2019}
\Month{February}
\Vol{62}
\No{2}
\DOI{10.1007/s11433-017-9186-1}
\ArtNo{029506}
\ReceiveDate{December 6, 2017}
\AcceptDate{February 9, 2018}
\OnlineDate{August 30, 2018}

\title{Observatory science with \extp}{Observatory science with \extp}

\author[]{\parbox[t]{17.5cm}{
Jean J.M. in 't Zand$^{  1}$,
Enrico Bozzo$^{  2}$,
Jinlu Qu$^{  3}$,
Xiang-Dong Li$^{  4}$,
Lorenzo Amati$^{  5}$,
Yang Chen$^{  4}$,
Immacolata Donnarumma$^{  6,  7}$,
Victor Doroshenko$^{  8}$,
Stephen A. Drake$^{  9}$,
Margarita Hernanz$^{ 10}$
% ...
}}{}

\address[]{\parbox[t]{18cm}{
$^{  1}$SRON Netherlands Institute for Space Research, Sorbonnelaan 2, 3584 CA Utrecht, The Netherlands,
$^{  2}$Department of Astronomy, University of Geneva, 1290 Versoix, Switzerland,
$^{  3}$Key Laboratory for Particle Astrophysics, CAS Institute of High Energy Physics, 19B Yuquan Road, Beijing 100049, China,
$^{  4}$School of Astronomy and Space Science, Nanjing University, Nanjing 210023, China,
$^{  5}$INAF-IASF Bologna, via P. Gobetti 101, 40129 Bologna, Italy,
$^{  6}$ASI, Via del Politecnico snc, 00133 Roma, Italy,
$^{  7}$INAF - Istituto di Astrofisica e Planetologie Spaziali. Via Fosso del Cavaliere 100, 00133 Roma, Italy,
$^{  8}$Institut f\"ur Astronomie und Astrophysik T\"ubingen, Universit\"at T\"ubingen, Sand 1, 72076 T\"ubingen, Germany,
$^{  9}$NASA Goddard Space Flight Center, Code 662, Greenbelt, MD 20771, USA,
$^{ 10}$ICREA \& Institute of Space Sciences (IEEC-CSIC), Campus UAB, Carrer de Can Magrans, s/n 08193 Barcelona, Spain,
% ...
}}{}


\AuthorMark{in 't Zand J.J.M., Bozzo E., Li X., Qu J., et al.}

\AuthorCitation{in 't Zand J.J.M., Bozzo E., Li X., Qu J., et al.}


\abstract{In this White Paper we present the potential of the
  \textit{enhanced X-ray Timing and Polarimetry} (eXTP) mission for
  studies related to \textit{Observatory Science} targets. These
  include flaring stars, supernova remnants, accreting white dwarfs,
  low and high mass X-ray binaries, radio quiet and radio loud active
  galactic nuclei, tidal disruption events, and gamma-ray bursts.
  {$\,$  \extp will be excellently suited to study one common aspect of
    these objects: their often transient nature.}  Developed by an
  international Consortium led by the Institute of High Energy Physics
  of the Chinese Academy of Science, the \extp mission is expected to
  be launched in the mid 2020s.}



\maketitle


\begin{multicols}{2}

\section{Introduction}

The enhanced X-ray Timing and Polarimetry mission (eXTP) is a mission
concept proposed by a consortium led by the Institute of High-Energy
Physics of the Chinese Academy of Sciences and envisaged for a launch
in the mid 2020s. It carries 4 instrument packages for the 0.5--30 keV
bandpass, with the primary purpose to study conditions of extreme
density, gravity and magnetism in and around compact objects in the
universe. The mission concept provides for a low-Earth orbit with a
low inclination ($<15$\degr), incurring earth obscurations every one
and a half hour for targets near the celestial equator such as the
Galactic center region. The scientific payload of \extp consists of:
the Spectroscopic Focusing Array (\lfa), the Polarimetry Focusing
Array (\gpd), the Large Area Detector (\lad) and the Wide Field
Monitor (\wfm).

The \extp instrumentation is discussed in detail in
\citet{WPinstrumentation}, which includes a comparison with other
instruments. We summarize as follows.  The \lfa is an array of nine
identical X-ray telescopes covering the energy range 0.5--20~keV {$\,$
  with a spectral resolution of better than 180 eV (full width at half
  maximum, FWHM) at 6 keV}, and featuring a total effective area from
$\sim$0.5~m$^2$ at 6~keV to {$\sim 0.7$~m$^2$} at 1~keV. The \lfa
angular resolution is required to be less than
$\sim1$\arcmin\ (half-power diameter; HPD) and its sensitivity reaches
$10^{-14}$~erg~s$^{-1}$~cm$^{-2}$ (0.5--20 keV)
for an exposure time of $10^{4}$~ks. In the current baseline, the \lfa
focal plane detectors are silicon drift detectors that combine
CCD-like spectral resolutions with very small dead times, and
therefore are excellently suited for studies of the brightest cosmic
X-ray sources at the smallest time scales.

The \gpd consists of four identical X-ray telescopes that are
sensitive between 2 and 8 keV, have an angular resolution better than
30\arcsec\ (HPD) and a total effective area of $\sim 900$~cm$^2$ at
2~keV (including the detector efficiency). The \gpd features Gas Pixel
Detectors (GPDs) to allow polarization measurements in the X-rays. It
reaches a minimum detectable polarization (MDP) of 5\% in 100~ks for a
source with a Crab-like spectrum of flux
$3\times10^{-11}$~erg~s$^{-1}$~cm$^{-2}$. The spectral resolution is
1.1 keV at 6 keV (FWHM).

The \lad has a very large effective area of $\sim 3.4$~m$^2$ at 8~keV,
obtained with non-imaging silicon drift detectors, active between 2
and 30~keV with a spectral resolution of about 260 eV and collimated
to a field of view of 1~degree (FWHM). The \lad and the \lfa together
reach an unprecedented total effective area of more than $4$~m$^{2}$,
compared to 0.7~m$^2$ for the largest flown instrument
\citep[Proportional Counter Array on the Rossi X-ray Timing
  Explorer;][]{jahoda2006}.

The science payload is completed by the \wfm, consisting of 6
coded-mask cameras covering 4~sr of the sky at a sensitivity of
4~mCrab for an exposure time of 1~d in the 2 to 50~keV energy range,
and for a typical sensitivity of 0.2~mCrab combining 1~yr of
observations outside the Galactic plane. The instrument will feature
an angular resolution of a few arcminutes and will be endowed with an
energy resolution of about 300~eV (FWHM). The \wfm pairs the largest
duty cycle of 30\% for an imaging monitoring 2-30 keV device
\citep[e.g., compared to 8\% for BeppoSAX-WFC;][]{jager1997} with the
largest spectral resolution (usually about 1 keV at 6 keV).

The baseline for the observatory response time to targets of
opportunity within the 50\% part of the sky accessible to \extp is
4--8 hours. Depending on the the outcome of mission studies, this may
improve to 1--3 hours. The launch is currently foreseen in the mid 2020s.

\begin{figure*}[ht!]
\includegraphics[width=\textwidth,angle=0]{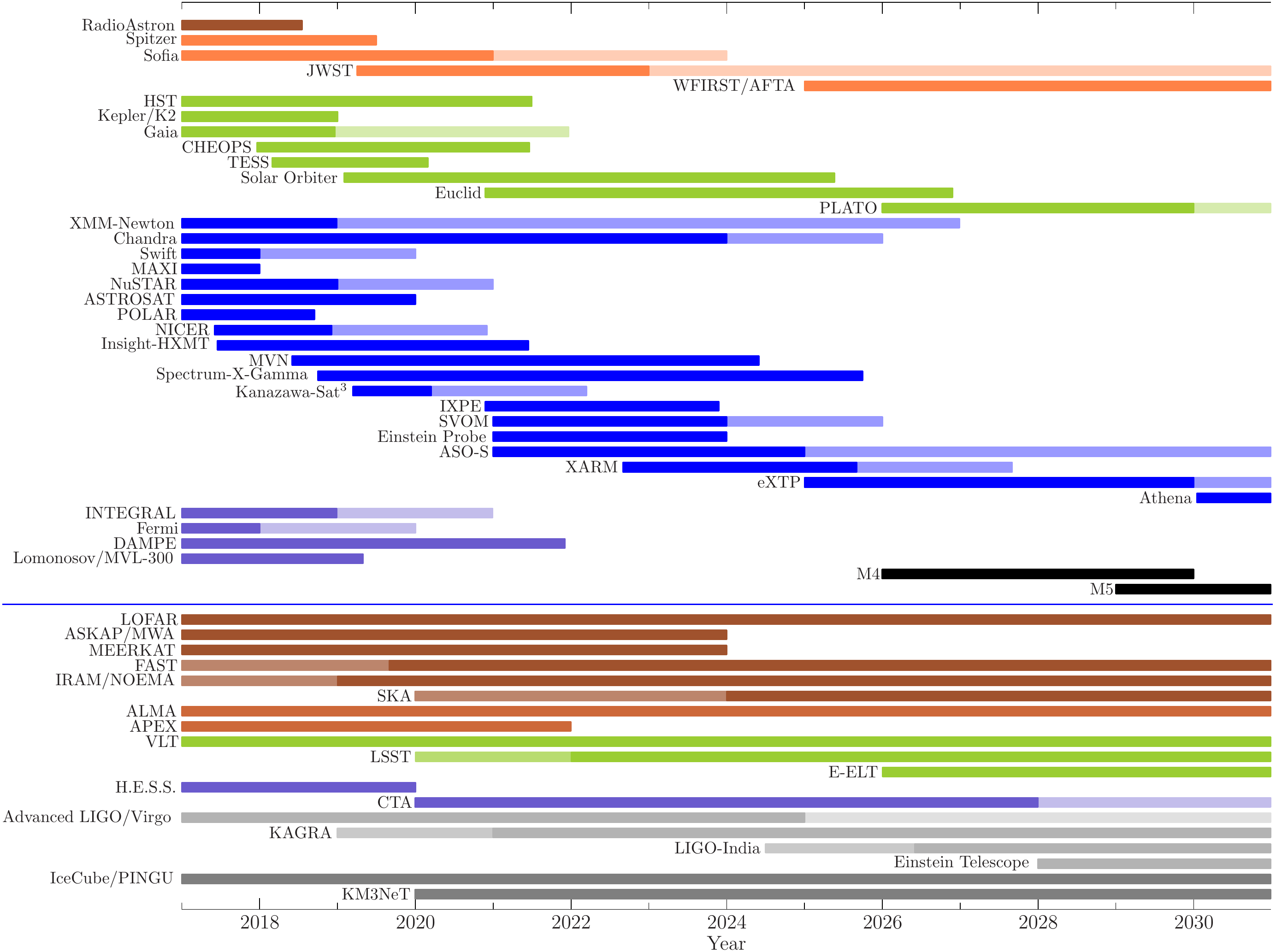}
\caption{ Multi-wavelength and multi-messenger facilities relevant to
  \extp\ with timelines as known in January 2018. The thin line
  separates space-based (top) from ground-based (bottom)
  facilities. Colors indicate similar wavebands from the radio (brown)
  via IR (red) and optical (green) to X-rays (blue) and gamma rays
  (purple). Grey bands: gravitational wave and neutrino
  detectors. Dark colors: funded lifetime, light colors: expected
  lifetime, where known, independent of funding decisions. Missions
  might last longer. }
 \label{figfacilities}
\end{figure*}

With a uniquely high throughput, good spectral resolution,
polarimetric capability and wide sky coverage, \extp is an observatory
very well suited for a variety of studies complementing the core
science of strong gravity \citep{WP_SG}, dense matter \citep{WP_DM}
and strong magnetism \citep{WP_SM}. The \lfa will provide high
sensitivity in a soft but wide bandpass, the \gpd will provide X-ray
polarimetry capability, the \lad will provide the best timing and
spectroscopic studies ever for a wide range of high energy sources
brighter than $\sim$1 mCrab in the 2 to 30 keV band, and the \wfm,
with its unprecedented combination of field of view and imaging down
to 2 keV, makes \extp a discovery machine of the variable and
transient X-ray sky. The \wfm will reveal many new sources that need
to be followed up with the \lfa, \gpd, \lad and other
facilities. Experience from the past decades shows that newly
discovered, unforeseen types of sources will provide unexpected
insights into fundamental questions. The \wfm will also be monitoring
daily hundreds of sources, to catch unexpected events and provide
long-term records of their variability and spectroscopic
evolution. Several targets of the \extp core program (e.g., low-mass
X-ray binaries) allow one to address scientific questions beyond those
related to the key mission science goals. Thus, they can be used to
fulfill the observatory science goals without requiring additional
exposure time.  Other targets are instead uniquely observed as part of
the observatory science program (e.g., accreting white dwarfs,
blazars, high mass X-ray binaries), in turn providing useful
comparative insights for the core science objectives.

\extp will be a unique, powerful X-ray partner of other new
large-scale multi-messenger facilities across the spectrum likely
available in the 2020s, such as advanced gravitational wave (GW) and
neutrino experiments (Advanced LIGO and Virgo, KAGRA, LIGO-India,
Einstein Telescope in GWs and KM3NeT and Icecube in neutrinos), SKA
and pathfinders in the radio, LSST and E-ELT in the optical, and CTA
at TeV energies (Fig.~\ref{figfacilities}). First \extp and 4 years
later Athena will be the major facilities in the X-ray regime and as
such cover the hot and energetic non-thermal phenomena of any cosmic
object, Athena focusing on diffuse constituents in the early Universe
and \extp focusing on bright, variable and transient phenomena. \extp
will be particularly suitable for electromagnetic (EM) studies of GW
sources, an exciting field of research whose birth we witnessed
through GW170817/GRB170817A \citep{grb170817A,gw170817}. The combined
EM/GW signal is expected to provide us new insights of neutron stars
through their merger events, see Sect.~\ref{sec:grbs} and
\citet{WP_DM}.  Steady GW sources related to spinning neutron stars and
ultracompact X-ray binaries may also be detected and identified in the
coming decades, providing new diagnostics for these multi-wavelength
sources that are particularly bright in the radio and X-ray bands, see
Sect.~\ref{sec:lmxbs} and \citet{WP_DM}.

To explain the science case for \extp as an observatory in a
kaleidoscopic manner in this White Paper, material is drawn from
eleven White Papers that were written for the \loft study by
scientists from the community at large
\citep{wpamati,wpdemartino,wpdonnarumma, wpdrake, wpintzand,
  wpmaccarone,wpmarisaldi,
  wpmignani,wporlandini,wprossi,wpderosa}. \loft was an ESA mission
concept with two of the three primary science goals of \extp and
larger versions of the \lad and \wfm. The outstanding capabilities of
all four \extp instruments will enable to answer key questions that
could not be addressed for many years, due to a lack of facilities
with sufficient sensitivity. These questions, across a broad swath of
topics, will not be answered by any other planned future X-ray
facility. The \extp science involves a wide range of objects, from
normal stars to supermassive black holes in other galaxies. These
objects have a wide range of fluxes. The wide range in capabilities of
the instrument package allows one to deal with this range in fluxes:
bright sources with the \lad, \wfm, and \gpd; faint sources mainly
with the \lfa. In this White Paper, the main questions are briefly
discussed per category of objects, in order of approximate distance,
that can be addressed with \extp and that are not dealt with in the
other \extp White Papers on strong gravity, dense matter and strong
magnetism. Due to the breadth of the science, this paper can only
summarize those questions.

\section{Flare stars}

\begin{figure*}[ht!]
\includegraphics[trim=0 0 0 0,height=0.5\textwidth,angle=-90]{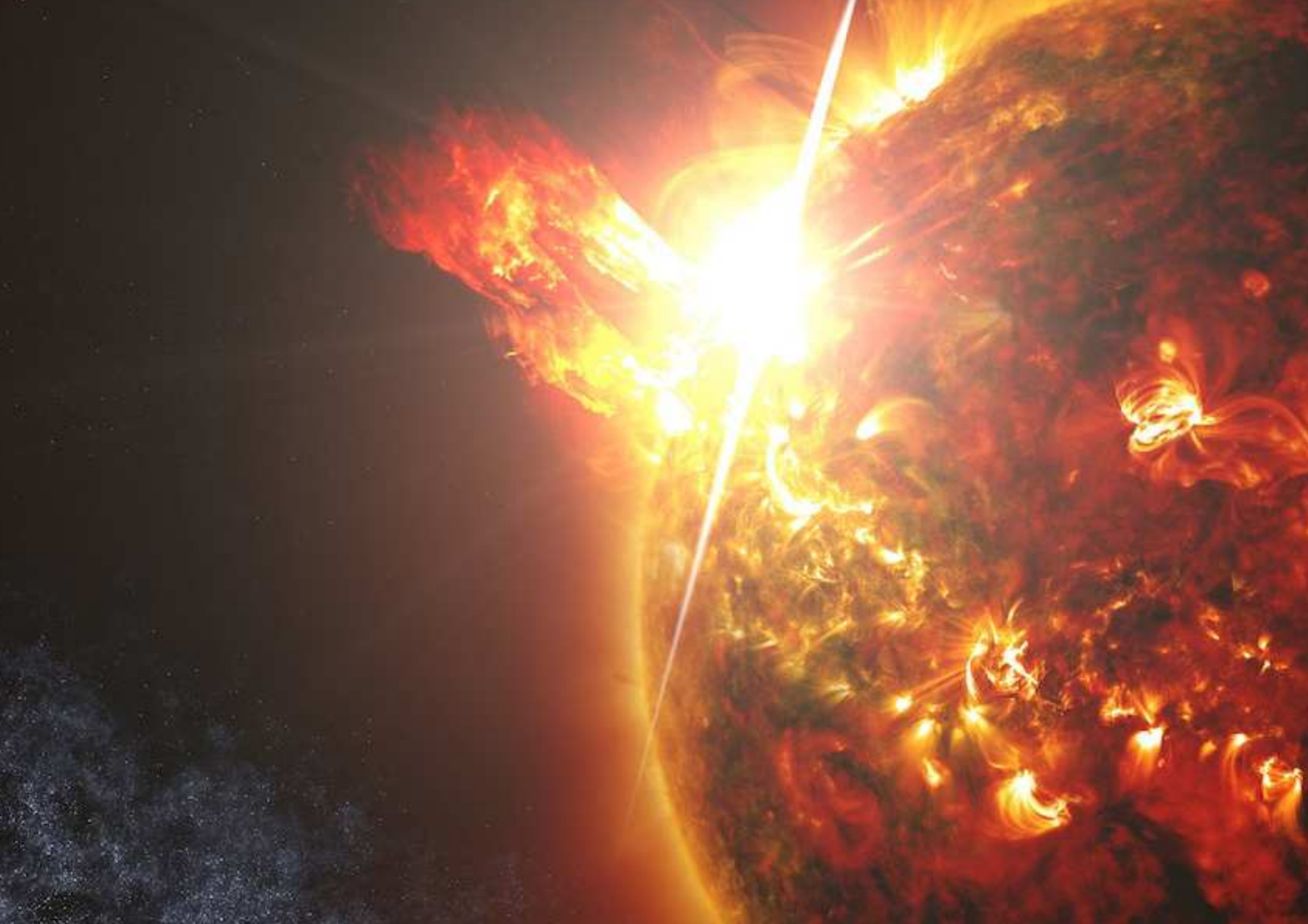}
\includegraphics[width=0.35\textwidth,angle=-90]{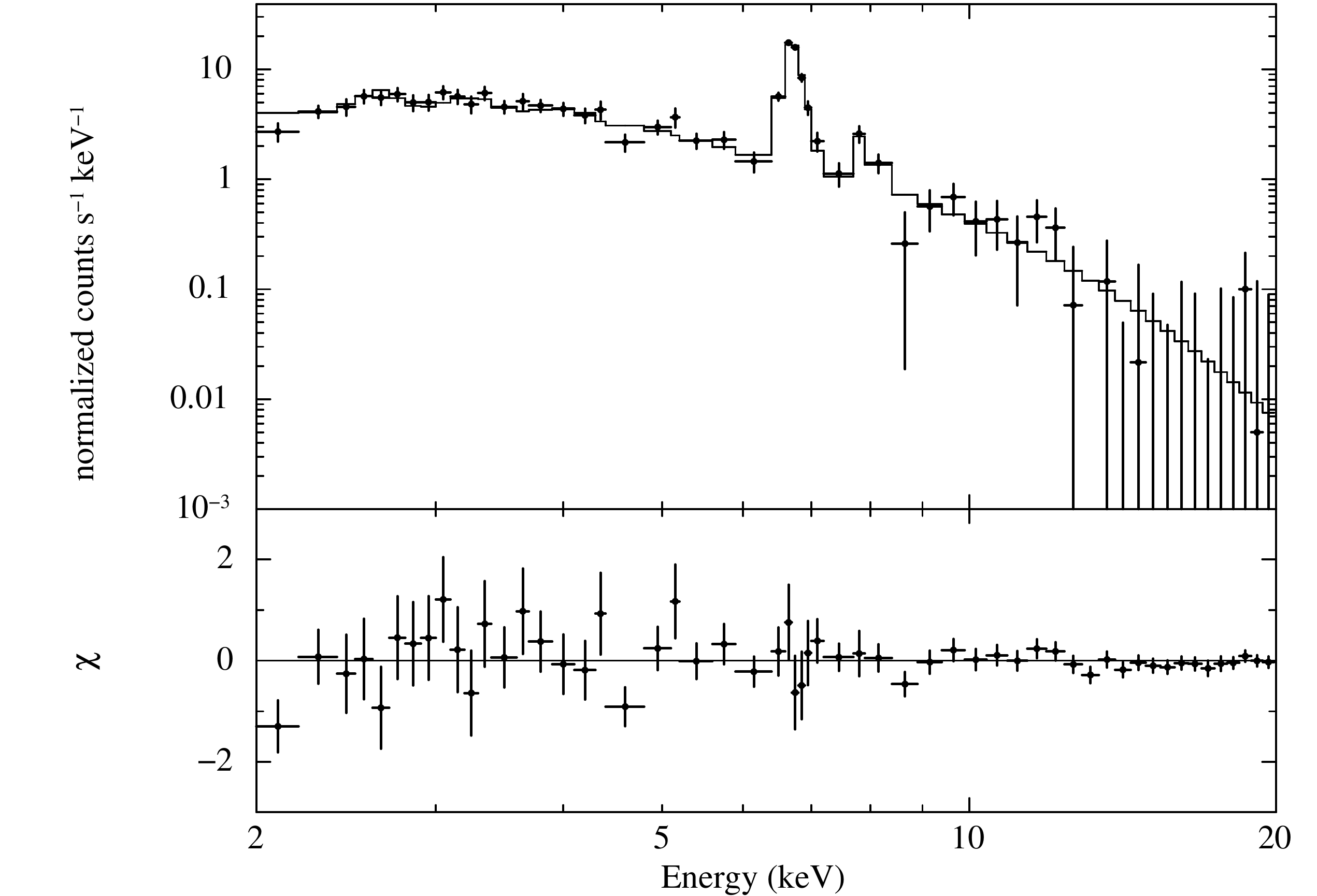}

\caption{\textit{Left:} artist's impression of the 2014 April 23
  "superflare" of one of the stars in the wide M dwarf binary DG CVn
  (credit Scott Wiessinger of the NASA/GSFC Scientific Visualization
  Studio). \textit{Right:} Simulated \lad spectrum of a 1000\,s time
  segment of a flare with similar flux and spectrum to the second,
  slightly less powerful flare of DG CVn 3 hours after the first
  outburst. Notice the well-exposed He-like Fe lines at 6.7\,keV and
  7.9\,keV which would enable an accurate coronal Fe abundance to be
  derived \citep[e.g.,][ and references therein]{phillips2006}
  .}\label{figDrake3}
\end{figure*}

On the Sun, flares occur in close proximity to active regions that are
characterized by boosted localized kG-strength magnetic fields
\citep{carrington59,guo06,guedel09,benz10}. Loops from these regions
extend into the solar corona and, as the footpoints of these loops are
jostled by solar convective motions, they are twisted and distorted
until magnetic reconnection occurs. There is then a sudden release of
energy, resulting in the acceleration of electrons and ions in these
loops up to MeV energies, emitting non-thermal radio (gyrosynchrotron)
and non-thermal hard X-ray emission (bremsstrahlung and Compton back
scattering). These energetic particles both stream away from the Sun
and down the loops where they deposit substantial amounts of energy to
the lower solar atmosphere (the chromosphere), producing the observed
intense hard X-ray emission at the loop footpoints
\citep{benz10,huang11}. Flares are also observed on other late-type
stars, with luminosities up to 10$^{33}$~erg~s$^{-1}$ and durations
from less than an hour to several days or longer
\citep{nordon08,he15}. The high temperatures result in spectral peaks
in the soft to hard X-ray regime, which make them particularly visible
with wide-field high duty cycle monitoring devices such as the \wfm.

\extp can contribute greatly to our understanding of stellar flares.
Expanding our knowledge of stellar flaring is crucial to examining the
influence of transient sources of ionizing radiation from a host star
on exoplanet systems, important for habitability concerns and space
weather which other worlds might experience
\citep[e.g.][]{segura2010}. Such studies extend the solar-stellar
connection in determining the extent to which solar models for
conversion of magnetic energy into plasma heating, particle
acceleration and mass motions apply to the more energetic stellar
flares. Extrapolating from lower efficiency high-energy monitoring
observatories, we estimate that \extp will detect at least 35
superflares per year from stars exhibiting the extremes of magnetic
activity. The anticipated source types include young stars, fully
convective M dwarfs, and tidally locked active binary systems. The
wide-field monitoring capability of \extp will also likely enable the
detection of flares from classes of stars hitherto not systematically
studied for their flaring, and will be important for expanding our
understanding of plasma physics processes in non-degenerate stellar
environments. Four important questions that \extp can address are:

\begin{itemize}

\item \textit{What are the properties of the non-thermal particles
  responsible for the initial flare energy input?} The study of the
  flare distribution with energy and of their temporal behavior, e.g.,
  spikes, oscillations, etc., will be enabled by \extp's sensitive hard
  X-ray capability provided by the \lad, far in excess of previous
  missions.

\item \textit{What are the physical conditions of the thermal plasma whose
  emission dominates the later stages of stellar flares?} Through its
  soft X-ray capabilities, \extp will enable us to study the variation
  of properties such as the temperature and elemental abundances as a
  function of time during flares.

\item \textit{What is the maximum energy that stellar flares can
  reach?} The wide-field, broad X-ray band and high-sensitivity
  capabilities of \extp will enable us to detect the rare
  ``superflares'' that have been only sporadically detected by previous and
  current missions: better understanding of the prevalence of such
  powerful events is important both for the physics of stellar flares
  and for the implications of their effect on potential planets
  orbiting these stars (see~Fig.~\ref{figDrake3}). 
  
\item \textit{What is the origin of flares from stars in nearby young
  associations?}  \extp sensitivity will allow one to detect flares from
  stars in nearby young associations, such as the TW Hya group, at
  distances up to 60~pc. As discussed by \citet{benz10}, young stars
  with accretion disks like TW Hya are expected to have a second type
  of flares due to interactions between the magnetic fields of the
  active regions and the disk.  It is an open question as to whether
  these flares have different properties than the better known in situ
  stellar flares (especially due to the very limited detections of
  these events in the X-rays).

\end{itemize}

\section{Supernova remnants}

Supernova remnants (SNRs), including pulsar wind nebulae (PWNe), are
important Galactic producers of relativistic particles and emitters of
diffuse X-ray synchrotron radiation \citep{rey07,rey08}. \extp is
expected to provide breakthroughs in some longstanding problems in the
studies of shell SNRs and PWNe via fine measurement of the
distribution of magnetic fields and polarized emission.

The strong shocks driven by supernova explosions are commonly regarded
as the main factory of the cosmic rays (CRs) with energy up to
$3\times10^{15}$ eV (the so called ``knee''), and accumulating evidence
has also pointed to this notion \citep{blasi15,caprioli15}. Despite
the well established theory of diffusive shock acceleration for CRs,
many crucial problems in the acceleration process remain unsolved,
such as the injection mechanism of particles and energy, the
efficiency of acceleration, the size of the acceleration region, and
the upper limit of particle energy \citep{hillas05,blasi13}.  Clues to
solve the problems can be provided by the non-thermal X-ray emission
of the relativistic electrons arising from the strong shocks of SNRs.

\begin{figure*}[ht!]
\includegraphics[trim=0 0 0 00,width=0.51\textwidth]{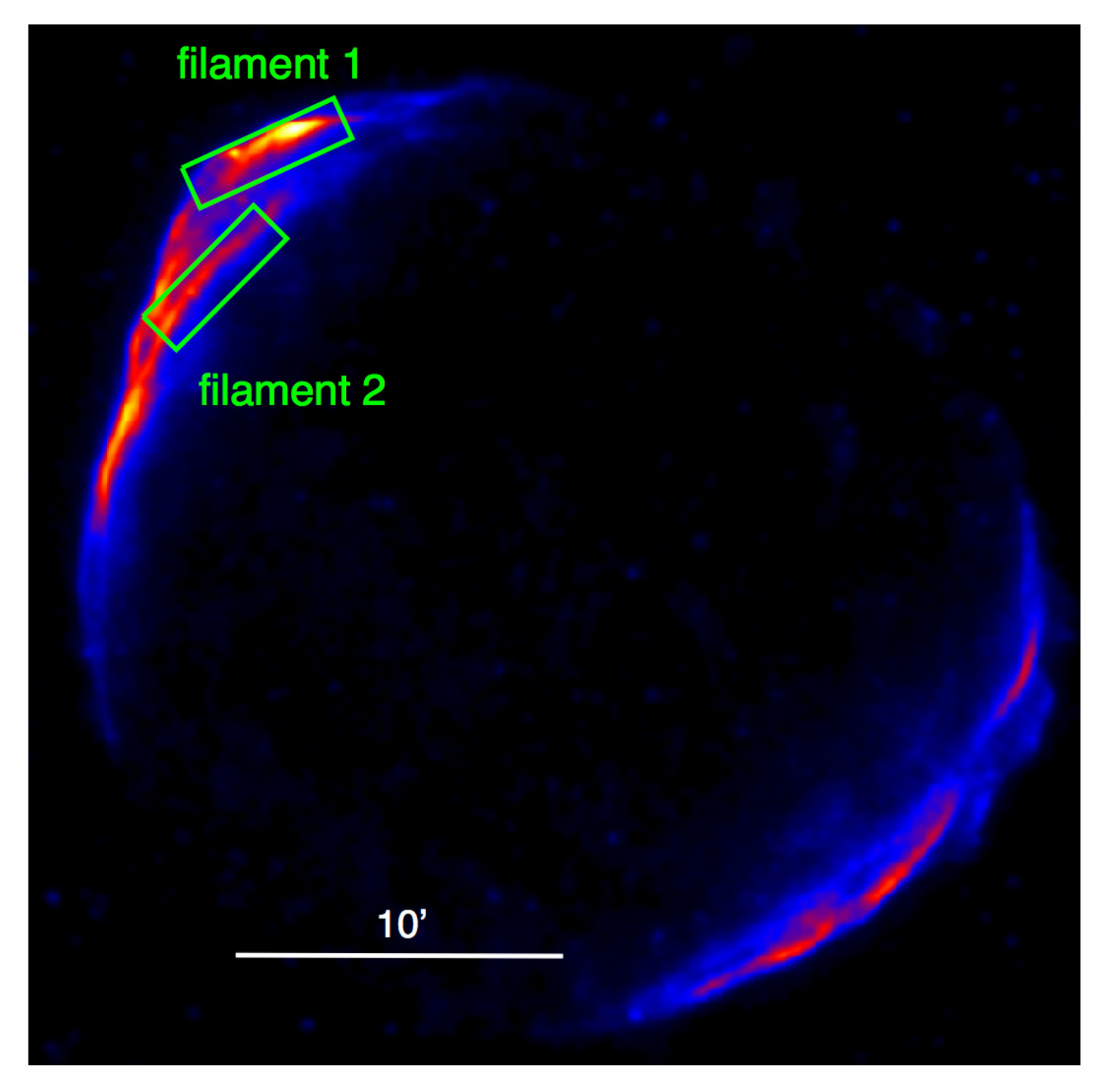}
\includegraphics[width=0.65\textwidth]{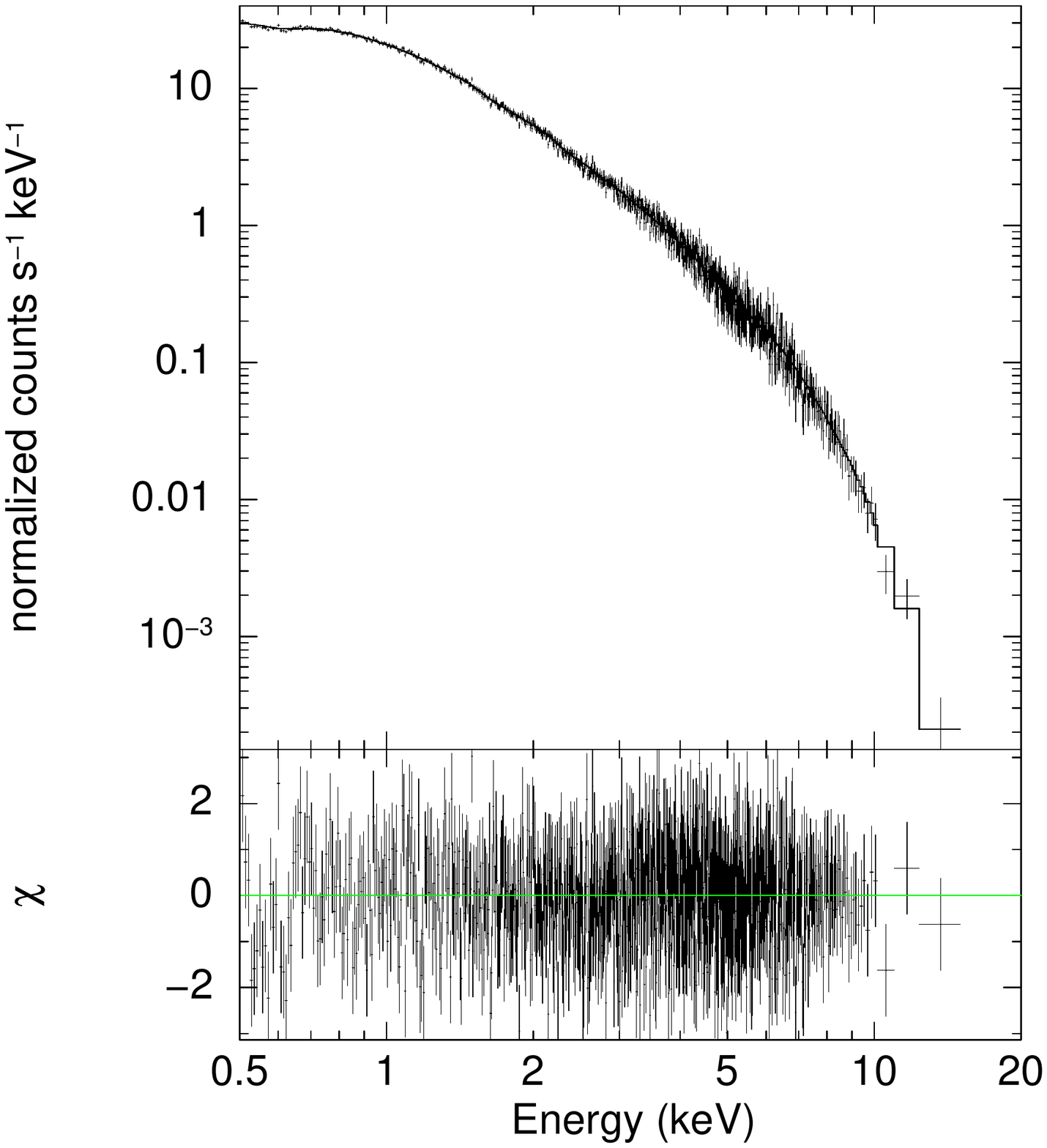}
\caption{ (Left) The 2--8 keV image of SN1006 convolved with the \gpd
  angular resolution (15\arcsec). For the boxes around filaments 1 and
  2, the polarization can be measured with accuracies of $\sigma_{\rm
    dop}=1.8$\% and 2.1\% and $\sigma_{\rm PA}=2.9$ and 3.2~deg for an
  exposure time of 0.5~Ms, respectively. (Right) Simulated \lfa
  spectrum of region 'filament 1' in the northeast of SN1006 with a
  10~ks observation. A model is employed of synchrotron radiation from
  an exponentially cutoff power-law distribution of electrons with the
  parameters hydrogen absorption column density $N_{\rm
    H}=6.8\times10^{20}$~cm$^{-2}$, power-law photon index of -0.57
  and cut-off energy of 0.5 keV.
\label{figsimsnr}}
\end{figure*} 

In particular, \extp will be powerful in addressing the following
questions about SNRs and PWNe:

\begin{itemize}

\item \textit{What effect does the orientation of injection with
  respect to the direction of the magnetic field have on the shock
  acceleration efficiency in SNRs?} In the acceleration mechanisms, it
  remains unclear how the orientation of injection with respect to the
  magnetic lines affects the acceleration efficiency
  \citep[``quasi-parallel scenario'' or ``quasi-perpendicular
    scenario'';][]{fulbright1990}. SN1006, a SNR with a bilateral-like
  non-thermal X-ray morphology, is an ideal example to study this
  problem. The most recent radio polarization observation favors the
  case of quasi-parallel injection \citep{reynoso2013}. \extp can
  resolve the X-ray structures in the SNR (30' in diameter) and
  provide spatially resolved physical information for them. The \gpd
  polarimeter can measure the polarization degree and angle in the
  post-shock region of SN1006 with 15'' angular resolution (see
  Fig.~\ref{figsimsnr}-left). The measurement accuracy will be about
  2\% for the polarization degree ($\sigma_{\rm dop}$; assuming a
  polarization degree of 17\%) and 3\degr\ for the position angle
  ($\sigma_{\rm PA}$) after a 0.5~Ms observation of filaments 1 and 2.
  With a 10 ks observation, the \lfa can obtain spatially resolved
  spectra of the non-thermal emission of the shell (see
  Fig.~\ref{figsimsnr}-right for the simulated spectra of 'filament 1'
  in the northeast of the SNR). The \extp polarization measurement and
  spectral analysis are thus expected to give a precise answer to a
  specific physical problem.

\item \textit{What is the intensity of the synchrotron radiation and
  the orientation of the magnetic field in the forward shock and,
  especially, the reverse shock in SNRs?} While CR acceleration has
  been extensively studied at the forward shocks of SNRs, how the
  reverse shocks accelerate CRs became an intriguing question in
  recent years. Using a 1~Ms \textit{Chandra} observation towards Cas
  A, most of the non-thermal emission from the SNR was found in the
  reverse shock \citep{helder2008,uchiyama2008}. \extp will be able to
  test this result and compare the polarization parameters between the
  forward and reverse shocks. With an exposure time of 2~Ms, the \gpd
  polarimeter can measure the polarization degree and angle of
  0.5\arcmin$\times$1\arcmin\ sub-structures in Cas A (4.2-6~keV) with
  accuracies of $\sigma_{\rm dop}$$\sim$3\% and $\sigma_{\rm
    PA}$$\sim$10\degr\ for the forward shock (assuming a polarization
  degree of 10\%), and $\sigma_{\rm dop}$$\sim$1.5\% and $\sigma_{\rm
    PA}$$\sim$8\degr\ for the reverse shock (assuming a polarization
  degree of 5\%).  The spatially resolved polarization measurements
  with \extp will provide crucial information on the magnetic field at
  the reverse shock where CRs are effectively accelerated.

\item \textit{What is the spatial distribution of the non-thermal
  X-ray emitting components with respect to that of the thermally
  emitting regions in SNRs?} Since the first detection in SN1006
  \citep{koyama1995}, non-thermal X-ray emission has been detected at
  the rims of several SNRs. This is believed to be due to synchrotron
  radiation of the CR electrons accelerated by the SNR shocks
  \citep[e.g.][]{ballet2006}.

  In addition to the non-thermal X-ray dominant SNRs (such as SN1006
  and RX J1713-3946) and a few young SNRs (such as Cas A and Tycho) in
  which non-thermal and thermal X-ray emission can be
  spatially/spectroscopically resolved, there are several SNRs in
  which potential non-thermal X-ray emission can hardly be
  distinguished from the thermal emission. These SNRs include N132D,
  G32.4+0.1, G28.6-0.1, CTB 37B, W28, G156.2+5.7,
  etc. \citep[e.g.][]{nakamura2012}. Thanks to the large collecting
  area of \extp in the hard X-ray band, short exposures of \lfa
  towards these SNRs can collect enough photons for spectral analysis
  to clarify whether the hard X-ray emission is non-thermal. The
  enlarged sample of non-thermal X-ray emitting SNRs in various
  environments and at different ages will allow one to study the
  mechanism of CR electron acceleration and to constrain the leptonic
  emission, acceleration efficiency, magnetic field and other
  interstellar conditions.

\item \textit{Which are the mechanisms of amplification of the
  magnetic field at the SNR shocks?} Several high-energy observations
  of SNRs are consistent with a magnetic field at the shock far
  exceeding the theoretically predicted shock-compressed strength. The
  joint capability of \extp to measure X-ray emission and polarization
  will allow one to test current models of amplification of magnetic
  fields by SNR shocks. It is believed that such shocks propagate into
  a generally turbulent interstellar medium \citep{armstrong95}. The
  turbulence can be pre-existing or caused by the shock itself via,
  e.g., a non-resonant streaming instability of cosmic-rays as first
  found by \citet{bell04}. In the former case, the density
  inhomogeneities traversed by the shock front are expected to produce
  corrugation of the shock surface; such a surface, as it propagates
  further, triggers differential rotation of over-densities during
  their shock-crossing generating vorticity, therefore amplifying the
  magnetic field via a small-scale dynamo process in the downstream
  fluid \citep{giacalone07,inoue12,fraschetti13}.  The field growth is
  exponentially fast, as short as years or months \citep[see, e.g.,
    the rapid changes in hot spots brightness observed in the SNR
    $RXJ1713.7-3496$ and in Cas A;][]{uchiyama07,patnaude09}, likely
  due to synchrotron electron cooling in a strong magnetic field.

  \extp's polarimetric capability will be a valuable asset here in the
  sense that it will provide, for the first time, the polarization in
  those hot spots, leading to estimates of the coherence scale of the
  magnetic turbulence. This will be of the order of the Field length
  \citep[i.e., about 0.01 pc;][]{field65} if the vortical mechanism
  dominates. Indeed, the growth time scale of the magnetic turbulence
  is approximately equal to the ratio of the Field length to the local
  shock speed. As described above, a 1 Ms exposure on Cas A would
  allow one to measure the polarization degree (angle) with an
  accuracy better than $1\%$ (3\degr) for the forward shock.  The
  proposed analysis requires relatively high SNR shock speeds (not too
  high to have a growth time at least of a few months) colliding with
  dense clouds or propagating into a region with significant
  clumpiness. By observing the X-ray emission during the early phases
  of the shock/cloud collision, \extp will allow one to constrain the
  growth time scale of the magnetic field.

\item \textit{What is the configuration of the magnetic field in PWNe
  and what physical insight can we infer from that?} PWNe, such as the
  Crab nebula, are high-energy nebulae composed of relativistic
  particles and magnetic fields created by the central pulsars. The
  energy input of pulsars, the particle acceleration at the terminal
  shocks, the PWN structures (including the equatorial ring and
  bipolar jets) and other physical properties (such as particle
  transport and magnetization factor) are not fully understood yet
  \citep[e.g.,][]{reynolds2017,kargaltsev2015}. \extp can perform
  X-ray polarization measurements towards the Crab nebula with high
  spatial resolution, and obtain the detailed distribution of magnetic
  field and polarization degree. The \gpd polarimeter can measure the
  polarization degree and angle of the sub-structures (15\arcsec) in
  the Crab nebula with a mean accuracy of about 1.7\% ($\sigma_{\rm
    dop}$, assuming an polarization degree of 19\%) and
  2.6\degr\ ($\sigma_{\rm PA}$), respectively (we considered an
  exposure time of 10~ks).  The unprecedented X-ray polarization
  imaging of the Crab nebula will play an important role in the study
  of PWNe.  Another PWN worth following up with \extp is MSH~15--52
  which has a size of 10\arcmin\ and a total flux of
  $\sim5.5\times10^{-11}$~erg~s$^{-1}$cm$^{-2}$. \extp mapping will
  provide polarization accuracies of $\sigma_{\rm dop}$=2\% and
  $\sigma_{\rm PA}$=3.7\degr\ for 10 pointings with a 300 ks exposure
  (assuming a photon index of -2 and polarization degree of 15\%).

\end{itemize}

\section{Accreting White Dwarf Binaries}

White dwarfs (WDs) accreting from a companion star in a binary exhibit
themselves as cataclysmic variables \citep[CVs; e.g., review
  by][]{knigge2011}. CVs represent the most common end products of
close binary evolution.  These binaries display a wide range of time
scales in variability (from seconds-minutes up to months-years) that
make them ideal laboratories for the study of accretion/ejection
processes in the non-relativistic regime.  Disk instabilities, the
disk-jet connection and magnetic accretion are common processes
occurring in a wide variety of astrophysical objects. Accreting WDs
provide an excellent perspective of these phenomena.  Furthermore,
thermonuclear runaways giving rise to novae \citep[e.g., review
  by][]{jose2016} have the potential to probe the accretion/ignition
mechanisms, outflows and subsequent chemical enrichment of the
interstellar medium, as well as to address the question of whether
accretion dominates over outflow, possibly bringing the WD mass over
the Chandrasekhar limit yielding type Ia supernovae. This is a key
aspect for understanding Type Ia supernova progenitors.  In addition,
population studies of galactic X-ray sources have recently recognized
a crucial role of accreting WD binaries in the origin of the Galactic
Ridge X-ray Emission (GRXE), one of the great mysteries in X-rays
\citep{Warwick14,perez2015,hailey2016,hong2016,koyama2017}.

\begin{figure*}[ht!]
\center
\includegraphics[width=0.6\textwidth,angle=0]{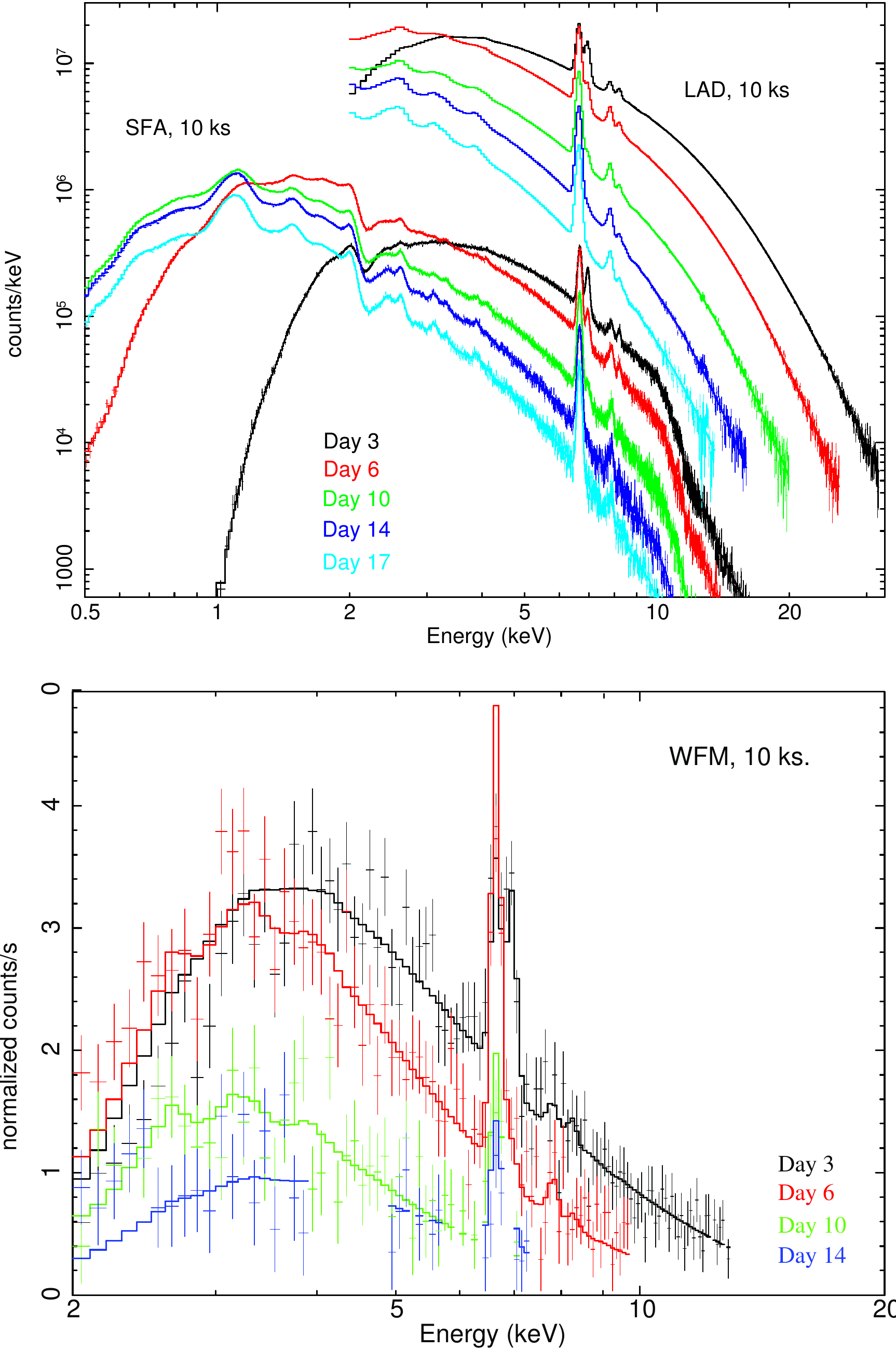}
\caption{Simulations of the spectral evolution of the symbiotic
  recurrent nova RS Oph over its outburst, as seen by the \lad and
  \lfa (upper panel) and by the \wfm (lower panel).  The spectra
  include a thermal plasma model ({\it mekal}) with solar abundances,
  evolving from kT= 8.2 keV and a flux of $\rm
  3.2\times10^{-9}$~\cgsflux\, (0.5-20 keV) on day 3, down to kT= 2.5
  keV and a flux of $\rm 4.3\times10^{-10}$~\cgsflux\, (0.5--20 keV)
  on day 17 \citep[after][]{Sokoloski2006}.  The high X-ray flux
  during the first 17 days would trigger the \wfm in full resolution
  mode and a detailed study of the spectral evolution will be possible
  with the \lad and \lfa .}\label{figRsOph}
\end{figure*}

The X-ray domain is crucial to understand the physical conditions
close to the compact star \citep[e.g., review
  by][]{kuulkers2006}. Progress in many open questions is hampered by
the low quiescent luminosities ($\rm L_{X}\sim
10^{29-33}\,{erg}\,{s}^{-1}$) and the unpredictability of large-scale
variations that can reach luminosities as low as $\rm \sim
10^{32}-10^{34}\,{erg}\,{s}^{-1}$ in dwarf novae (DNe) outbursts and
as high as $\sim 10^{38}\,\mathrm{erg}\,\mathrm{s}^{-1}$ in nova
explosions. The majority of persistent, steadily nuclear-burning
accreting white dwarfs thought to exist in the Milky Way, its
satellites, and M 31 have yet to be detected
\citep{distefano94,woods2016}. These ``supersoft X-ray sources'' are
candidate Type Ia supernova progenitors, and can exhibit oscillations
on very short timescales ($\lesssim$ 100 s) for as yet unknown reasons
\citep{Ness15}.

The unique combination in \extp of large effective area, fair spectral
resolution and excellent sensitivity of the \lad and \lfa instruments,
as well as the wide field of the \wfm instrument, can uncover the wide
range of variabilities to understand better the underlying physical
processes.  The three main questions that can be addressed are:

\begin{itemize}

\item
{\it How does matter accrete onto white dwarfs and mix with the CO/CNe
  substrate?}  Fast (seconds to minutes) X-ray variability, both
periodic and aperiodic, allow one to trace the accretion flow close to
the WD surface. In magnetic CVs, periodic coherent pulsations are a
signature of magnetically funneled accretion and their spectral study
allows one to infer the physical conditions of the post-shock region,
the hot spot over the WD surface as well as complex absorption from
pre-shock material.  Furthermore, quasi-periodic oscillations (QPOs)
at a mean period of a few seconds are expected to arise from shock
oscillations \citep{Wu2000}, but so far these were only detected in a
handful of systems at optical wavelengths.  The high shock
temperatures ($\sim$20-30\,keV) imply that QPO searches need to be
extended into the hard X-rays.

In non-magnetic CVs, when accreting at high rates, a variety of low
amplitude variability has been detected. DN oscillations with a period
of a few seconds were found in the optical and soft X-rays, changing
frequency during outburst. Slower (with a time scale of minutes)
low-coherence QPOs were detected in a few systems in the hard X-rays
during the decline \citep{Warner04}.  Their appearance is related to
the transition of the boundary layer from optically thick to optically
thin and thus have a great potential to infer details of boundary
layer regimes.  Their frequencies stay in ratio, like the low and high
frequency QPOs seen in low-mass X-ray binaries.  Furthermore, the
presence of a break in the power spectra during quiescence can allow
one to infer disk truncation \citep{Balman12}. The striking
similarities with X-ray binaries make CVs interesting for a unified
accretion scenario.

Peak temperatures reached during a nova explosion are constrained by
the chemical abundance pattern inferred from the ejecta and do not
seem to exceed $4 \times 10^8$ K, so it is unlikely that the large
metallicities reported from the ejecta can be due to thermonuclear
processes. Instead, mixing at the core-envelope interface is a more
likely explanation. The true mixing mechanism has remained elusive for
decades. Multidimensional models of the explosion suggest that mixing
may result from Kelvin-Helmholtz instabilities at the core-envelope
interface \citep[e.g.,][]{casanova2011}.

\extp can address all this by studying selected samples of magnetic
and non-magnetic systems, to detect for the first time low-amplitude
fast aperiodic and periodic variabilities over the broad energy range
offered by the \lad and \lfa and to perform time-resolved
spectroscopy.

\item 

{\it How does mass ejection work in nova explosions?}  Novae are,
after thermonuclear supernovae, the second most luminous X-ray sources
among accreting WDs with emission ranging from the very soft to the
hard X-rays. The soft X-rays probe the continuous nuclear burning on
the WD surface, while the hard X-rays are believed to trace the early
shocks inside the ejecta and between the ejecta and the circumstellar
material, as well as the onset of resumed accretion. The details of
the mass outflow and the shaping and evolution of the ejected matter
are still poorly known.  A recent challenge is the unprecedented
detection by {\it Fermi} LAT of high-energy gamma rays ($E>100$ MeV)
in an increasing number of novae \citep{Ackermann14,cheung2016}
indicating that these systems are the site of particle
acceleration. This was predicted for RS Oph \citep{Tatischeff07}, a
symbiotic recurrent nova where the early strong shock between the nova
ejecta and the red giant wind was traced in hard X-rays by RXTE
\citep{Sokoloski2006}. In classical novae there is not a red giant
wind to interact with the nova mass outflow, because the companion is
a main sequence star; therefore, particle acceleration is more
challenging to explain \citep{metzger2015}. \extp prompt observations
of nova explosions would be crucial to better understand such
processes in both nova types (see Fig.~\ref{figRsOph}).

It is very important to observe over the broadest X-ray range
the spectral evolution of these explosions from the early hard X-ray
onset to the later super-soft X-ray phase
\citep{Schwarz2011,Sokoloski2006}.  Furthermore X-ray variability and
periodicities on many different time scales, from days to less than a
minute are still to be understood. Short oscillations in the soft
X-rays can originate from H-burning instabilities \citep{Osborne11}
while those in the hard X-rays still remain to be explored.

Thanks to the \wfm,\ \extp will be able to catch the early X-ray
emission from bright novae to be followed-up by the \lfa and \lad
instruments to monitor over an unprecedented energy range the X-ray
spectral evolution including the onset of hard X-ray emission and of
the super-soft X-ray phase as well as to characterize for the first
time fast variability over \extp's unique broad range.  An example of
the \extp potential is shown in Fig.~\ref{figRsOph}.

\item

{\it What causes DN outburst diversity and what are the conditions for
  disk-jet launching?}  DN outbursts are believed to be due to disk
instabilities but there is still lack of knowledge of what changes
occur at the inner boundary layer. The boundary layer is optically
thin during quiescence and becomes optically thick during outburst
\citep{wheatley2003}. Hard X-ray suppression during optical outbursts
was believed to be a general behavior.  Instead, the few DNe observed
so far showed great diversity and not all of them show X-ray/optical
anticorrelation \citep{Fertig2011}.  The understanding of what
fundamental parameters divide the different regimes of the boundary
layer is crucial.  A further challenge is the detection of radio
emission from a DN and from high mass accretion rate CVs
\citep{Koerding2008,Koerding2011,coppejans2016} interpreted as the
presence of a radio jet. CVs were believed to be unable to launch
jets. It was recently shown that nova-like CVs can have optically thin
inefficient accretion disk boundary layers \citep{Balman14}, allowing
the possibility of retaining sufficient energy to power jets.  If this
were the case, a radio/X-ray correlation, similar to that observed in
neutron star and black hole binaries \citep{Coriat11}, should be
present in these systems.
 
\extp\ can address these challenging issues by following a few DNe
with adequate multi-outburst coverage through time-resolved
spectroscopy and fast timing in unprecedented detail.
\end{itemize}

\vspace{5mm}\noindent
Thus \extp will be crucial to measure the spectral and temporal
properties of the poorly known hard X-ray tails in these systems and
to correlate them with the soft X-ray emission.  When \extp is
expected to be operational, wide field area surveys will have provided
statistically significant samples to allow detailed investigation in
coordination with ground-based optical and radio facilities foreseen
in the post-2020 time frame (see Fig.~\ref{figfacilities}).

\section{Binary evolution}
\label{sec:lmxbs}


\begin{figure*}[t]
\centering
\vspace{0.cm}
\includegraphics[width=0.47\textwidth,angle=0]{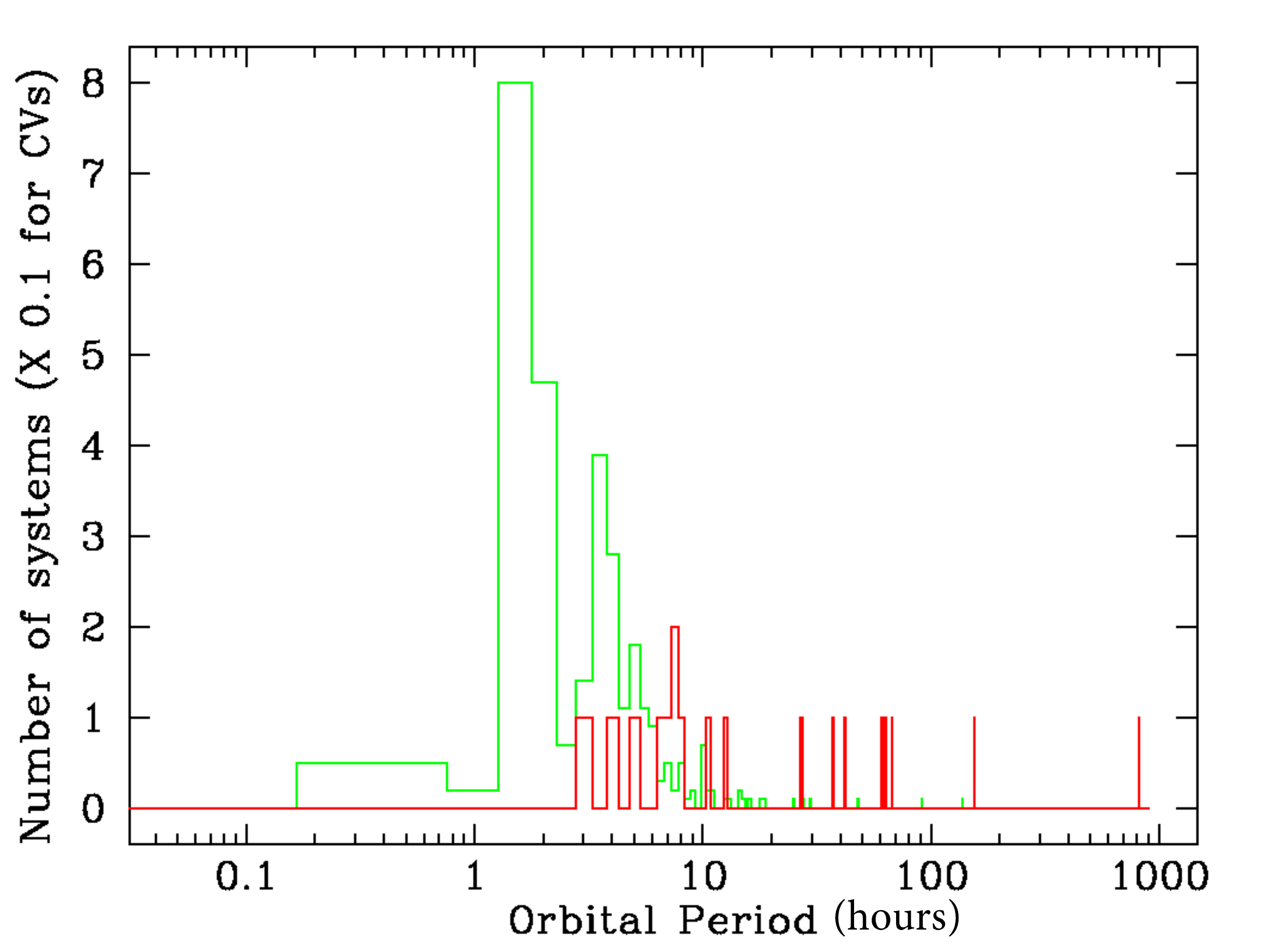}
\includegraphics[width=0.47\textwidth,angle=0]{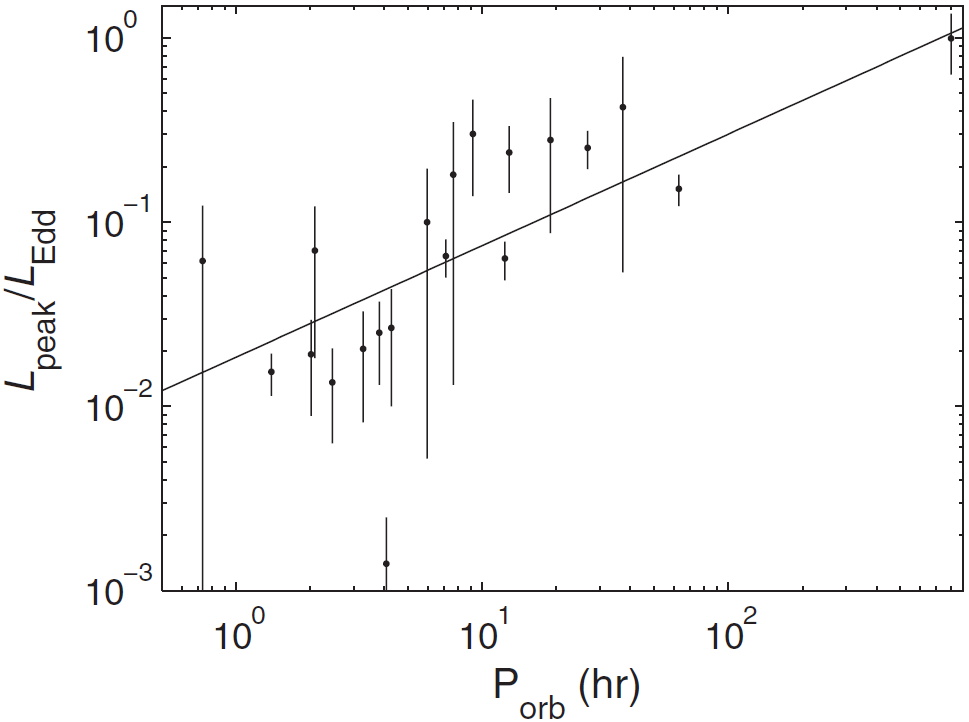}
\caption{\textit{(Left)} The orbital period distributions of
  dynamically confirmed black hole X-ray binaries (red curve), and
  cataclysmic variables (green curve, with the numbers multiplied by
  0.1 to fit the two distributions well on the same scale), taken from
  the Ritter catalog \citep{ritter2003}. It is apparent that the
  distributions are very different from one another.  \cite{knevitt14}
  find a probability of $0.0012$ that these two distributions are
  drawn from the same parent distribution.  \textit{(Right)} Peak
  luminosity plotted versus orbital period for the black hole X-ray
  transients seen with RXTE, with a linear fit. Reproduced by
  permission of the AAS from \citet{wu2010}.  Similar relations are
  also seen for cataclysmic variables \citep{warner1987,wpmaccarone}
  and accreting neutron stars \citep{wu2010}.  }\label{figMaccarone1}
\end{figure*} 

The evolution of binary stars is one of the most important and
challenging problems in modern stellar astronomy. Binary evolution --
even when restricted merely to the evolution of binaries with compact
objects -- touches on a range of key questions
\citep{bhattacharyya1991,han08,tauris2006,tauris2017}: the formation
of Type Ia supernovae, the formation of merging neutron stars and
black holes, the rate of production of millisecond pulsars, and the
production rate of a variety of classes of stars with abundance
anomalies. The challenges in understanding the populations of binary
stars stem from the wide range of physics that goes into them. In
addition to normal stellar evolutionary processes, the following must
be understood: the stability and efficiency of mass transfer, the
common envelope stage, the tides and angular momentum loss mechanisms
that keep systems tight (i.e., magnetic braking and circumbinary
disks), the spin-up of the accreting star, irradiation effects on the
donor star, the details of the supernova explosions of (ultra)stripped
stars, the kicks applied to neutron stars and black holes at birth,
the spin-down due to the propeller effect, possible gravitational wave
emission mechanisms, and the initial parameter distributions of binary
and multiple star systems.

Proper testing of theories of binary evolution must come from
exploration of substantial samples of close binaries, and good
estimates of their system parameters. Because most X-ray binaries,
especially with black hole primaries, are transients, the ideal way to
detect such objects is with all-sky monitoring. The \wfm 
represents a major step forward in the capabilities of all-sky
monitors relative to present monitors.

\extp will make essential contributions to several major aspects of
compact binary populations: the very faint X-ray transient problem
\citep{cornelisse2002,degenaar2010,king06}; the evolutionary link
between low-mass X-ray binaries and radio millisecond pulsars; the
orbital period distribution of X-ray binaries, by finding the short
period systems selected against by shallower all sky monitors
\citep{wu2010,knevitt14}; the ``mass gap'' between black holes and
neutron stars that has been tentatively observed in existing data
\citep{ozel10,kreidberg12,wang16} and not observed in microlensing
data \citep{wyrykowski2016}; understanding the formation of
millisecond pulsars in compact binaries \citep{roberts2013} as well as
the details of their spin evolution \citep{haskell17}.

\begin{figure*}[t!]
\centering
\includegraphics[trim=0 0cm 0 0cm,width=1.\textwidth,angle=0]{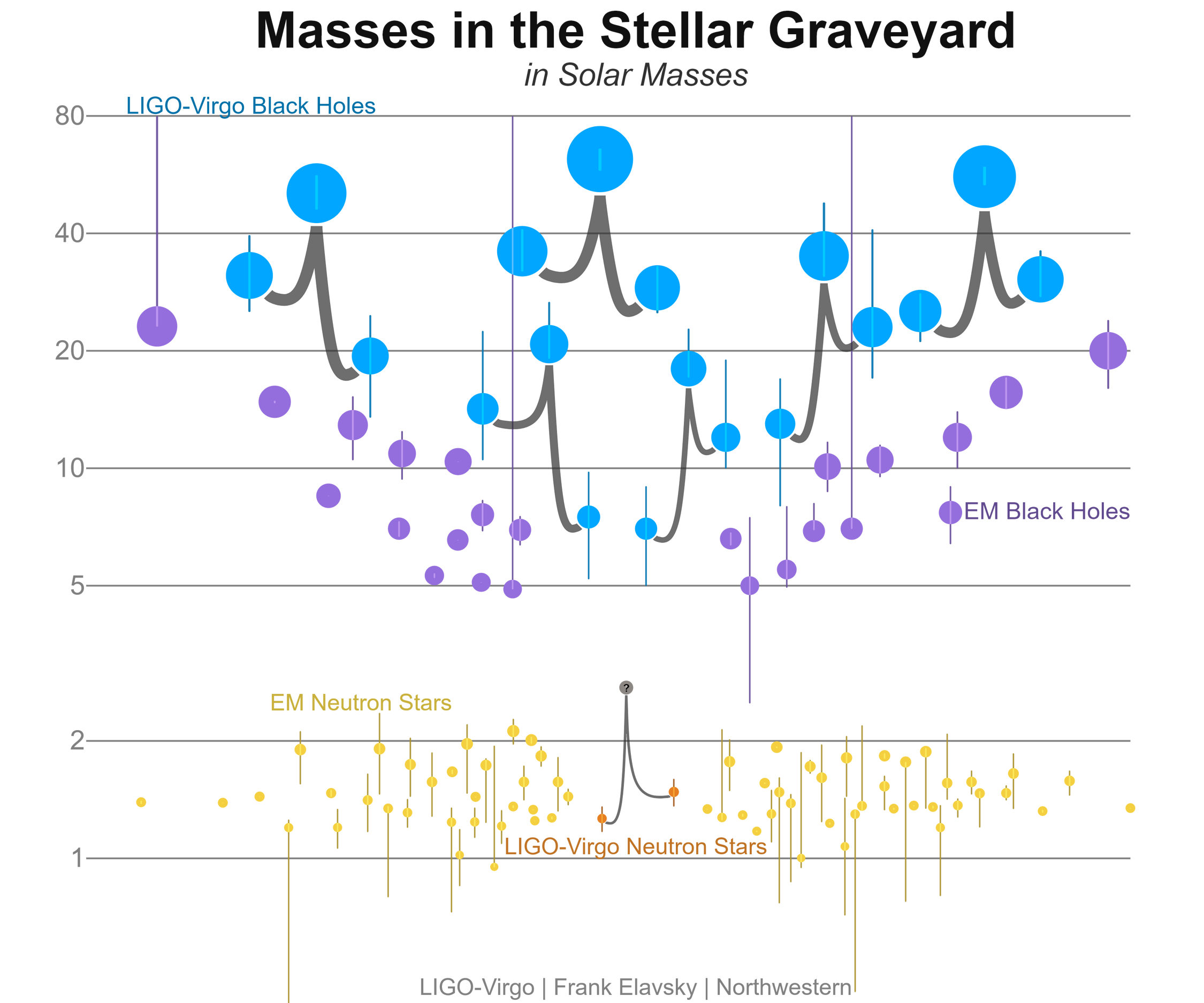}
\caption{Graphic showing neutron star and black hole mass measurements
  along the vertical axes (in units of solar mass)
  \citep[from][]{wiktorowicz2014,oezel2016,casares2016}, including the
  16 black holes and 2 probable neutron stars in GW150914, LVT151012,
  GW151226, GW170104, GW170608, GW170814, GW170817
  \citep{abbott1,abbott2,abbottlvt,abbott3,abbott4,gw170817,gw170608}. The
  GW measurements are indicated with curved line pointing to the
  merger event. There is a clear gap between 2 and 5 solar
  masses. Credit: Frank Elavsky, LIGO-Virgo, Northwestern
  University. \label{figMaccarone2}}
\end{figure*} 

\begin{figure*}[ht!]
\centering
\includegraphics[width=0.49\textwidth,angle=0]{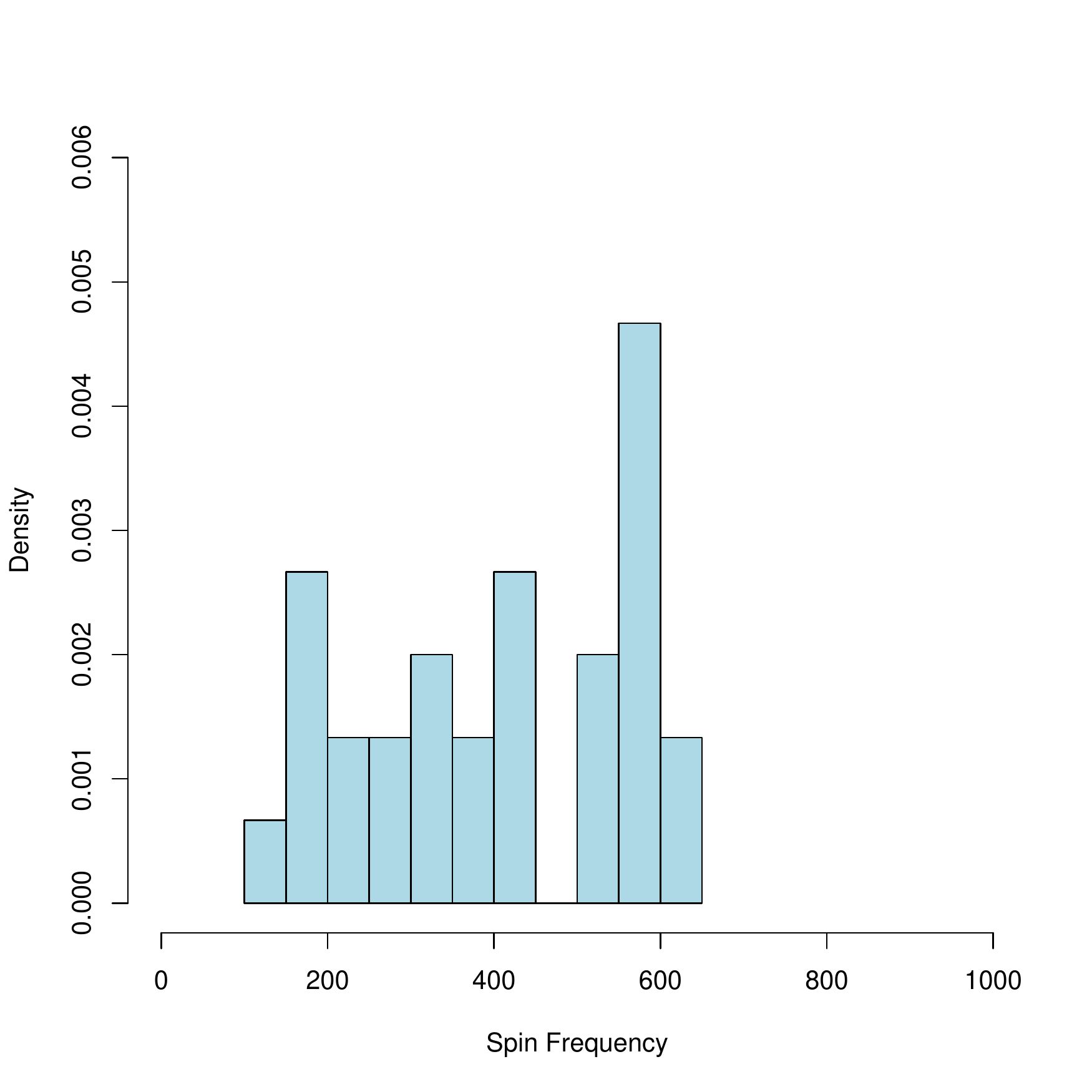}
\caption{The spin distribution for the fastest known ($\nu>100$~Hz)
  accreting neutron stars \citep[adapted from][]{patruno2017}.
}\label{figMaccarone1b}
\end{figure*} 

\extp will present a range of new opportunities to study binary
stellar evolution processes. These are:
\begin{itemize}
 
\item \textit{What are the characteristics of the population of Very
  Faint X-ray Binaries?} The \wfm will have the necessary sensitivity
  and large field of view to detect flares and outbursts from the Very
  Faint X-ray Binaries with luminosities below 10$^{36}$~erg~s$^{-1}$,
  allowing a census of these sources across a large fraction of the
  Galaxy. Furthermore, the \wfm's large field of view will easily pick
  up the rare X-ray bursts (see Sect.~\ref{sec:xraybursts}) that very
  faint X-ray transients exhibit if the accretor is a neutron star
  \citet{cornelisse2002}. By establishing a substantial sample of such
  objects through the wide field coverage and good sensitivity of the
  \wfm \citep{wpmaccarone}, it should be possible to engage optical
  follow-up on enough of them to uncover their nature, which is still
  poorly understood.

\item \textit{What is the evolutionary link between low-mass X-ray
  binaries and radio millisecond pulsars?}  \extp will provide
  detailed timing studies of known low-mass and ultracompact X-ray
  binaries \citep[with an orbital period of less than $\sim$1 hr,
    e.g.][]{heinke13,vanhaaften12} which will help us to better
  understand their evolutionary link to radio millisecond pulsars,
  such as the so-called redbacks and black widow systems
  \citep{chen13}.

  A particular class of objects of interest for \extp studies is that
  of the transitional millisecond pulsars. These represent the most
  direct evidence for an evolutionary link between radio millisecond
  pulsars and low-mass X-ray binaries, since they are observed to
  transit between both states. Three, possibly four, are currently
  known \citep{archibald2010,papitto2013,bassa2014,bogdanov2015}. Two
  of these have relatively low luminosities in their X-ray states of
  10$^{33-34}$~erg~s$^{-1}$ while the third had an ordinary X-ray
  outburst as an accretion-powered millisecond pulsar peaking at
  10$^{36}$ erg~s$^{-1}$. There may be many more of these
  systems. Detecting them in larger numbers, and mapping out the
  statistical properties (e.g., pulse and orbital periods) of the
  resulting sample will enable a better understanding of the binary
  evolution. Good candidates may be found among the so-called
  redbacks, which are radio millisecond pulsars in binaries with an
  ablated companion star of mass around 0.1 M$_\odot$. Observations
  more sensitive than with XMM-Newton and Chandra are needed to study
  the common flaring behavior of these systems, for instance how the
  pulsar behaves on short time scales, to understand better the
  process behind switching between X-ray and radio states. \extp would
  make this observational progress possible.
  
  It is worth mentioning the puzzling fact that fully recycled radio
  millisecond pulsars are seen in systems with orbital periods up to
  200 days, but only 3 pulsar-containing low-mass X-ray binaries are
  known with orbital periods longer than 1~day and the neutron stars
  hosted are relatively slow pulsars (0.5-4.0 sec). \extp will be able
  to probe pulsations down to the ms range with an unprecedented
  sensitivity, hopefully allowing one to reveal the currently missing
  fast rotators.

\item \textit{What is the heaviest neutron star and the lightest black
  hole?} \extp will allow one to determine whether the orbital period
  distribution of black hole X-ray binaries follows the predictions of
  binary evolution. At the present time, the severe selection biases
  against black hole X-ray binaries with orbital periods less than
  about 4\,hours (Fig.~\ref{figMaccarone1}-left) lead to a deficit of what
  is likely to be the largest group of systems. Such systems are
  likely fainter, see Fig.~\ref{figMaccarone1}-right. Possibly, these contain the
  lighter black holes, especially if one includes high-mass X-ray
  binaries \citep[e.g.,][]{lee2002}. By enhancing the sample of
  objects for which masses can be estimated, combined with much more
  sensitive upcoming optical telescopes E-ELT and TMT to obtain
  dynamical mass estimates, \extp will also help verify or refute the
  ``mass gap'' between neutron stars and black holes
  (Fig.~\ref{figMaccarone2}) which has profound implications for
  understanding how supernovae actually explode.

\item \textit{How efficient is the process of neutron star spin up?}
  The process of spin-up of neutron stars in binaries
  (Fig.~\ref{figMaccarone1b}) can be probed in two ways: by using \lad
  measurements to expand the sample of neutron stars with good spin
  period measurements and by using the combination of \wfm and \lad to
  discover and study the population of objects that are in the act of
  transitioning from low mass X-ray binaries into millisecond radio
  pulsars.The study of accreting neutron stars and the torques acting
  on them is important for probing disk-magnetosphere interactions
  \citep{rappaport04,bozzo09,tauris2012,dangelo17} as well as
  investigating continuous GW emission from the spin of asymmetric
  neutron stars and the prospects of detecting these sources using
  second (Advanced LIGO and Virgo, KAGRA, LIGO-India) or
  third-generation \citep[Einstein Telescope in Europe, Cosmic
    Explorer in USA][]{abbott2015} GW detectors
  \citep{haskell15,haskell17}.  Among the promising sources for dual
  observations in both X-rays and gravitational waves is, for example,
  Sco~X-1, provided its neutron star spin rate is detected
  \citep{watts08,ligosco1,ligosco2} which is where \extp can help
  through the detection of a weak pulsation. Recent calculations
  \citep{bhattacharyya2017} suggest that transiently accreting
  millisecond pulsars are promising as well.

\item \textit{How can \extp help find sources for LISA?}  The very
  tight X-ray binaries with orbital periods of less than one hour, the
  ultracompact X-ray binaries, are ideal sources to be detected in GWs
  induced by the orbital motion by the upcoming space-borne GW
  detector LISA \citep{nelemans01} and LISA-like missions like TianQin
  \citep{luo2016} and TAIJI \citep{gong2011}, missions with envisaged
  launch dates in the first half of the 2030s. It is expected that
  \extp will discover a number of new X-ray binaries and may thus find
  additional ultracompact X-ray binaries (neutron star / white dwarf
  or double white dwarf binaries) that will provide important and
  guaranteed sources of GWs for detection within the frequency band
  ($\sim$mHz) of LISA and LISA-like missions.

\end{itemize}

\section{Thermonuclear X-ray bursts}
\label{sec:xraybursts}

While thermonuclear ('type-I') X-ray bursts are well understood as a
thermonuclear runaway in the upper freshly accreted layers of a
neutron star in a low-mass X-ray binary, many important questions are
unanswered \citep[e.g.,][see reviews
  by]{lewin1993,strohmayerbildsten2006,wpintzand,jose2016,
  galloway2017}. Some of those are fundamental, pertaining to the
structure of the neutron star and exotic nuclear processes, while
others are related to interesting physical phenomena such as neutron
star spins, convection and radiation transport under strong-gravity
circumstances, the geometry of accretion and magnetic field, and
unusual stellar abundances. Since \extp is particularly well suited to
observe short bright phenomena, it will be able to answer many of
these questions.

About 20\% of all bursts exhibit oscillations with a frequency very
near that of the neutron stars spin (11-620 Hz). The oscillations
during the burst rise can be explained by a hot spot expanding from
the point of ignition
\citep{strohmayer1997,spitkovsky2002,bhattacharyya2006, cavecchi2013,
  chakraborty2014,cavecchi2015}, although this still requires further
study.  The oscillations during the tail have not yet been
explained. They could be caused by non-uniform emission during the
cooling phase \citep[e.g., because of different depths of fuel over
  the surface or cooling wakes;][]{zhang2013}, but an interesting
alternative involves global surface modes that may be excited by the
motion of the deflagration front across the surface. This was first
suggested by \citet{heyl2004}, and further theoretical work was
performed by (among others) \citet{heyl2005}, \citet{cumming2005},
\citet{lee2005}, \citet{piro2005} and \citet{berkhout2008}.

The primary nuclear processes in thermonuclear bursts are the CNO
cycle to burn hydrogen ($\beta$-limited 'hot' CNO cycle above
$\sim8\times10^7$\,K), the triple-$\alpha$ process to burn helium
(above several $10^8$\,K), the $\alpha$p-process above
$5\times10^8$\,K to produce elements like Ne, Na and Mg, and the rapid
proton capture process (`rp-process', above $10^9$\,K) to burn
hydrogen into even heavier elements
\citep[e.g.,][]{fujimoto1981,wallace1981,woosley2004,fisker2008,jose2010}. In
cases of hydrogen-free helium accretion or where the $\beta$-limited
CNO cycle burns all hydrogen to helium before ignition, bursts are due
to ignition in the helium layer.  If the CNO burning is not
$\beta$-limited, it can be unstable leading to a temperature increase
that subsequently ignites helium. The ignition of helium may be
delayed in that regime \citep{peng2007,cooper2007}, resulting in pure
hydrogen flashes. If the CNO burning is $\beta$-limited and this
burning has no time to burn away the hydrogen, helium ignition occurs
in the presence of hydrogen, a situation that leads to the
rp-process. Hundreds of proton-rich isotopes are produced during
thermonuclear burning, particularly by the rp-process, yielding rare
isotopes, whose reaction rates have not yet been quantified in
laboratories \citep{cyburt2016}.

In the past one and a half decades, two new kinds of X-ray bursts have
been discovered that are long and rare: `superbursts'
\citep{cornelisse2000a,kuulkers2002b,strohmayer2002a} and
`intermediate duration bursts' \citep{zand2005,cumming2006a} that
ignite at $10^2$ to $10^4$ larger column depths than ordinary
bursts. It has been proposed that superbursts are fueled by carbon
\citep{cumming2001a,strohmayer2002a}, but it is unclear how the carbon
can survive the rp-process or ignite
\citep{schatz2003,zand2003,keek2008a,schatz2014,deibel2016} and
whether superbursts are sometimes not fueled by carbon
\citep{kuulkers2010}. In contrast, it seems clear that intermediate
duration bursts are fueled by helium on relatively cold neutron
stars. The helium may accumulate by slow and stable burning of
hydrogen into helium as described above, or by direct helium accretion
at low accretion rate
\citep[e.g.,][]{zand2005,cumming2006a,chenevez2007,falanga2008}. Due
to their larger ignition depth, these long bursts can serve as probes
of the thermal properties of the crust, one of the outstanding
questions of neutron star research.  Superbursts have a property which
is intriguing in light of this dependence: they ignite sooner than
expected on the basis of the measured mass accretion rate and presumed
crust properties \citep{keek2008a}. It looks as though the thermal
balance in the crust (determined by nuclear reactions in the crust and
neutrino cooling in the core) is different from expectations
\citep[see also][]{schatz2014}. Better measurements are needed to
constrain more accurately the recurrence time and the time history of
the accretion between superbursts. Currently, 26 superbursts have been
detected in the last 20 years \citep{intzand2017}. \extp can double
that in a few years.

\extp will provide the opportunity to address the outstanding
questions on X-ray bursts. A typical peak photon count rate with the
\lad will be $10^5\,\mathrm{s}^{-1}$. Cases with ten times as high
rates will also occur. These observations will provide unprecedented
detailed insight at processes with sub-ms time scales. With the
anticipated observation program, it is expected that the \lad will
detect at least several hundreds of X-ray bursts, with about 5 times
as much effective area (i.e., 3.5\,$\mathrm{m}^2$) and 5 times finer
spectral resolution (i.e., 260 eV) than the previous burst workhorse
RXTE-PCA \citep{jahoda2006} and the currently operational
Astrosat-LAXPC \citep{paul2009} and about 50 times as much effective
area at 6 keV than the currently operational NICER
\citep{gendreau2016}. The coded-mask imaging \wfm, thanks to its 4\,sr
field of view (one third of the sky) will detect as many as ten
thousand X-ray bursts in 3 years, which is of the same order of
magnitude as all bursts that were detected throughout the history of
X-ray astronomy. It is clear that \extp will provide a rich and novel
data set on X-ray bursts. While the primary science motives for X-ray
burst observations with \extp are dense matter physics and gravity in
the strong field regime, the secondary motives relevant to observatory
science concern their fundamental understanding, in connection with
nucleosynthesis, hydrodynamics, flame spreading and accretion flows
\citep[see][ and reference therein]{wpintzand}. As a benefit, the
outcome of these studies will help reduce the risk of any systematic
errors in studies in the primary science goals.

\begin{figure*}[ht!]
\centering
\includegraphics[width=0.49\textwidth]{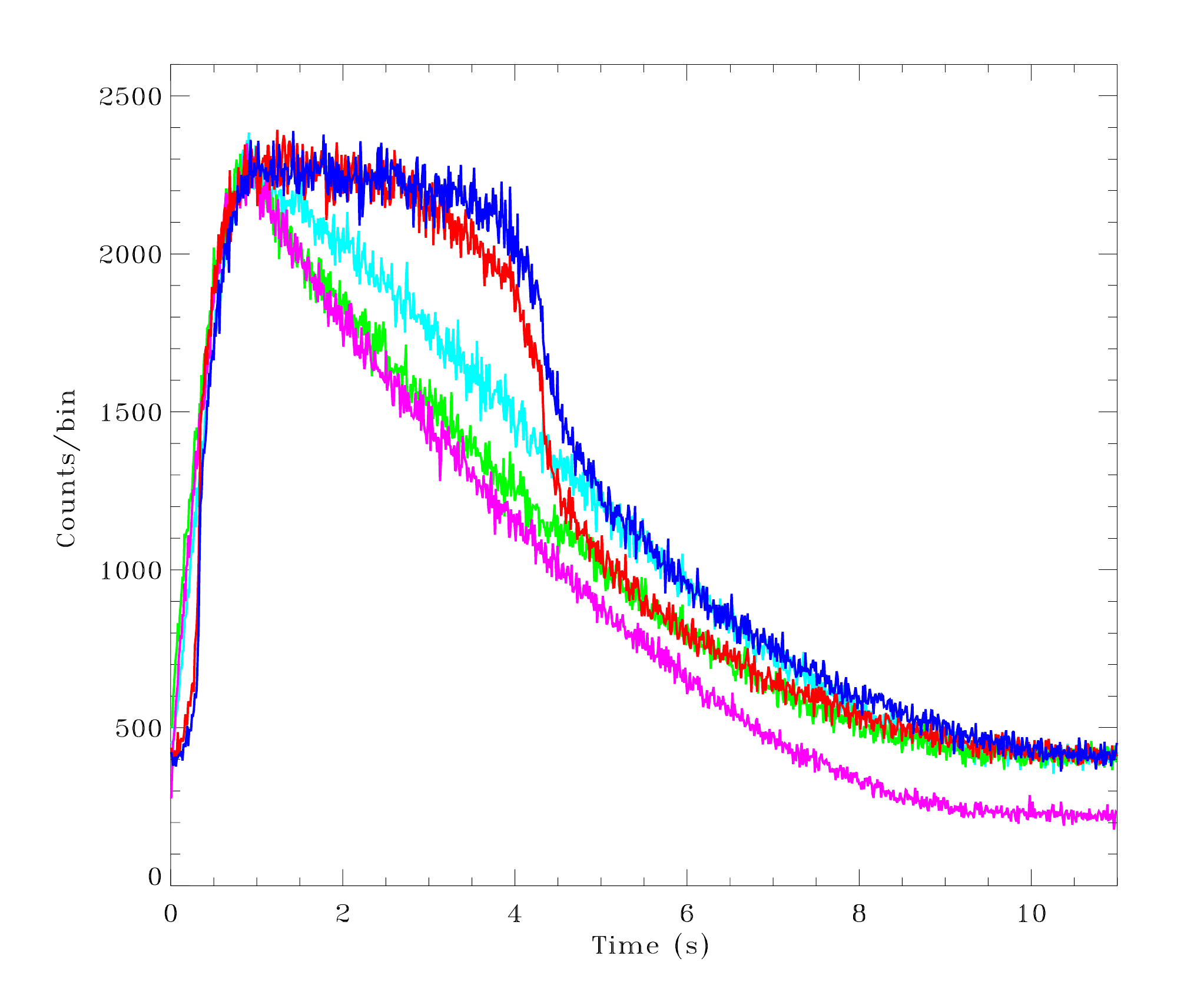} 
\includegraphics[width=0.49\textwidth]{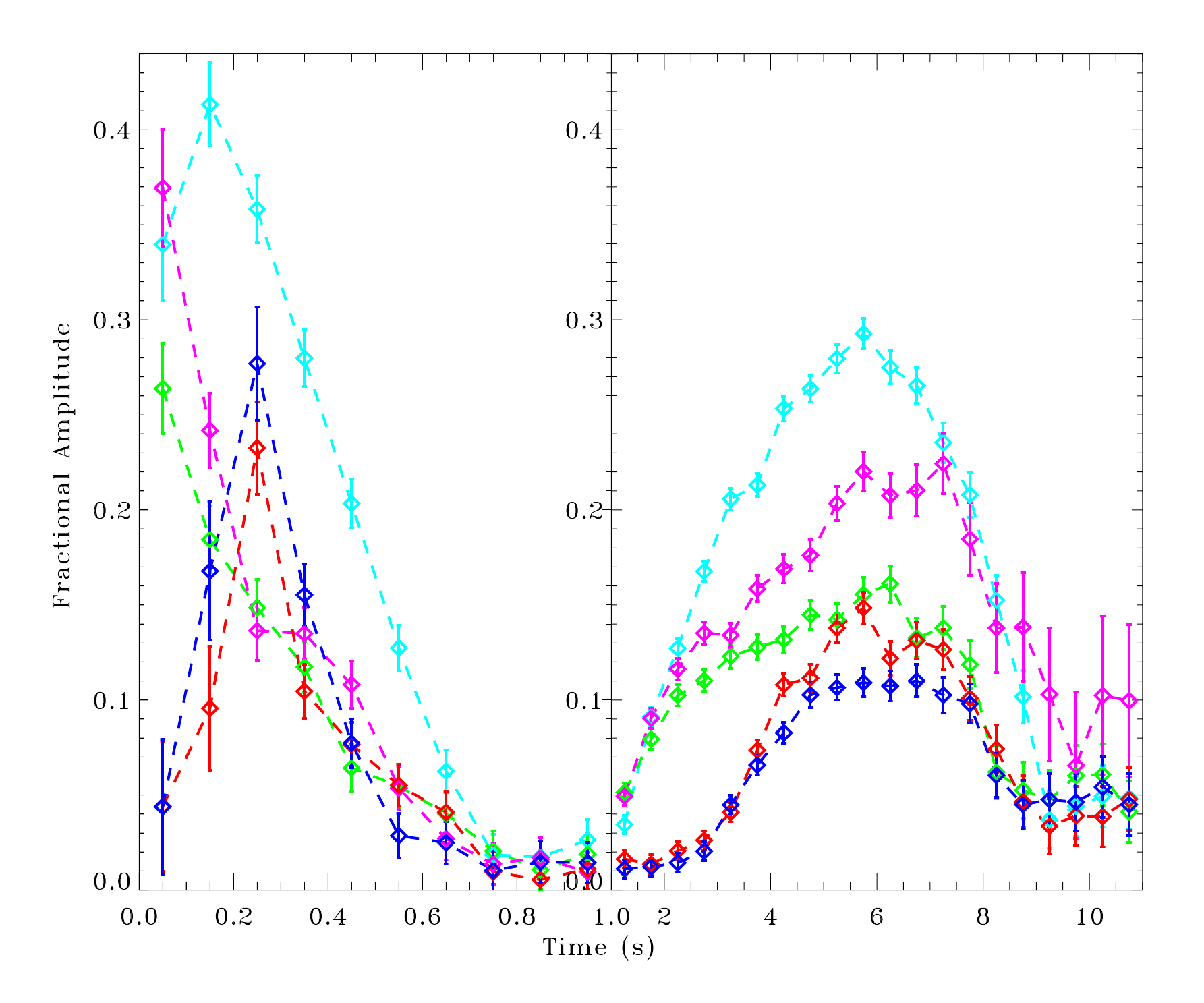}
\caption{\lad simulations of X-ray burst oscillations for a bright
  burster similar to 4U 1636-536 that is bursting faithfully for at
  least 40 years already. For details on the model, see
  \citet{simin16}. The simulations were carried out for a neutron star
  of $M=1.4M_{\odot}$, $R$=10~km and $\nu$=400~Hz, for an observer's
  inclination angle of $i=70^{\circ}$ (measured from the spin vector)
  and with a time resolution of 0.125 ms. The left panel shows the
  smoothed light curves for the rise and decay of the bursts (with 11
  ms time bins). The red, green and blue curves show simulated light
  curves for ignition co-latitudes (i.e., also measured from the spin
  vector) $\theta_s= 30^{\circ}, 85^{\circ}$ and $150^{\circ}$,
  respectively. The hot spot temperature is assumed to be $T_h=3$ keV
  and the temperature outside the hot spot $\Delta T=1.5$ keV
  lower. The flame spreading velocity goes as $v_{\rm flame}\propto
  1/\cos\theta_s$, consistent with strong frictional coupling
  between different fuel layers \citep{spitkovsky2002}. The magenta
  and cyan curves are similar to the green curve ($\theta_s=85
  ^{\circ}$) but with $\Delta T$=2 keV and $v_{flame}\propto
  1/\sqrt {\cos\theta_s}$, respectively, the latter referring
  to weak frictional coupling. The right panel shows the simulated
  \extp measurements of the fractional oscillation amplitudes for the
  same models as on the left panel. Note that the fractional
  amplitudes are shown on an expanded scale up to $t=1$~s (burst
  rise). These simulations show that \extp is capable of diagnosing
  ignition co-latitude and frictional coupling. They also indicate
  \citep[see][]{simin16} that they can constrain the cause of the
  burst oscillations in burst tails.}\label{burstoscil}
\end{figure*}

\begin{figure*}
 \centering
    \includegraphics[width=17.6cm]{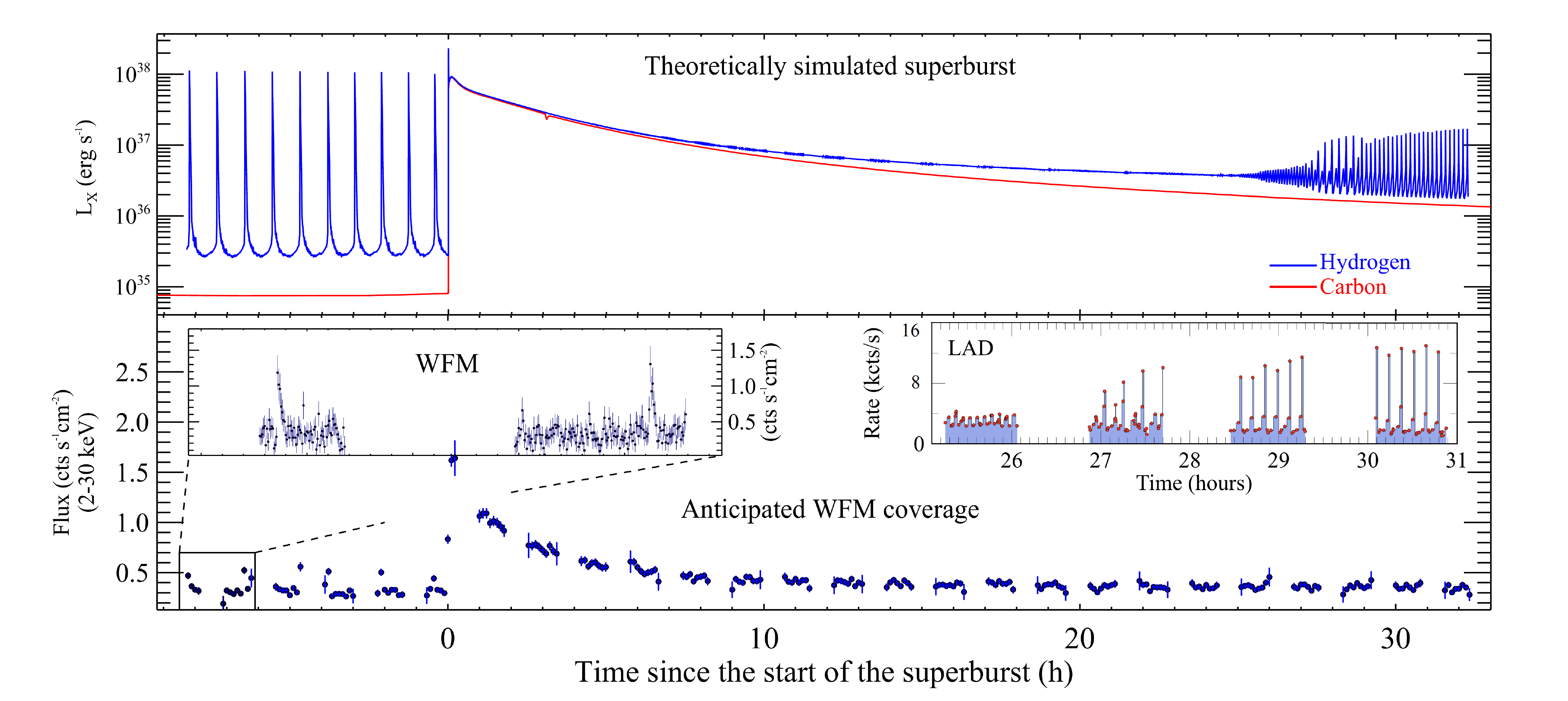}
  \caption{Simulation of the ordinary bursting behavior (blue curve)
    around a superburst (red curve) in a hydrogen-rich low-mass X-ray
    binary \citep[from][]{keek2012}. The superburst quenches ordinary
    bursts for one day after which a regime of marginal burning
    commences, characterized by mHz QPOs. The lower panel shows
    possible \wfm and \lad measurements of such a superburst, if it is
    at 8 kpc, on-axis, placed at 45\degr from the Galactic center and
    if the accretion rate is 10\% of the Eddington
    limit.}\label{fig:superburst}
\end{figure*}

\extp will enable substantial advances in the understanding of
thermonuclear X-ray bursts, on particularly the following questions:
\begin{itemize}

\item \textit{What is the nature of burst oscillations?}  Oscillations
  in burst tails can be studied at much better sensitivity than with
  previous missions and at photon energies below 2 keV it will improve
  on \textit{NICER} \citep{gendreau2016} by having a four times larger
  effective area. This will increase the percentage of bursts with
  detected oscillations, enable the detection of shorter oscillation
  trains and faster frequency drifts, probe fainter amplitudes and
  make possible spectro-timing analyses. Thus, \extp will enlarge our
  knowledge about burst oscillations including the possible
  persistence of hot spots during the burst decay and the potential
  roles of surface $r$ and $g$ modes
  \citep[see][]{watts2012a}. Simulations of measurements of burst
  oscillations are provided in Fig.~\ref{burstoscil}, under different
  model assumptions \citep[for details, see][]{simin16}. These
  simulations show that \extp has a diagnostic capability for the
  cause of burst oscillations in burst tails of a common type of X-ray
  bursts.

\item \textit{Is there a preferred ignition latitude on the neutron
  star (as expected for fast spinning neutron stars with anisotropic
  accretion); how do Coriolis, thermal and hydrodynamic effects
  combine to control flame spread?}  A thermonuclear burst is expected
  to be ignited at one point on the neutron star. The latitude of that
  point may not be random, but influenced by Coriolis forces. A
  thermonuclear flame should subsequently spread over the stellar
  surface in a fashion that is determined again by rotation and by the
  thermal properties of the ocean in which this happens and the power
  of the thermonuclear flash \citep{spitkovsky2002,cavecchi2013,
    cavecchi2015}.  Observational studies of the rising phases of
  X-ray bursts show diverse and interesting details
  \citep[e.g.,][]{maurer2008,bhattacharyya2007,
    chakraborty2014,intzand2014}. With the much larger area of the
  \lad, thermonuclear flame spreading can be studied at an
  unprecedented level of detail and in many more bursts. For example,
  simulations show (see Fig.~\ref{burstoscil}) that \extp measurements
  would enable constraining the latitude of the burst ignition.

\item \textit{What is the composition of the ashes of nuclear burning;
  how large is the gravitational redshift on the neutron star
  surface?} The \lfa and \lad combined will be uniquely sensitive to
  details of the continuum spectrum that are induced by Compton
  scattering in the atmosphere and to absorption features from
  dredged-up nuclear ashes such as iron and nickel which may enable
  direct compositional studies \citep[e.g.,][]{bildsten2003}.
  Low-resolution evidence for such features has been found in RXTE
  data of bursts that generated strongly super-Eddington nuclear
  powers yielding strong photospheric expansion due to radiation
  pressure \citep{intzand2010,kajava2017}, as predicted by
  \citet{weinberg2006}. Surface absorption features will be
  gravitationally redshifted, which can provide additional constraints
  on the compactness of neutron stars if rest wavelengths are
  identified. In this respect, the \lfa would provide unique
  data. Most burst data currently have no coverage of sub- 2 keV
  photon energies. Those that do and are not susceptible to detector
  saturation, taken for instance with the grating spectrometers on
  \textit{Chandra} and \xmm, are of low statistical quality due to
  effective areas less than 150~cm$^2$. The \lfa is anticipated to
  have pile-up fractions of a few percent for typical burst peak
  fluxes of a few Crab units and has an effective area one order of
  magnitude larger, at a fair spectral resolution to detect absorption
  edges. This provides excellent opportunity for the study of
    narrow spectral features, particularly absorption edges for
    systems at known distance with Eddington-limited bursts from a
    slowly rotating neutron star, like IGR~J17480-2446 in the globular
    cluster Terzan 5 \citep[e.g.,][]{strohmayer2010,linares2012}.

\item \textit{What characterizes the transition from stable to
  unstable nuclear burning?}  A long standing issue in X-ray burst
  research is the threshold between stable and unstable helium burning
  \citep[][]{vanparadijs1988,cornelisse2003,zamfir2014,stevens2014,
    keek2016b,cavecchi2017}. Thermonuclear burning becomes stable once
  the temperature becomes so high that the temperature dependence of
  the energy generation rate levels off to that of the radiative
  cooling rate (which is proportional to $T^4$). More than 99\% of all
  X-ray bursts are due to ignition of helium through $3\alpha$ burning
  and $\alpha$-captures. It is predicted that the ambient temperatures
  in the burning layer only become high enough ($\gtrsim
  5\times10^8$\,K) for stable burning when the mass accretion rate is
  above the Eddington limit \citep[e.g.,][]{bildsten1998}. However,
  one finds evidence of stable burning (i.e., absence of X-ray bursts
  and presence of mHz QPOs indicative of marginally stable burning)
  already at 10--50\% of Eddington \citep[e.g.,][]{revnivtsev2001,
    cornelisse2003,heger2007,altamirano2008,linares2012}.  A different
  perspective on this issue may be provided by \extp, by following
  extensively the aftermath of superbursts for about a week. The
  superburst is another means to increase the ambient
  temperatures. While the neutron star ocean cools down in the hours
  to days following a superburst, models predict \citep{keek2012} that
  it is possible to follow the transition from stable to unstable
  burning on a convenient time scale, through sensitive measurements
  of low-amplitude ($\sim0.1\%$) mHz oscillations and the return of
  normal, initially faint, bursting activity (see
  Fig.~\ref{fig:superburst}).

\item \textit{What burns in superbursts; what determines the
  thermodynamic equilibrium of the crust and ocean; what determines
  the stability of nuclear burning; what is the nature of surface
  nuclear burning at low accretion rates?} \wfm observations will
  measure more precisely the superburst recurrence time and time
  profile \citep[e.g.,][]{intzand2017}. This will constrain the
  thermodynamic balance in the ocean and crust, and probe in
  unprecedented detail the boundary between stable and unstable
  thermonuclear burning \citep{cumming2006a}. \wfm will accumulate
  large exposure times on all bursters (including yet unknown ones)
  and thus detect rare classes of events, such as the most powerful
  intermediate-duration bursts at very low accretion rates.

\item \textit{How do bursts impact and influence the surrounding
  accretion flow; can bursts yield insights into the underlying
  accretion physics?} X-ray bursts can deliver a powerful impulse to
  the surrounding accretion disk \citep[e.g., review by
  ][]{degenaar2018}. The response of the disk to the bursts could
  include (i) an increase in the accretion rate due to
  Poynting-Robertson drag, (ii) radiative driven outflows, (iii) an
  increasing scale height due to X-ray heating or (iv) a combination
  of all these effects \citep{ballantyneeverett2005}. Unique insights
  into the properties of accretion disks could be obtained if this
  response can be measured for bursts with a wide range of
  durations. The response of the disk can be measured by detecting the
  X-ray reflection features produced by the burst interacting with the
  accretion flow
  \citep[e.g.,][]{ballantynestrohmayer2004,keek2014}. The \lfa and
  \lad will be able to detect reflection features from bursts with
  integration times as short as a few seconds
  \citep[e.g.,][]{keek2016}. As the evolution of the burst is easily
  measured, changes in the reflection features can directly map the
  changes in the accretion disk over the course of the burst.

\end{itemize}

\noindent
X-ray burst science is also a high priority for the nuclear science
community. X-ray bursts are unique nuclear phenomena, and identified
as key problems for nuclear science in the US National Academies
report NP2010 "An Assessment and Outlook for Nuclear Physics", the
2015 US Nuclear Science Long Range Plan, and the 2017 NuPECC European
Long Range Plan. Major investments in a new generation of radioactive
beam accelerator facilities, in part motivated by X-ray burst science,
are now being made around the world, including FRIB in the US, FAIR in
Germany, and HIAF in China. New instruments such as the recoil
separator SECAR at FRIB or storage rings at FAIR or Lanzhou (China)
are being developed to enable nuclear reaction rate measurements for
astrophysical explosions such as X-ray bursts.

With a simultaneous advance in X-ray observational capabilities
through \extp there is a unique opportunity to combine accurate
nuclear physics and precision light curve data to determine stellar
system parameters, to identify different nuclear burning regimes, to
validate astrophysical 1D and 2D models, and to constrain neutron star
properties. In addition, the detection of spectral features from heavy
elements on the neutron star surface would be of high importance for
nuclear science, as it would provide direct insight into the nuclear
processes that power X-ray bursts. The Joint Institute for Nuclear
Astrophysics (JINA-CEE) is now fostering the close connections between
observational and nuclear science communities that are needed for
nuclear physics based model predictions to guide the interpretation of
observations, and for new observations to motivate nuclear
experiments.

\section{High-mass X-ray binaries}

Massive supergiant, hypergiant, and Wolf-Rayet stars
($M\gtrsim$10~$M_{\odot}$) have the densest, fastest, and most
structured winds. The radiatively accelerated outflows trigger star
formation and drive the chemical enrichment and evolution of Galaxies
\citep{kud2002}. The amount of mass lost through these winds has a
large impact on the evolution of the star. The winds can give rise to
an extremely variable X-ray flux when accreted onto an orbiting
compact object in a high-mass X-ray binary (HMXB).  The understanding
of the relation between the X-ray variability and stellar wind
properties has been limited so far by the lack of simultaneous large
collecting area and good spectral and timing resolution.

\begin{figure*}[ht!]
\centering
\includegraphics[height=\textwidth,angle=90]{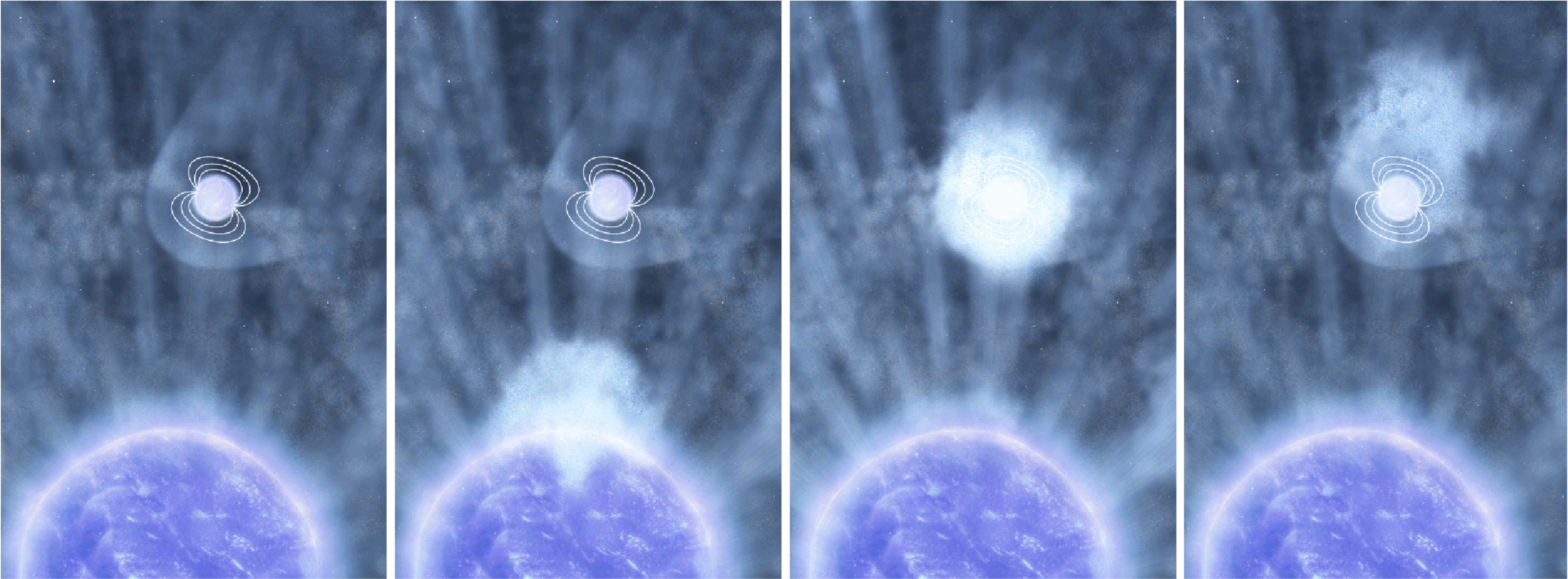}
\hspace{5mm}\includegraphics[width=0.42\textwidth]{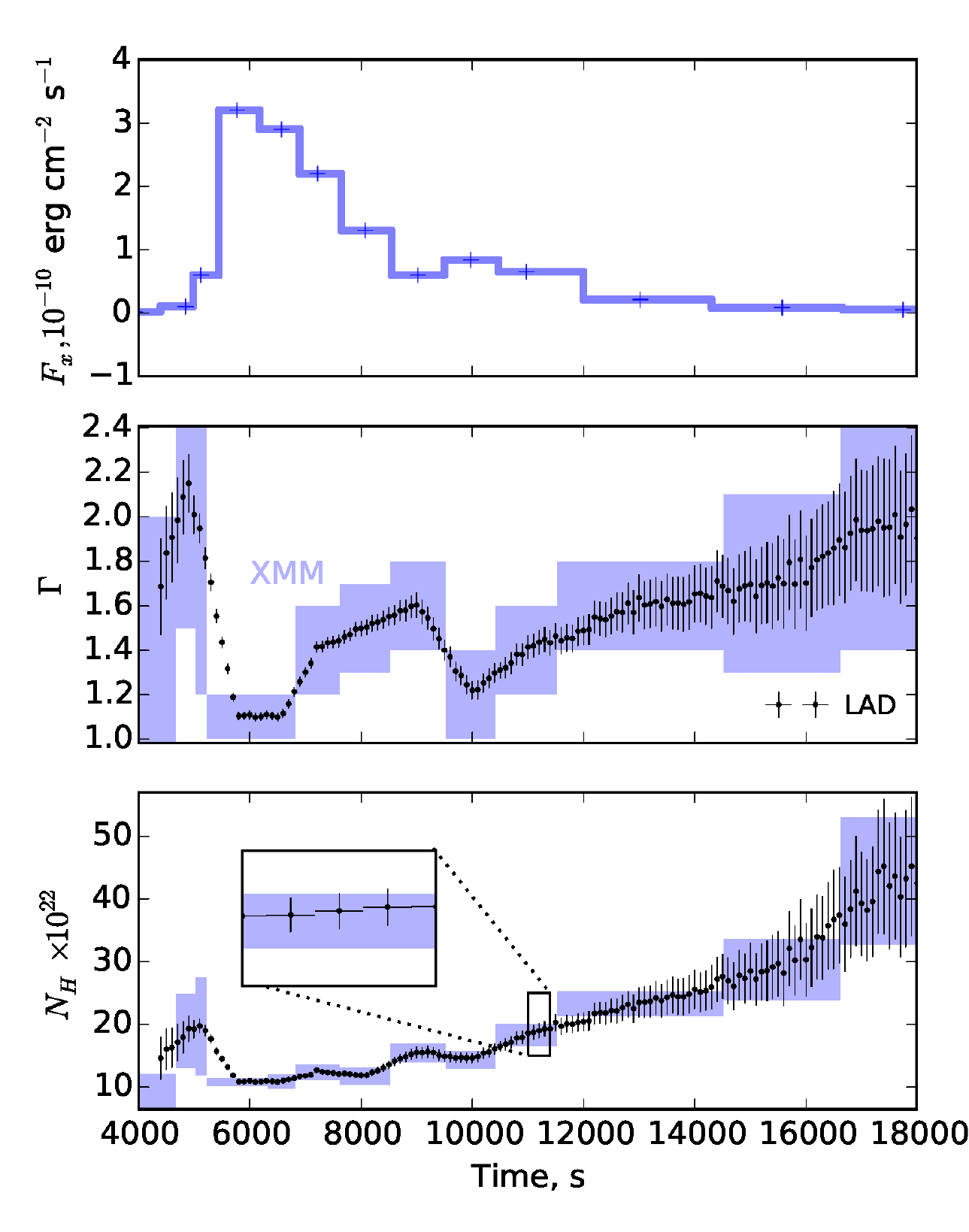}
\includegraphics[width=0.50\textwidth]{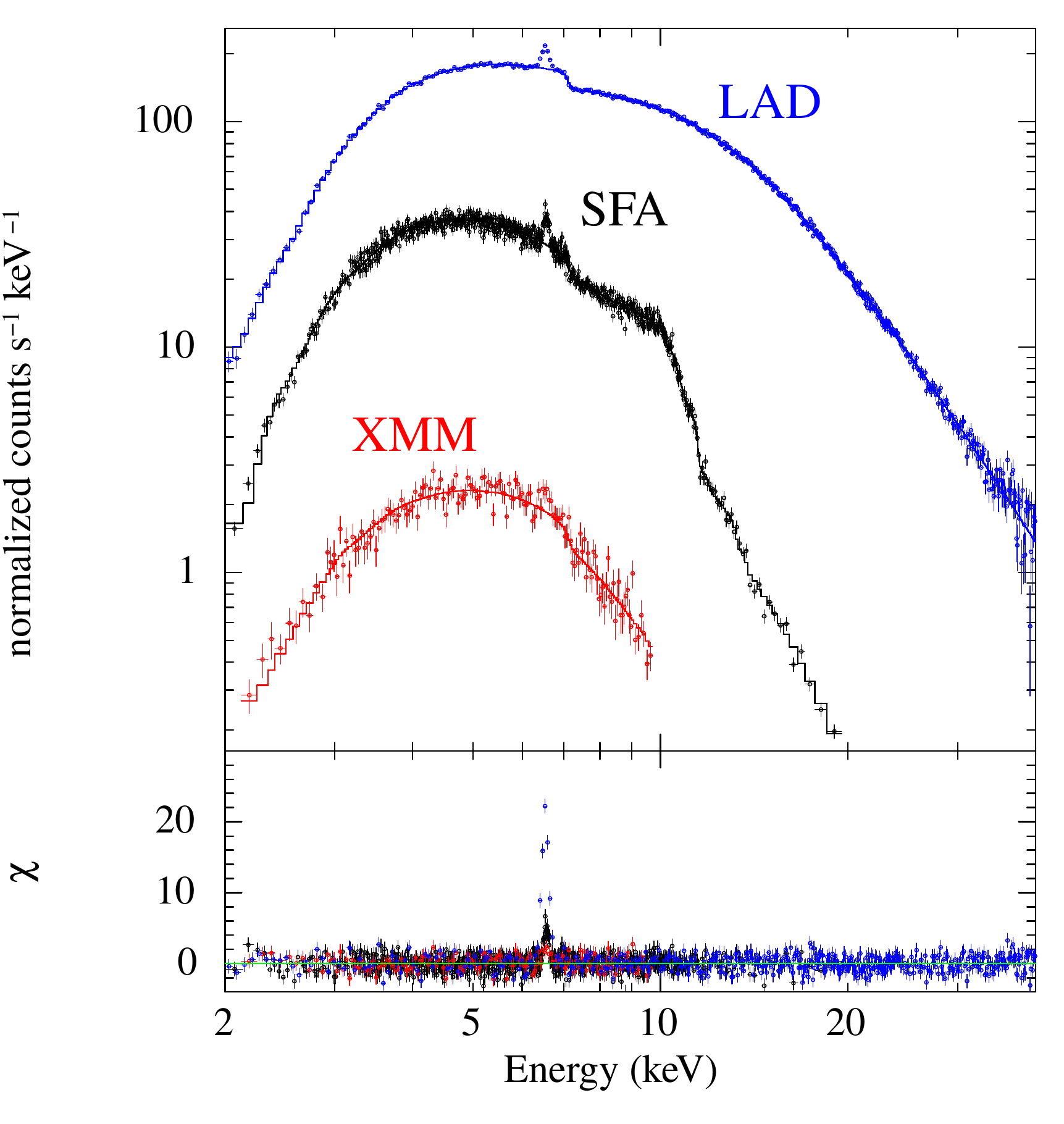}
\caption{{\it Top}: Sketch of the accretion by the upper neutron star
  of a clump ejected by the lower supergiant star (credits: ESA). {\it
    Bottom left}: Changes in the flux and spectral parameters during
  the flare recorded by \xmm, from the SFXT IGR J18410-0535
  \citep{Bozzo2011}. Violet boxes represent the measurements obtained
  with \xmm, while black points represent values obtained from the
  simulated \lad+\lfa spectra with exposure times as short as 100 s.
  Compared to \xmm, the dynamic process of the clump accretion can be
  studied in much more details and fast spectral changes can be
  revealed to an unprecedented accuracy (we remark also that the
  pile-up and dead-time free data provided by the \lad and \lfa
  greatly reduce the uncertainties affecting the \xmm, data obtained
  so far from HMXBs during bright flares). {\it Bottom right:} Shown,
  for comparison, is an example of a \xmm, spectrum extracted in a
  1~ks-long interval (source flux 3$\times$10$^{-10}$~\cgsflux) during
  the decay from the flare shown on the left and the corresponding
  simulated \lfa and \lad spectra.  The Fe-K line at 6.5~keV, used to
  probe the clump material ionized by the high X-ray flux, is barely
  visible in \xmm, but very prominently detected in both the \lfa and
  \lad spectra.}
\label{fig:flare_hmxb} 
\end{figure*}

\begin{figure*}[ht!]
\centering
\includegraphics[trim=0 9cm 0 11cm,width=1.0\textwidth]{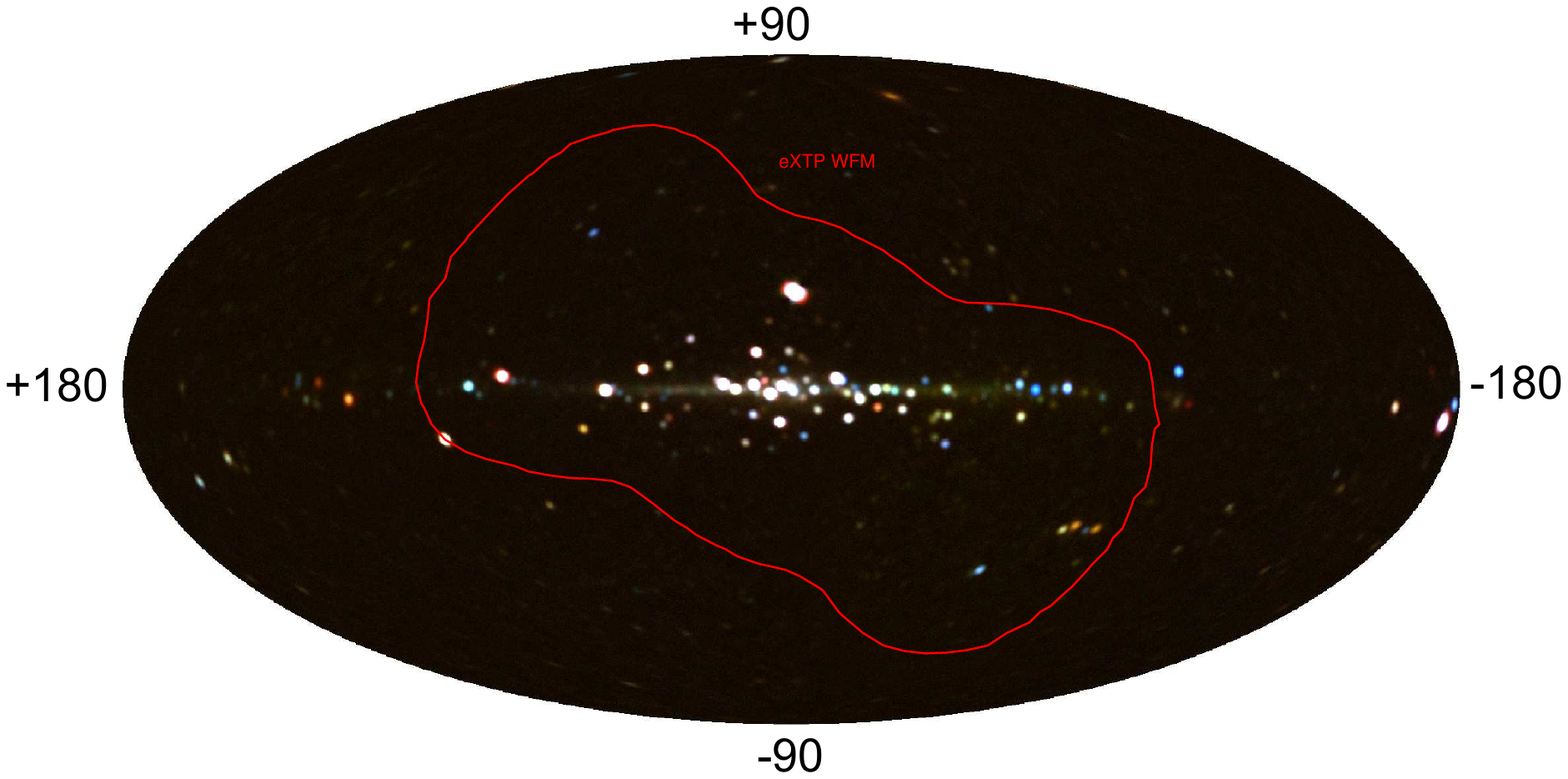}
\caption{Comparison of \wfm's field of view with the 2-20 keV map, in
  Galactic coordinates, of the X-ray sky as observed with MAXI
  (courtesy of T. Mihara, RIKEN, JAXA, and the MAXI team). The \wfm
  will have the largest field of view and monitor simultaneously a
  large fraction of all currently known HMXBs and low-mass X-ray
  binaries.}
\label{fig:wfm_hmxb} 
\end{figure*} 

In the past decades, observational evidence has been growing that
winds of massive stars are populated by dense `clumps'.  The presence
of these structures affect the mass loss rates derived from the
optical spectroscopy of stellar wind features, thus leading to
uncertainties in our understanding of their evolutionary paths
\citep{Puls08}.  HMXBs were long considered an interesting possibility
to probe clumpiness \citep{sako03}. As X-rays released by accretion
trace the mass inflow rate to the compact object, an HMXB provides a
natural in-situ probe of the physical properties of the massive star
wind including its clumpiness
\citep{zand2005a,Walter15,martinez2017}. The so-called `Supergiant
Fast X-ray Transients' (SFXTs) form the most convincing evidence for
the presence of large clumps. In X-rays, the imprint of clumps is
two-fold: 1) clumps passing through the line of sight to the compact
object cause (partial) obscuration of the X-ray source and display
photo-electric absorption and photo-ionization; 2) clumps lead to
temporarily increased accretion and X-ray flares.  A number of
hours-long flares displayed by the SFXTs could be convincingly
associated with the accretion of dense clumps \citep{Bozzo2011}.

The compact objects in HMXBs are usually young neutron stars (typical
age 10$^{6-7}$~yr) that are known to possess strong magnetic fields
($\gtrsim$10$^{12}$~G). These channel the accreting material from
distances as large as $\sim$10$^3$-10$^4$~km down to the surface of
the compact object, yielding X-ray pulsations, and change the spin of
the compact object.  Large positive and negative accretion torques
have been measured in both wind and disk-fed HMXBs on time scales
between years and decades. These torques result from the yet unknown
coupling between the neutron star magnetic field and the accreting
material \citep{Perna2006,bozzo08,postnov15}. Extending the time scale
of long-term monitoring of the behavior of pulsars in HMXBs is
essential to make progress in understanding this phenomenon.  Previous
large field-of-view instruments provided the opportunity to monitor
many X-ray pulsars in HMXBs, symbiotic X-ray binaries (SyXBs) in which
the donor star is a red giant, and in intermediate systems between
HMXBs and low-mass X-ray binaries.  Particularly striking were the
findings for GX\,1+4 and 4U\,1626-67: after about 15~yr (for GX 1+4)
and 20 yr (for 4U\,1626-67) of spin-up, both systems showed a torque
reversal, which made them switch to a spin-down phase
\citep{bildsten1997,sahiner2012}. Other systems, such as Cen\,X-3,
Vela\,X-1, 4U~1907+09 and Her X-1, often showed torque reversals,
sometimes superimposed on a longer term trend of either spin-down or
spin-up \citep{bildsten1997,sahiner2012}.

At low mass inflow rates, the accretion might be inhibited completely
through the onset of so-called centrifugal and/or magnetic barrier,
which could drive the rapid variability observed in the SFXTs and
other HMXBs \citep[see, e.g.,][and references therein]{bozzo08}.  The
observed variability is thus often used to probe the strength of the
dipolar component of the neutron star magnetic field.  This also
shapes the geometry of the emission region in X-ray pulsars, which, in
turn, influences their observed properties. Most notably, for
luminosities above the critical value $\sim10^{37}$\,erg\,s$^{-1}$,
the local accretion rate at the neutron star polar areas might exceed
the Eddington limit, leading to the formation of an extended accretion
columns \citep{basko76}. The transitional luminosity setting the
threshold between sub- and super-critical accretion is directly
related to the compactness of the neutron star and its magnetic field
strength, and can thus be used to constrain these parameters.  The
presence of an accretion column is expected to alter dramatically the
intrinsic beaming of the pulsar X-ray emission, as well as its
spectral energy distribution and polarization properties
\citep{basko76}. Observing these changes can thus be used to obtain a
direct measurement of the critical luminosity from the observations
\citep{2017MNRAS.466.2143D}.

The propagation of X-ray photons emerging through the strongly
magnetized plasma within the neutron star emitting region also affects
the observed spectra. The electron scattering cross section in strong
magnetic field is largely enhanced around the electrons
gyro-frequency, leading to the so-called cyclotron resonance
scattering features (CRSFs) often detected in the X-ray spectra of
young accreting pulsars \citep[for a recent review see,
  e.g.,][]{Walter15}. The scattering cross section strongly depends on
the angle between the direction of the photon propagation and the
magnetic field orientation. This implies that the polarization degree
and angle, the X-ray continuum spectrum, and the CRSF parameters
depend on pulse phase. Modeling these changes allows us to reconstruct
the configuration of the magnetic field and its orientation with
respect to the observer and the neutron star.

The mass accretion rate for black holes in HMXBs tends to be less
variable compared to the NS case. However, most black holes in HMXB
show some hard-to-soft state variability, similarly to what is
observed in BH low-mass X-ray binaries.  The most studied and observed
BH HMXBs are Cyg X-1 \citep[][and references
  therein]{grinberg2013,grinberg2014} and LMC X-3
\citep{wilms2001,boyd2001}. These sources proved particularly useful
to ``X-ray'' the stellar wind of their donor stars, complementing
similar studies carried out on NS HMXBs and discussed above. For
example, the Galactic black hole Cyg X-1 shows a periodic modulation
of the absorption column, which is due to the absorption of X-rays
from the black hole in the massive stellar wind of its donor star, HDE
226868. As shown by \citet{grinberg2015}, the large scatter in
individual measurements of the absorption column to the black hole can
be used to study the size distribution of clumps in the stellar wind,
which is of great interest for understanding the physics of the winds
of massive stars \citep[e.g.,][]{sundqvist2013}.

The instruments on-board \extp will dramatically open up perspectives
for research in all above mentioned fields. In particular the
following questions can be addressed:

\begin{figure*}[ht!]
    \includegraphics[width=0.99\columnwidth]{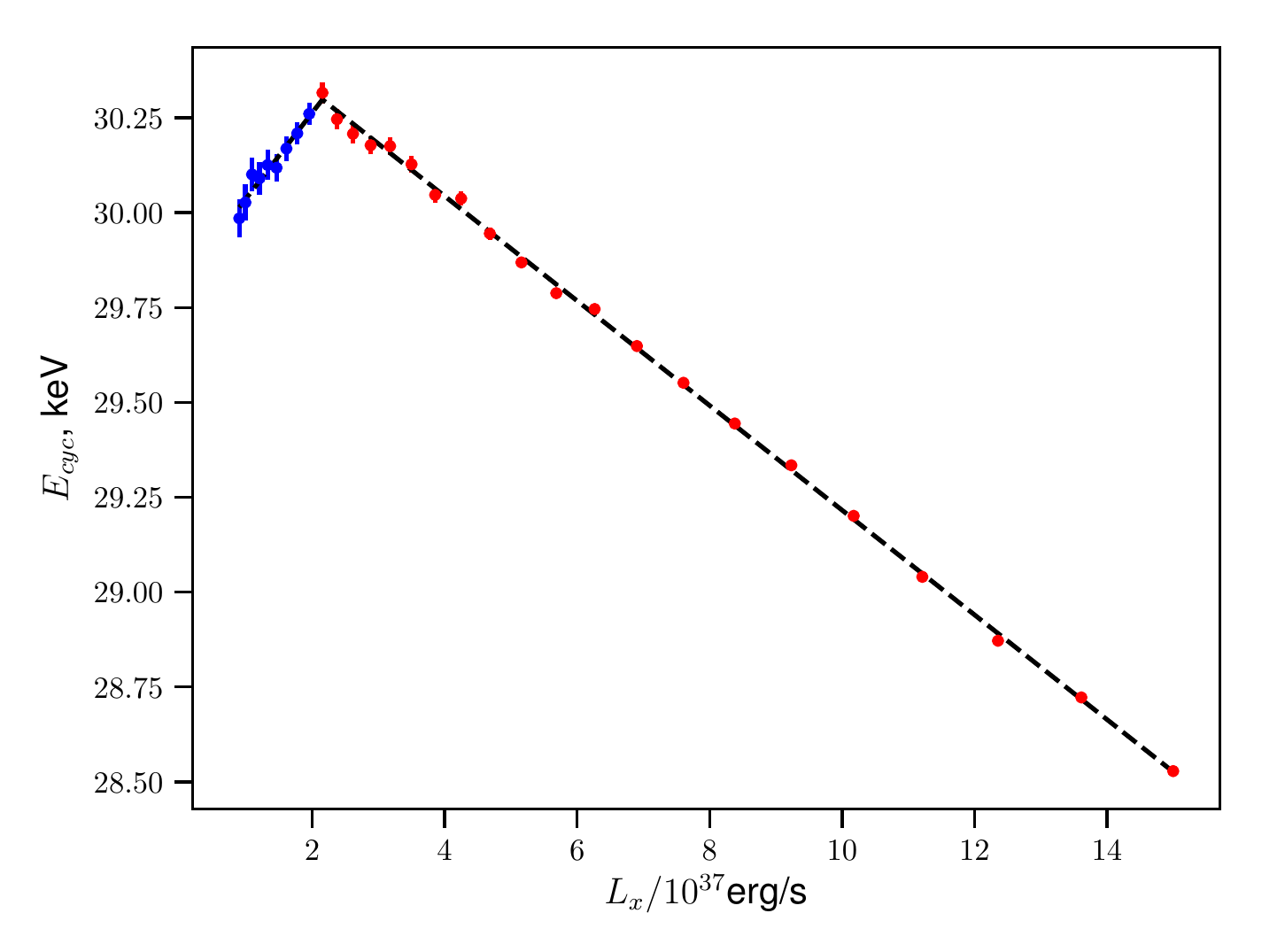}
    \includegraphics[width=0.99\columnwidth]{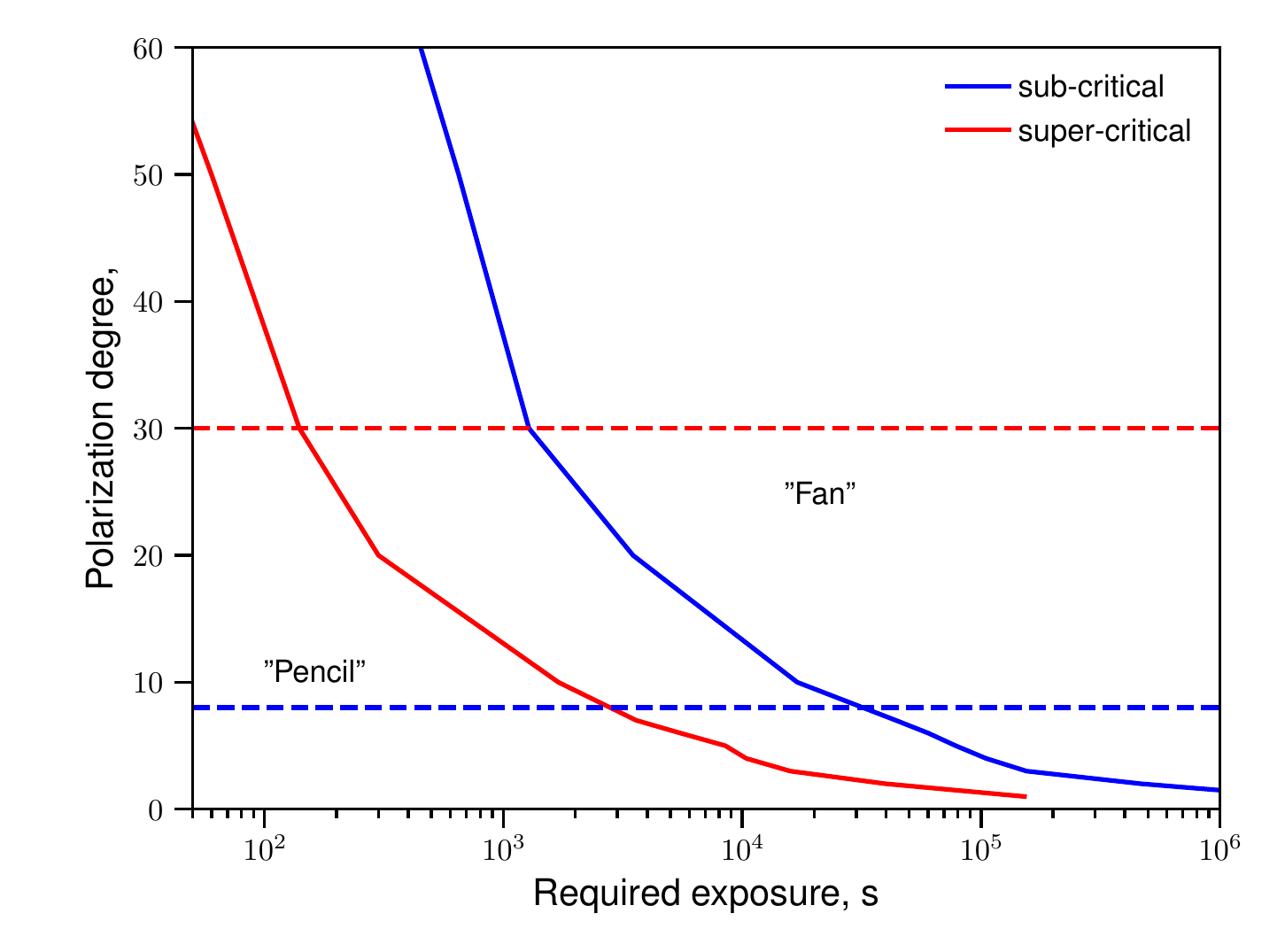}
    \caption{The combination of the \lad large effective area and
      energy resolution will allow us to monitor the
      luminosity-related changes of the observed CRSF energy in
      transient X-ray pulsars with unprecedented accuracy. The left
      panel of the figure shows a simulation of thirty 10\,ks \lad
      observations of the X-ray pulsar V~0332+53, where a transition
      from super- to sub-critical accretion has been recently detected
      through a luminosity-dependent change of the observed CRSF
      energy \citep{2017MNRAS.466.2143D}. \extp will allow us to
      investigate the same effect with a much larger accuracy and to
      investigate changes in pulse profile shape and polarization
      signatures associated with similar transitions.  The right panel
      of the figure shows the exposure required to measure the
      polarization of a given strength at a 3$\sigma$ significance
      level for the typical super-critical and sub-critical fluxes
      observed from V~0332+53. The polarization signal strength
      predicted by \citet{1988ApJ...324.1056M} in the two cases are
      also shown. The \extp \gpd will allow us to detect and compare the
      polarized emission in both the sub- and super-critical regime
      using relatively short exposure times, performing also studies
      of the dependence of the polarization properties as a function
      of the pulse phase .}\label{fig:v0332}
\end{figure*}

\begin{itemize}

\item {\it What are the physical properties of massive star wind
  structures and how do these affect the mass loss rates from these
  objects?}  With current instruments, integration times of several
  hundreds to thousands of seconds are needed to get a rough estimate
  of clump properties and only an average picture of the clump
  accretion process can be obtained.  Similar observations performed
  with the \lad and \lfa on-board \extp will dramatically improve our
  present understanding of clumpy wind accretion and winds in massive
  stars in general, by studying (spectral) variability on time scales
  as short as a few to tens of seconds (see Fig.~\ref{fig:flare_hmxb}).
  This will permit a detailed investigation of the dynamics of the
  clump accretion process and obtain more reliable estimates of the
  clump mass, radius, density, velocity, and photo-ionization
  state. In turn, it will improve our understanding on the mass loss
  rates from massive stars. As structured winds are not spherically
  symmetric, they are also predicted to emit polarized X-ray
  radiation. Measurements with the \gpd can yield additional
  information to constrain the stellar mass loss rate
  \citep{kallman15}.  The improved capabilities of all \extp
  instruments compared to the current facilities will permit to
  conduct also similar studies on the SyXBs \citep{enoto14}, thus
  exploring the structure and composition of the still poorly known
  winds of red giants stars (which might be accelerated through the
  absorption of the stellar radiation by dust grains rather than heavy
  ions).  Finally, it is important to understand the consequences of
  inflated envelopes of massive giant stars \citep{sanyal15} on the
  observational properties of HMXBs in which the donor stars are
  filling, or close to filling, their Roche lobes.  The ability of the
  \extp instruments (especially the \wfm) to detect and monitor
  transient sources in a very broad range of mass accretion rates will
  allow us to study the spectral and temporal behavior of accreting
  neutron stars with strong magnetic fields.  Follow-up observations
  exploiting the high sensitivity of the \lfa will allow us to observe
  the so-called propeller effect in dozens of X-ray pulsars.  For the
  sources with measured cyclotron energies it will make possible to
  determine the configuration of the neutron star magnetic
  field. Monitoring observations of long-period pulsars at low mass
  accretion rates will constrain the physical mechanisms regulating
  the interaction of plasma in a low ionization state with the neutron
  star strong magnetic field \citep{arXiv:1703.04528}.

\item{\it What are the geometry and physical conditions within the
  emission region of X-ray pulsars, and their relation to the observed
  pulse profiles, spectra, and polarization in the X-ray band?} Pulse
  phase-resolved spectroscopy of accreting pulsars has been one of the
  main tools to diagnose their emission region properties since
  decades. \extp will bring such investigations to an entirely new
  level. The large effective area of the \lfa and the \lad will provide
  spectra and pulse profiles of unprecedented statistical quality even
  within comparatively short integration times. It will be possible
  for the first time to study the spectral and pulse profile evolution
  on very short timescales comparable with the spin period of the
  neutron star. The energy resolution of the \lad in a broad energy
  band will allow us to perform detailed studies of the complex CRSF
  shapes observed in some sources \citep{2015ApJ...806L..24F}, and
  investigate dependence of the CRSF parameters on the pulse phase,
  which is essential to constrain the line formation models and the
  emission region geometry.

In addition, \extp will provide for the first time highly complementary
simultaneous X-ray polarization data. The expected strong polarization
and high X-ray fluxes of the accreting pulsars make these sources a
prime target for X-ray polarimetry studies. The polarization degree
and angle carry information on the geometry of their emission region,
magnetic field configuration, and orientation of the pulsar with
respect to the observer \citep{1988ApJ...324.1056M}. A major change of
the observed polarization signatures is expected with the transition
from the sub- to super-critical accretion, i.e. at the onset of
extended accretion columns. \extp observations can be used to detect
such transitions and the corresponding changes in the polarization
properties of X-ray pulsars, as illustrated in Fig.~\ref{fig:v0332}.

\item {\it What are the mechanisms triggering torque reversals in
  HMXBs and what leads to orbital and super-orbital modulation of the
  X-ray emission from these systems?} For systems with well known
  distances, simultaneous measurements of fluxes and spin-rates can
  provide constraints on the magnetic field \citep{klus14}. If the
  pulsar magnetic field is also known from cyclotron line
  measurements, then the relationship between the spin rate and the
  luminosity can be used to constrain the neutron star parameters,
  probing not only accretion physics but also fundamental physics in
  these highly magnetized accreting systems.  The \wfm is a perfectly
  suited instrument for these studies in that it is a wide field
  monitor capable of simultaneously measuring the source flux and
  accurately determining its spin frequency. Its lower energy
  threshold will make it the most sensitive monitor for these types of
  systems that has ever flown.  Measurements from the \wfm of pulsed
  flux, unpulsed flux, frequency and frequency derivatives of
  accreting X-ray pulsars will contribute to the solution of the
  questions mentioned above, and it will also trigger target of
  opportunity observations with the \lad/\lfa at the onset of torque
  reversals. Early detections of these state changes are critical to
  initiate early observations by the \lad/\lfa in order to understand
  the associated spectral/timing variation 
  As a byproduct of the monitoring observations, it will be possible
  to carry out detailed spectroscopic analyses as a function of
  orbital, superorbital and spin phase. The orbital dependence, for
  instance, can be used to map long-lived structures surrounding the
  neutron stars in these systems, like accretion wakes.

\end{itemize}

Last but not least, it is worth mentioning that many HMXBs are
transient and thus the discovery of new sources or new outbursts from
previously known objects in this class will greatly benefit from the
capabilities of the \wfm. Figure~\ref{fig:wfm_hmxb} shows that a large
fraction of all these sources can be efficiently monitored during each
single pointing of the \wfm, let alone in sequential different
pointings.

\section{Radio-quiet active galactic nuclei}

\begin{figure*}[ht!]
\centering
\includegraphics[trim=0 1cm 0 0,width=0.7\textwidth,angle=270]{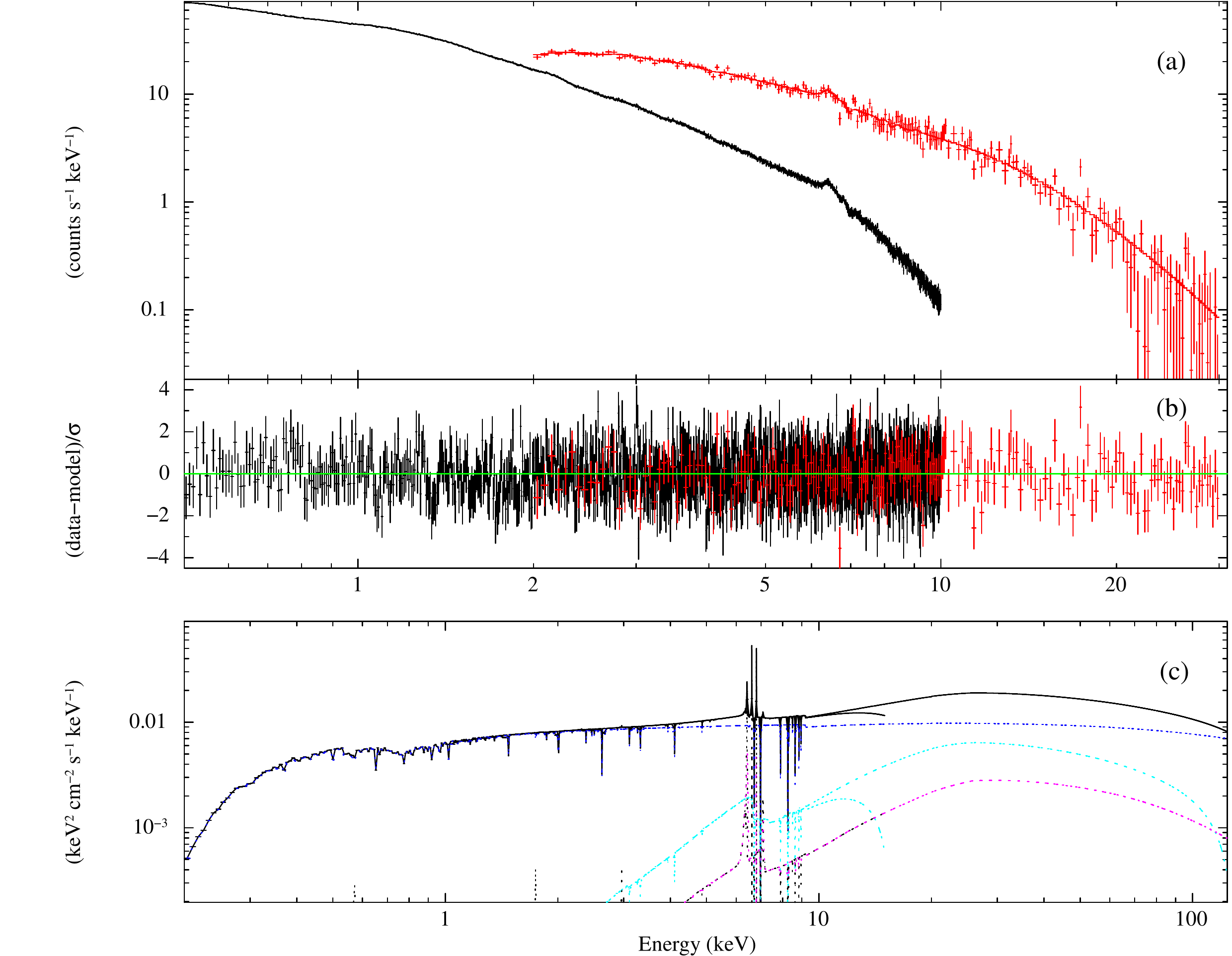}
\caption{Panel a: \extp (\lfa in black, \lad in red) spectrum. Panel
  b: residuals as obtained by a 100 ks integration of a bright 1 mCrab
  AGN composed of: the continuum, three components of ionized
  absorbers, the cold reflection and narrow Fe line K$\alpha$, Fe
  K$\beta$ and Ni K$\alpha$, the ionized lines FeXXV and FeXXVI and
  the blurred ionized reflection component (black hole spin
  $a=0.7$). The model complexity we used for the simulation is similar
  to the case of the AGN SWIFT\,J2127.4+5654
  \citep{marinucci14,miniutti07} and is shown in panel c where
  separate spectral components are indicated with different colors (in
  blue the primary cut-off power--law, in cyan the blurred reflection
  component originated in the innermost regions nearby the central BH
  and in magenta the contribution of the a cold reflection component
  originated in a medium distant from the central BH.}
\label{figderosa} 
\end{figure*} 

\begin{figure*}[ht!]
\centering
\includegraphics[trim=2cm 8cm 2cm 8cm,width=0.7\textwidth]{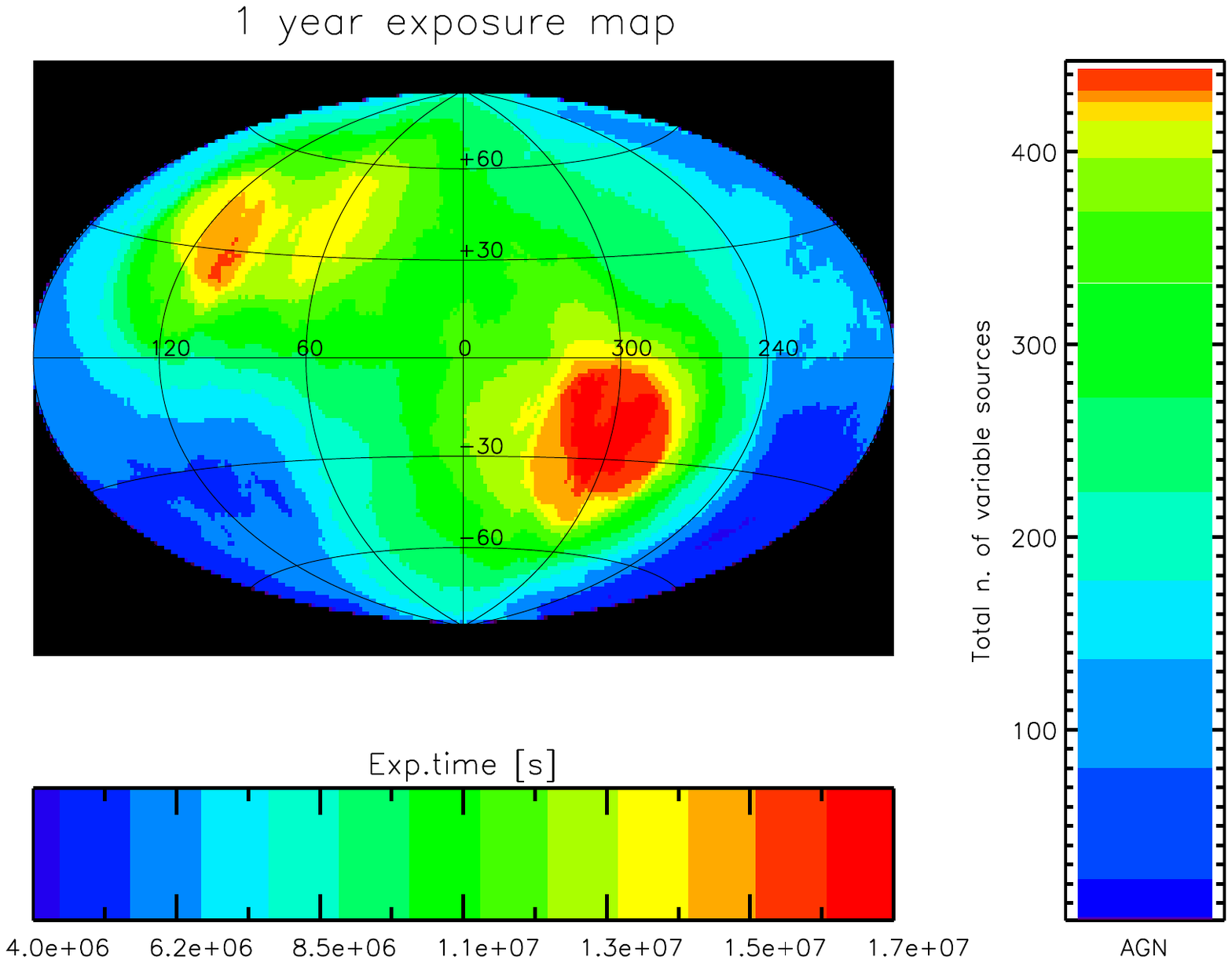}
\caption{Number of AGN per 4$\pi$ sr as a function of the position in
  the sky that are bright enough to be accessible to the \wfm for
  studies of X-ray variability. For these sources, a light curve with
  at least 10 time bins in a 1-year period can be built, achieving in
  each bin a signal-to-noise ratio of at least 3. The lateral bar
  reports the total number of variable AGN observed on the sky,
  assuming an example of a 1-yr pointing plan. Only regions of the sky
  covered with more than 10~cm$^2$ of net detector area have been
  considered.}
\label{figderosa2} 
\end{figure*} 

The standard model of active galactic nuclei (AGN) predicts a
supermassive black hole (SMBH) surrounded by a geometrically thin,
optically thick accretion disk. Like the case of X-ray binaries, the
seed photons from the disk are Compton scattered into the X-ray band
in a hot corona, the main difference being that the accretion disk in
AGN emits at optical/UV instead of X-ray temperatures. The UV and
optical seed photons from the disk are Compton scattered into the
X-ray band in a hot corona above the accretion disk which produces the
hard power-law spectrum extending to energies determined by the
electron temperature in the hot corona, with a cut-off energy ranging
from 100 to 200 keV
\citep{perola2002,derosa2012,molina2013,marinucci2014,ballantyne2014}.
This primary X-ray radiation in turn illuminates the inner disk and is
partly reflected towards the observer's line of sight
\citep{haardt1994}, producing a Fe-K line that will be heavily
distorted, with a pronounced red wing - a clear signature of strong
field gravity effects on the emitted radiation. The phenomenology of
AGN is quite similar to stellar black holes accreting from companion
stars in X-ray binaries, except for a difference in time scales and a
difference in typical temperatures of spectral thermal components from
the accretion disk that scales with the black hole mass because the
size of the innermost stable circular orbit scales with mass. The
inner structure in AGN is surrounded by rapidly rotating clouds of gas
- the Broad Line Region (BLR). A toroidal absorber is located at a
distance of a parsec from the black hole and within the Narrow Line
Region (NLR) clouds. These regions contribute with additional features
to the primary continuum: (1) a narrow Fe-K line at 6.4 keV produced
in the BLR, in external regions of the accretion disc or in the
molecular torus; (2) a reflection hump above 10 keV, due to the
reprocessing of the continuum from the optically-thick medium (such as
the torus and/or the accretion disk); and (3) the absorption and
variable features produced in the ionized and cold material around the
inner region (up to the NLR distance scales). The typical broadband
spectrum as may be observed by \extp is shown in Fig.~\ref{figderosa}.

Thanks to the large effective area and the CCD-class spectral
resolution of the \lad and \lfa, coupled with the large sky coverage
of the \wfm, \extp will make a decisive step forward in our
understanding of the inner structure of AGN. The unprecedented
\lfa\ spectral sensitivity will allow one to observe hundreds of AGN
down to a flux of 10$^{-13}$~\cgsflux\, in the local universe with a
very good accuracy (i.e., a signal-to-noise ratio in excess of
100). Thanks to the \lad effective area, the brightest AGN
($3\times10^{-11}$~\cgsflux) can be characterized with extreme
accuracy. Moreover, the \wfm will be capable of continuous X-ray
monitoring of large numbers ($\gtrsim$100) of AGN on time scales of a
few days to weeks. The \wfm will thus provide an essential synergy
with those observatories that will, in parallel to \extp, be opening
up AGN time domain science on those time scales at other wavebands,
for example with CTA (TeV energy domain), Meerkat/ASKAP/SKA (radio
wavelengths) and LSST (optical band, see
Fig.~\ref{figfacilities}). Here, the focus is on three key aspects of
AGN physics (related to the accretion properties and variability),
which will be specifically addressed by the unique capabilities of
\extp:

\begin{itemize}
 
\item \textit{What are the emitting materials around the central
  supermassive black hole?} The origin of the narrow Fe emission line
  observed in AGN is not completely clear, while this is a fundamental
  piece of information about the SMBH central engine and its
  environment. The width of the line is less than few thousand km/s
  \citep[in terms of full width at half maximum;][]{yaqoob04} and
  suggests an origin within the BLR or the putative molecular torus
  (at pc scale) or within the BLR closer to the BH. A powerful method
  to study the origin of the narrow Fe line is to measure the time lag
  between the variation of the continuum and the line
  \citep{blandford82}. The reverberation mapping methodology is used
  with the optical and UV broad emission lines with great success
  \citep{peterson04} and, recently, in X-rays to investigate the
  origin of the broad Fe line \citep{fabian09,uttley14}. However, the
  uncertainties on the Fe line flux (about 10\%) and the low sampling
  frequency of the X-ray continuum, prevent us applying this method to
  a large sample of AGN, especially on the daily time scale expected
  for a BLR origin \citep{markowitz09,chiang00,liuy10}. \extp will
  allow one, for the first time, to perform X-ray reverberation of the
  narrow core and broad component \citep[see the \extp white paper on
    strong gravity;][]{WP_SG} of the Fe line with a single mission.

  The very large field of view of the \wfm is a powerful tool to
  monitor the continuum intensity of virtually all the sources in the
  sky. To demonstrate the unique capabilities of the \wfm we used the
  1-year exposure sky map shown in Fig.~\ref{figderosa2} as a good
  approximation of the adopted pointing strategy. The \wfm can produce
  continuous light curves with daily $3\sigma$ detections for bright
  sources (i.e., brighter than $\sim1\times10^{-10}$~\cgsflux\, in the
  7--50\,keV band, i.e., above the Fe K edge). Weaker objects
  ($\sim5\times10^{-11}$~\cgsflux) will be detectable in weekly
  exposures. These time scales are perfectly suited for the narrow Fe
  line reverberation analysis, since the expected time scales are from
  days to weeks to years \citep{markowitzetal03a}, depending on the
  location and geometry of the material and the intrinsic variability
  of the sources (therefore, on the black hole mass and the accretion
  rate). At the same time, the large area available with the \lfa and
  \lad will allow one to perform well-sampled (weekly to monthly)
  monitoring of the Fe-K line and Compton reflection component with
  short targeted observations. The CCD-class energy resolution will
  easily allow one to disentangle the broad Fe line component from the
  narrow core. In fact, in only 1 ks a narrow Fe line of equivalent
  width 100 eV can be recovered with an uncertainty of 1-2\% for
  bright sources ($\sim1\times10^{-10}$~\cgsflux\ with $\sim10^3$ \lfa
  plus \lad counts in the line). The same uncertainties can be
  recovered for weaker sources ($\sim5\times10^{-11}$~\cgsflux) with 5
  ks exposures.  Thanks to the broad bandpass of the \lad, \extp will
  be able to recover the reflection continuum with 10\% accuracy in a
  1~ks exposure for the brightest sources
  ($\sim1\times10^{-10}$~\cgsflux).  This measurement, together with
  the \wfm daily monitoring of the continuum above 7 keV (even with a
  low signal-to-noise ratio), will be a instrumental in identifying
  the origin of the narrow distant reflection components in AGN.

\item \textit{What are the absorbing regions in the environment of
  supermassive black holes, from kpc to sub-pc scale?}  There is
  strong evidence of at least three absorption components on very
  different scales: on scales of hundreds of parsecs, on the parsec
  scale, and within the dust sublimation radius, on sub-parsec scale
  \citep[e.g.,][]{liuy10,bianchi2012,derosa2012}. The most effective
  way to estimate the distance of the absorber is by means of the
  analysis of the variability of its column density.  In particular,
  rapid (from a few hours to a few days) $N_\mathrm{H}$ variability
  has been observed in most bright AGN in the local Universe
  \citep{torricelli2014,maiolino2010,risaliti2013,sanfrutos2013,
    walton2014,markowitzetal14,rivers2015}, suggesting that
  obscuration in X-rays could due, at least in part, to BLR clouds or
  in the the inner regions of infrared-emitting dusty tori. These
  observations with \xmm and \textit{NuSTAR} have achieved a precision
  for the covering factor of $\sim$5--10\%. Detailed simulations show
  that such a measurement (assuming the same observed fluxes) will be
  obtained with \extp with significantly higher precision (1-2\%), on
  even shorter time scales (few ks instead of $\sim$100\,ks).  This
  would allow, for the first time, an extensive study of X-ray
  eclipses, which are frequently seen in local bright AGN
  \citep[e.g.,][]{markowitz2014}. Following the time evolution of the
  column density and of the covering factor over timescales of a few
  ks will provide unprecedented information on the structure, distance
  and kinematics of the obscuring clouds. In particular, close to the
  lower end of the BH mass range in AGN, with a single ~100 ks
  observation it will be possible to fully characterize the
  circumnuclear absorbing medium by sampling eclipsing events from
  multiple clouds \citep[e.g.,][]{nardini2011}.  which is now
  precluded by the short timescales involved. Moreover, the broad-band
  energy range offered by \extp will allow one to explore in detail
  also those clouds with larger column densities in Compton thick AGN,
  whose variability mostly affects the spectrum at high energies.

\item \textit{How does the central engine produce the observed
  variability?}  Our knowledge of the X-ray variability in AGN has
  advanced substantially in the last 15 years, thanks mainly to
  monitoring campaigns with RXTE and day-long \xmm observations of a
  few X-ray bright AGN \citep{markowitzetal03b,mchardyetal04,
    arevaloetal06,papadakisetal02,gonzalez-martin12}. AGN show red
  noise in their power spectral density functions (PSD) with a power
  law that decreases at high frequencies as a power law with index
  -2. Below some characteristic frequency the PSDs flatten and the
  break frequencies scale approximately inversely with the BH mass,
  from BH X-ray binaries to AGN \citep{mchardy06}. These
  characteristic frequencies seem linked to both the BH mass and the
  properties of the accretion flow. Although all the relevant
  accretion disk time scales depend on the mass of the central object,
  only a few of them may depend (indirectly) on the accretion rate as
  well. So far, our knowledge is based on the power-spectral analysis
  of a few objects. A substantial improvement of the analysis of X-ray
  variability in AGN would be possible by avoiding the problems
  introduced by large statistical noise and irregular time
  sampling. The \wfm plays a crucial role in our understanding of the
  X-ray variability in AGN.  The \wfm will survey large areas of the
  sky, allowing one to study long-term X-ray variability of
  bright/nearby AGN. Using the AGN number counts observed by previous
  missions, we computed the number density of sources that will be
  accessible to \extp in each region of the sky (assuming an average
  AGN spectrum and Galactic absorption). Using the typical AGN flux
  distribution, we derived the number of sources that are bright
  enough to build a light curve with a worst resolution of 1 month at
  signal-to-noise ratio of 3.  Fig.~\ref{figderosa2} shows the number
  density map of AGN that fulfill these criteria.  \citet{allevato13}
  demonstrated that it is possible to retrieve the intrinsic
  variability (within a factor of 2) even when using sparse light
  curves with low signal-to-noise ratio, such as the ones expected
  here for most of the faint AGN population in the \wfm
  sample. Depending on the exact monitoring pattern, much smaller
  uncertainties are expected for the brightest $20$ sources.  The
  \wfm, even in its first year of operations, is expected to return
  light curves with at least 10 data points per year for several
  \textit{hundreds of variable AGN}, allowing both individual (for the
  brightest) and ensemble studies of AGN variability on long (monthly)
  time scales. One will therefore be able to determine accurately the
  \textit{intrinsic} variability amplitude (on long time scales) for
  hundreds of AGN and study the variability-luminosity relation with
  an unprecedented accuracy. This relation will then be compared with
  similar relations for AGN at higher redshifts \citep[see, e.g., that
    derived from the \textit{Chandra} Deep Field-South
    survey;][]{yang16}, obtained, however, with much higher
  statistical uncertainties, due to the limited sample size, that
  prevent us to probe effectively the luminosity, redshift and
  timescale dependence of the intrinsic AGN variability)
  \citep[e.g.,][]{paolillo2017}.

\end{itemize}

\section{Radio-loud active galactic nuclei}

\extp is planned in the same time frame as other observatories that
are opening up a window to short time scale variability for AGN, such
as the Cherenkov Telescope Array (CTA) at TeV energies and the Square
Kilometer Array (SKA) and Atacama Large Millimeter Array (ALMA) in the
radio and (sub-)mm bands, respectively (see Fig.~\ref{figfacilities}).
This offers a unique opportunity to interpret short time scale
variability and connect accretion and jet physics in radio-loud AGN.

\extp will be able to follow up a large number of AGN. We estimate
this to be a few thousand with \lfa and a few hundred simultaneously
with the \lad. These numbers are derived from the 30--50 keV $\log
N-\log S$ reported in \citet{giommi2015}, after applying a flux
rescaling to match the \lad and \lfa energy bands.  This extrapolation
is well justified by the fact that a hard X-ray survey preferentially
selects Flat Spectrum Radio Quasars and low energy peaked BL Lacs due
to the hardness of their X-ray spectra, which are generally well
described by a single power law in this entire energy range.  In
addition, the \wfm can provide monitoring of bright AGN on time scales
of days to weeks thanks to its large sky coverage.

Among the radio-loud AGN, blazars are certainly the most promising in
the era of the time domain astronomy.  They are among the most
powerful persistent \citep{urry&padovani95} cosmic sources, able to
release (apparent) luminosities in excess of
$10^{48}\,\mathrm{erg}\,\mathrm{s}^{-1}$ over the entire EM spectrum,
from the radio to the very high energy $\gamma$-ray band. This intense
non-thermal emission is produced within a relativistic (bulk Lorentz
factor $\Gamma=10$--20) jet pointing toward the Earth. Relativistic
effects lead to the beaming of the radiation within a narrow cone of
semi aperture $\sim 1/\Gamma$ rad with strong apparent amplification
of the luminosity and shortening of the variability time scales in the
cases in which (as in blazars) the jet is closely aligned to the line
of sight.

Blazars come in two flavors: those that show optical broad emission
lines (typically observed in quasars that are called Flat Spectrum
Radio Quasars or FSRQs) and those that in general display rather weak
or even absent emission lines, collectively grouped in the BL Lac
object class. The latter generally show the most extreme variability,
with variations as fast as a few minutes, and the most energetic
photons (above 100 GeV).

The conventional scenario \citep[e.g.,][]{gt09} foresees a single
portion of the jet dominating the overall emission, at least during
high activity states. However, this simple idea has recently been
challenged by the observation of very fast (minutes) variability,
which requires the existence of very compact emission sites. The most
extreme example is PKS\,2155$-$304, which experienced two exceptional
flares in 2006 - with a recorded luminosity of the order of several
times $10^{47}\,\mathrm{erg}\,\mathrm{s}^{-1}$ - on top of which
events with rise times of $\sim$100\,s have been detected. Even
assuming that these strong events originated in very compact emission
regions, the physical conditions should be rather extreme and Lorentz
factors of $\Gamma \sim 50$--100 seem unavoidable
\citep{begelman08}. Physical mechanisms possibly triggering the
formation of these compact emission regions include magnetic
reconnection events \citep{giannios13} and relativistic turbulence
\citep{narayan12, marscher14}.

Alternatively, narrow beams of electrons can attain ultra-relativistic
energies ($\Gamma\sim 10^6$) in the black hole vicinity through
processes involving the black hole magnetosphere
\citep{rieger08,ghisellini09}.  In this case, it is expected that the
emission shows up only in the high energy $\gamma$-ray region, giving
no signals at lower frequencies (thus resembling the case of the
so-called ``orphan'' flares; different from the simultaneous X-ray and
TeV flaring observed from PKS\,2155$-$304 in 2006).

\begin{figure*}[ht!]
\centering
\includegraphics[trim=0 0 0 0,width=0.75\textwidth]{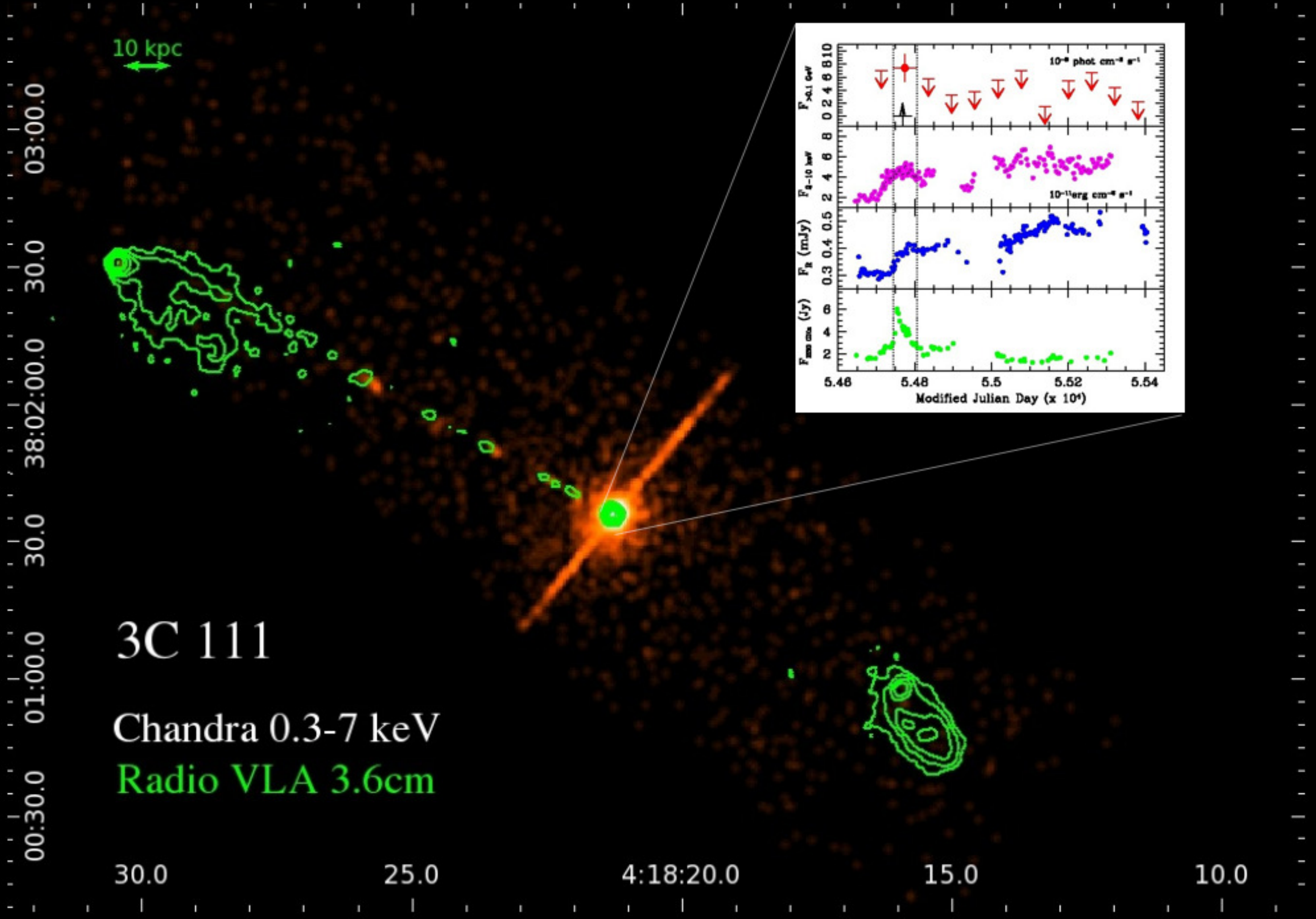}
\caption{\textit{Chandra} 0.3--7\,keV image of the FRII radio galaxy
  3C 111 with 3.6\,cm radio contours overlayed \citep{leahy1997}. In
  the inset the multi-wavelength light curve from mm to $\gamma$-rays
  is shown \citep[from][reproduced by permission of the
    AAS]{grandi2012}. The simultaneity of the flare is impressive: the
  core luminosity was increasing from millimeter to X-ray frequencies
  exactly when the greatest flux of $\gamma$-ray photons
  occurred. This is a clear indication of the cospatiality of the
  event. The outburst of photons is directly connected to the ejection
  of a new radio knot (the time of the ejection is indicated by the
  black arrow).}
\label{figblazars} 
\end{figure*} 

\extp can address the following questions with regards to radio-loud AGN:

\begin{itemize}

\item \textit{What causes the fast $\gamma$-ray flares?} Until now,
  ultra-fast variability has been recorded only in the $\gamma$-ray
  band, especially at TeV energies, at which the Cherenkov arrays are
  characterized by gigantic collection areas, required to probe such
  short time scales. Current instruments do not provide an analogous
  sensitivity in the key X-ray band, at which the low energy component
  of TeV emitting BL Lacs (the so-called HBLs) peaks and thus it is
  not possible to study potential ultra-fast variability events at
  these frequencies. Moreover, such events appear to be quite rare,
  with a duty cycle of less than 1\%. Therefore, monitoring of the
  sky is required to catch these events.

  The \lad, thanks to its timing capability, will allow one to detect
  the X-ray counterpart of the very fast variability observed at TeV
  energies for 2--10 keV fluxes above $2\times 10^{-11}$~\cgsflux,
  shedding light on the nature of the X-TeV connection in HBLs
  \citep{kang12}. Moreover, \extp polarimetric measurements of HBLs
  will constrain the nature of the electron population(s) responsible
  for the X-ray emission, providing us with the magnetic field
  orientation and its change during the flaring activity as compared
  with the average emission.  In these blazars, the polarimeter energy
  band will cover the region of the synchrotron peak, where flux and
  polarization degree demand sub-daily integration times. If one
  assumes a 2--10 keV flux of $2\times 10^{-10}$~\cgsflux, one can
  measure a minimum detectable polarization of $\sim 10\%$ in 5 ks
  with the \gpd. This value is in the range of polarization
  measurements for synchrotron emission as derived from optical
  measurements (between 3\% and 30\%).  A highly ordered magnetic
  field would imply higher polarization degrees.  In this respect,
  \extp will be able to probe the evolution of the magnetic field
  lines and their role in powering fast variations as detected by the
  \lad (flare rise over 100~s as detected in the TeV range).

\item \textit{What accelerates highly energetic electrons?}  A further
  potential of \extp in the study of HBLs is measuring the change in
  spectral curvature of their synchrotron spectra which will enable a
  direct study of the mechanism of acceleration of highly energetic
  electrons. These studies will benefit from the simultaneous
  observations of the \lad and \lfa that provide the broad band
  coverage (0.5--30 keV) of the synchrotron emission.

\item \textit{Where is the high-energy dissipation region?}  The \lfa
  sensitivity is very promising for the study of FSRQs with a
  0.5-10\,keV flux $\gtrsim 2\times 10^{-13}$~\cgsflux. For these
  objects, \extp will probe the rise of the inverse Compton (IC)
  emission whereas new generation radio and (sub-)mm instruments will
  provide excellent timing capabilities to study short time-scale
  correlations with the synchrotron spectral domain.  The nature of
  the photon seeds for the IC emission responsible for the high-energy
  emission is still strongly debated.  Tight constraints are expected
  to come from the investigation of time scales of variability and
  polarimetry.  In this respect, the \lfa will be crucial in a
  multi-frequency context providing: a) temporal investigation on
  second time scales; b) spectral trend investigation on minute time
  scales. With CTA, this may provide a multi-wavelength perspective to
  understand better the rapid (unexpected) TeV emission recently
  detected in some FSRQs. The \wfm will provide not only an excellent
  trigger for both CTA and (sub-)mm observatories but also long-term
  monitoring of a large sample of blazars.

  Moreover, polarimetric measurements performed in the 2--10 keV band,
  in comparison with polarization at lower frequencies, represent a
  powerful tool to solve the long-standing problem of the origin of
  the seed photons and therefore constrain the high-energy dissipation
  region for the brightest objects and flares on daily time scale. The
  average X-ray fluxes of FSRQs are actually much lower than those of
  high synchrotron-peak blazars (in the range
  10$^{-12}$-10$^{-11}$~erg~cm$^{-2}$~s$^{-1}$) and small amplitude
  variations have been detected during the flaring activity. This
  implies much longer integration times to achieve MDP values of the
  order of a few \%, important to disentangle among different leptonic
  models (external Compton versus synchrotron self Compton, SSC) by
  modeling the X-ray emission. In particular, in the case of FSRQs
  with flux $\sim 2\times 10^{-11}$~\cgsflux, a minimum detectable
  polarization of 7\% will be obtainable in 100 ks.

\item \textit{What is the disk-jet connection in misaligned AGN?} \wfm
  long term monitoring (monthly time scale) of radio galaxies hosting
  efficient accretion disks (mainly Fanaroff-Riley II sources,
  hereafter FRIIs; see Figure \ref{figblazars}) will provide an
  optimal tool to investigate the disk-jet connection by triggering
  both radio and \lad X-ray follow-ups. \extp will investigate the
  disk-jet connection in a similar manner as in X-ray binaries, that
  exhibit the same behavior \citep{wu13}.

  Recent detections of radio galaxies hosting inefficient accretion
  flows (Fanaroff-Riley I, hereafter FRI) at high and very high
  energies \citep[e.g., IC 310;][]{aleksic14} are providing a further
  sample of radio-loud sources well suited for follow-up observations
  with \extp. Combining \lfa and \lad with CTA simultaneous observations
  will be crucial in constraining the TeV emitting region, likely
  connected with the magnetosphere surrounding the central engine, as
  derived by the rapid variability observed by MAGIC
  \citep{aleksic14}. FRI behave like high synchrotron-peak
    blazars (with lower luminosities) and therefore simultaneous
  observations in X-rays and at TeV energies are the key to interpret
  their spectral energy distribution in terms of SSC rather than
  hadronic models (not ruled out so far).  Given the typical fluxes of
  TeV FRI (a few $\times$10$^{-12}$~\cgsflux), \lfa (combined with \lad
  for the brightest flares) observations will allow one to detect the
  X-ray counterpart of the fast TeV variations, providing tight
  constraints on the location of the TeV emitting region and the
  particle acceleration process \citep{ahnen17}.

\item \textit{How important is reflection in spectra of radio-loud
  narrow-line Seyfert 1 galaxies (NLSy1s)?} NLSy1s are often
  considered as a third class of AGN with relativistic jets
  \citep{foschini11,ammando15}.  \lad observations will be very
  promising not only for studying the physical mechanisms responsible
  for the higher energy emission (in $\gamma$-rays or at TeV energies)
  and then the blazar-like behavior, but also for determining the role
  of the thermal component.  The combination of \lfa and \lad
  follow-up of the high energy flares from NLSy1 detected in
  $\gamma$-rays or at higher energy $\gamma$-rays will allow one to
  follow the X-ray spectral variations of these sources in the broad
  energy range (0.5--30 keV) down to a $10^{-12}$~\cgsflux, unveiling
  the interplay between the Comptonized radiation from the corona and
  the non-thermal emission from the jet. The \lad observations
  extending up to 30 keV will play an important role in the detection
  and characterization of the Fe-K line, and test with high accuracy
  the reflection model applied to the X-ray spectra of NLSy1.  This is
  to investigate the possible link between accretion and jet
  properties.
\end{itemize}

\section{Tidal disruption events}

Observational evidence is mounting to support a scenario where most
galactic nuclei host SMBHs. Gas inflow causes a small ($\sim$1\%)
fraction of SMBHs to accrete continuously for millions of years and
shine as AGN. Most are expected to be quiescent, accreting if at all
at a very sub-Eddington rate \citep{shankar13}. Observationally, it is
therefore hard to assess the presence and mass of most SMBHs beyond
the local universe. Occasionally, a sudden increase of the accretion
rate may occur if a large mass of gas, for instance a star, falls into
the tidal sphere of influence of the black hole and finds itself torn
apart and accreted. One calls these events 'tidal disruption events'
\citep[TDEs, see][]{rees1988}. TDEs can result in a sudden increase of
EM emission. They can reach the luminosity of a quasar but they are
rare \citep[$\sim 10^{-4}$~yr$^{-1}$ per galaxy; in particular types
  of galaxies, it may be $\sim 10^{-3}$~yr$^{-1}$ per
  galaxy;][]{french2016,vanvelzen2017} and last several months or
years. TDEs are a multi wavelength phenomenon, that has so far been
detected in radio, optical, UV, soft and hard X-rays. They display
both thermal and non-thermal emission from relativistic and
non-relativistic matter \citep[for a recent review on observations,
  see][]{komossa2015}.

\begin{figure*}[ht!]
\centering
\includegraphics[width=\textwidth,angle=0]{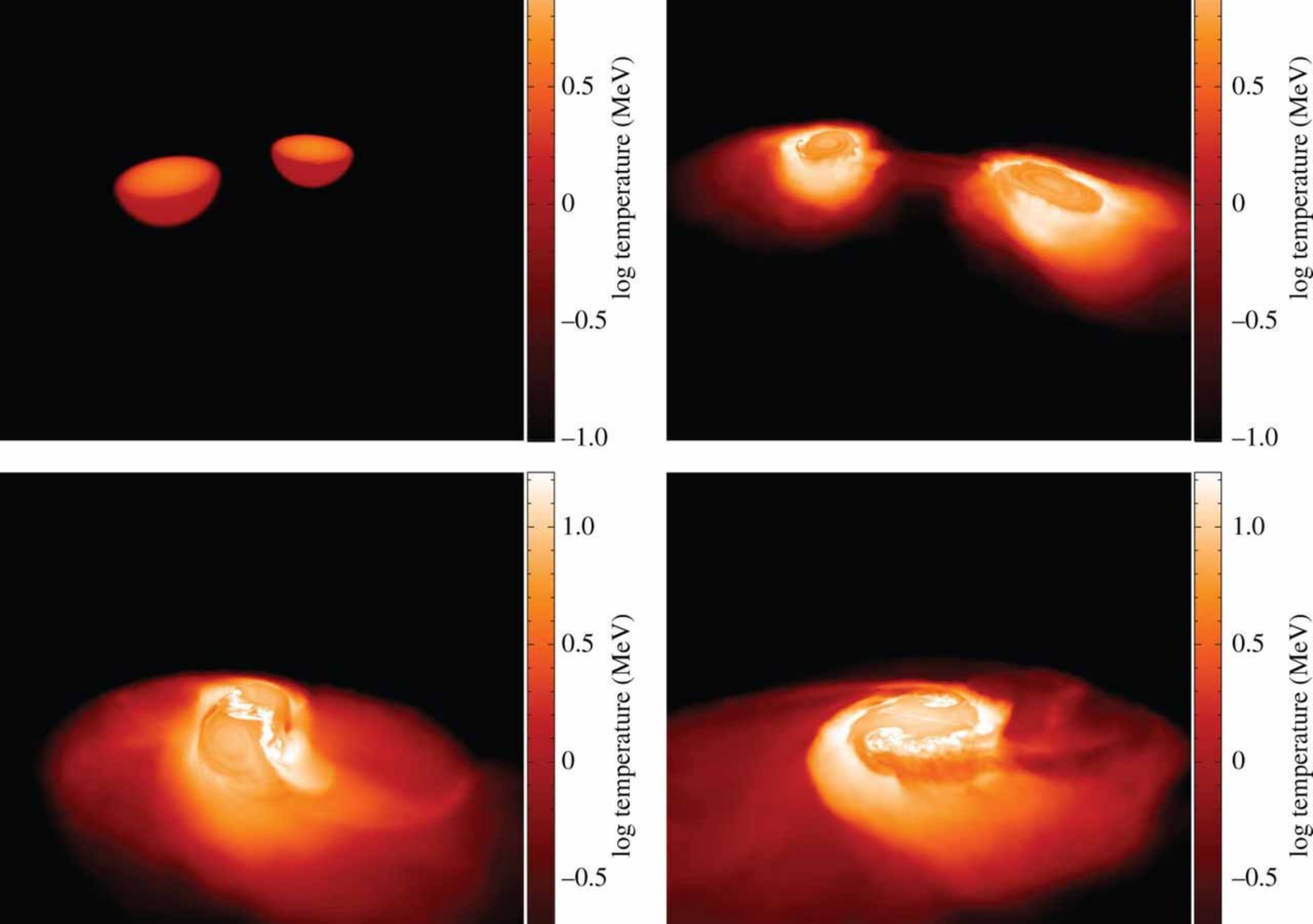}
\caption{Simulation of a neutron star - neutron star merger. The four
  frames are timed within 6.8 ms. From \citet{rosswog2013}.}
 \label{fignsns}
\end{figure*}

The detection and study of these flares can deliver other important
astrophysical information beyond probing the presence of a SMBH.
First, TDEs can allow one to discover intermediate mass black holes,
as their peak bolometric luminosity is expected to be inversely
proportional to the black hole mass. Second, TDEs are signposts of
supermassive {\it binary} black holes, as their light curves look
characteristically different in the presence of a second black hole,
which acts as a perturber on the stellar stream
\citep{liu09}. Recently, a disruption by a SMBH binary candidate was
claimed \citep{liu14}.  Thirdly, as also mentioned below, TDEs give us
the opportunity to study accretion and jet formation in transient
($\sim$month to year time scale) accretion episodes around SMBHs in
contrast to AGN.  Fourthly, dormant SMBHs represent the majority of
the SMBHs in our universe. X-ray flares from TDEs provide us with the
opportunity to probe the mass and spin distribution of previously
dormant SMBHs at the centers of normal galaxies as well as the
geometry of the debris flow of the accretion flare through X-ray line
reverberation \citep{zhang2015}. Finally, in X-rays TDEs represent a
new probe of strong gravity, for instance tracing precession effects
in the Kerr metric \citep[see, e.g.,][]{hayasaki16, franchini16}.

The first TDEs were discovered in \textit{ROSAT} surveys of the X-ray
sky \citep[e.g.,][]{KomossaBade99,Grupe99}, see \citet{komossa2002}
for a review. Later, \textit{GALEX} allowed for the selection of TDEs
at UV frequencies \citep{gezari2009, gezari2012}. Many of the most
recent TDE candidates are found in optical transient surveys
\citep{vanvelzen11a, cenko,chornock14}. An alternative method to
select TDE candidates is to look for optical spectra with extreme
high-ionization and Balmer lines \citep{komossa08, Wang12}. This class
of TDEs is called 'thermal', and the spectral energy distribution is
believed to be associated with phenomena that lead to mass accretion
onto the black hole.

Currently, the theoretical picture proposed to interpret the
observational properties of TDEs is as follows. After stellar
disruption, part of the stellar material is accreted onto the black
hole, causing a luminous flare of radiation. If the star is completely
disrupted, its debris is accreted at a decreasing rate according to
$\dot{M} \propto t^{-5/3}$ \citep{rees1988,phinney}. Therefore, TDEs
allow one to study the formation of a transient accretion disk and its
continuous transition through different accretion states.  The
super-Eddington phase - which occurs only for SMBH masses
$M<10^7$~M$_\odot$ - is theoretically uncertain, but it may be
associated with a powerful radiatively driven wind \citep{rossi09}
that thermally emits $10^{41}-10^{43}$~erg~s$^{-1}$ mainly at optical
frequencies \citep{strubbe09,lodatorossi}. The disk luminosity
($10^{44}-10^{46}$ erg~s$^{-1}$) peaks instead in the far-UV/soft
X-rays \citep{lodatorossi}. As an alternative scenario to thermal
emission from an accretion disk, the optical emission has been
suggested to come from shocks as the stellar debris self-crosses or
from reprocessed X-ray light \citep[e.g.,][]{piran2015}.

A few years ago, the \textit{Swift} Burst Alert Telescope (BAT)
triggered on two TDE candidates in the hard X-ray band
\citep{bloom,cenko}. A multi-frequency follow-up from radio to
$\gamma$-rays revealed a new class of \textit{non-thermal} TDEs. It is
widely believed that emission from a relativistic jet \citep[bulk
  Lorentz factor $\Gamma \approx 2$,][]{zauderer13,berger12} is
responsible for the hard X-ray spectrum (with a negative power-law
photon index between -1.6 and -1.8) and the increasing radio activity
\citep{levan2011} detected a few days after the trigger. More
recently, \citet{brown2015} reported another candidate relativistic TDE
(Swift J1112.2--8238). \citet{hryniewicz2016} found nine TDE candidates
in BAT data after searching the data from 53,000 galaxies out to 100
Mpc. Finally, \citet{alexander2016} and \citet{vanvelzen2016} discovered
the first radio outflow associated with a thermal TDE, suggesting that
jets may be a common feature of TDEs.

The best studied of the relativistic events is Swift J1644+57
\citep[Sw J1644 in short;][]{levan11,burrows11}. The two main features
that support the claim that Sw J1644 is a TDE are i) the X-ray light
curve behavior, that follows $L_{\rm X} \propto \dot{M}
(t-\tau)^{-5/3}$ after a few days ($\tau\approx 3$ day) from the
trigger \citep{donnarumma2015}, and ii) the radio localization of the
event within 150 pc from the center of a known quiescent galactic
nucleus \citep{zauderer}.

If one uses the Sw J1644 radio and X-ray light curves to estimate
prospects for any future planned missions, one obtains that
non-thermal TDEs can be discovered at a higher rate thanks to triggers
provided by radio surveys and to follow-ups at higher energies. The
wide field-of-view survey of SKA will effectively detect TDEs in the
very early phase, allowing follow-up X-ray observations in the very
early phase \citep{yu2015,donnarumma2015b}. In synergy with SKA in
survey mode, the \lfa can in principle successfully repoint to each
radio candidate in the sky accessible at any moment to \extp (i.e.,
50\% of the time), and measure its light curve decay index and
spectral properties.  The X-ray counterpart of radio-triggered events
can be followed up with the \lad too \citep[see Fig.~ 6
  in][]{donnarumma2015}, thus extending the energy coverage up to 30
keV with a better characterization of the non-thermal process and of
the jet energy budget. The X-ray polarimeter will definitely play the
major role. For events followed up within a few weeks from the
beginning, a fraction of $\sim$10\% of the radio-triggered events will
have a 2--10 keV flux greater than $10^{-10}$~\cgsflux. For these
events, it is possible to derive a minimum detectable polarization of
$\sim$10\% over a 5 ks time scale. This will allow one to test the
variation of magnetic field over a time scale of one hour and derive
the time needed for the transient jet to build up a large scale
coherent magnetic field. Moreover, if \extp were able to follow-up
TDEs within 1 day, with the trigger provided by LSST, SKA, the
Einstein Probe or \wfm and if the source is bright enough in 2--10 keV
($>10^{-10}$~\cgsflux), it will be possible to investigate also
polarization angle variations aimed at assessing the role of jet
precession in the early phase of the event. In this case, one foresees
a set of daily (50 ks) observations
\citep{saxton2012,tchekhovskoy2014} spread over 10 days in order to
appreciate angle variations within a few degrees.

\extp is therefore very well suited for the X-ray follow-up of jetted
TDEs in the SKA era. The expected rate is between a few to several
tens per yr (depending on the value of $\Gamma$).  The attractive
prospect is that of building for the first time a sample of well
studied TDEs associated with non-thermal emission from jets. This will
allow one to study disk-jet formation and their connection, in a way
complementary to other more persistent sources such as AGN and X-ray
binaries. In addition, jetted TDEs can uniquely probe the presence of
SMBHs in \textit{quiescent} galaxies well beyond the local universe
(redshift range 1.5-2).

Moreover, the \lfa energy band and sensitivity will enable the
follow-up of thermal TDEs, that are expected to be triggered in a
large number with LSST. These will be studied with eXTP in a tight
connection with the X-ray binary transients.  The combination of the
soft telescopes, the broad energy coverage and the polarimetric
capabilities makes eXTP a great X-ray mission to study all the variety
of tidal disruption events and to provide the unique opportunity to
answer the following questions:

\begin{itemize}

\item \textit{What is the rate of non-thermal TDEs?} The \lfa
  sensitivity will allow one to follow up and identify any radio
  triggered TDEs. The \wfm can allow the serendipitous detection of
  non-thermal TDEs in X-rays at a higher rate (between one and a few
  tens per year) than any current missions. The combination of the two
  will provide a unique tool to establish the rate of non-thermal TDEs
  and study the mechanisms powering the hard X-ray emission.

\item \textit{Do TDEs exhibit similar spectral states as X-ray binary
  transients?} The \lfa energy band and sensitivity will enable the
  follow-up thermal TDEs that are expected to be triggered in large
  numbers with LSST. Several thousands per year are expected to be
  triggered up to $z\sim1$ \citep{strubbe09}. The X-ray follow-up with
  \lfa will make it possible to study the evolution of the thermal TDEs
  and possibly see the transition from soft to hard spectra as
  expected at later stages due to the coronal Comptonized component
  \citep{lin17}.

\item \textit{What are the masses and spins of the previously dormant
  SMBHs at the centers of normal galaxies?} AGN cover only a small
  portion of the SMBHs in our universe. The distribution of the masses
  and the spins of these SMBHs is important to our understanding of
  the cosmological growth of mass and evolution of the spin of the
  majority SMBHs. At present and in the near future, we are only able
  to study the properties of these dormant SMBHs (also some
  intermediate-mass black holes in dwarf galaxies) through these
  flaring events. Among them, X-ray flaring TDEs allow us to probe the
  mass and spin with X-ray line reverberation
  \citep[e.g.,][]{zhang2015}. For future X-ray bright relativistic
  TDEs or thermal TDEs, it is possible to detect spectral features
  with 1 ks \lfa and \lad exposures.

\end{itemize}

\section{Gamma-ray bursts and supernovae}
\label{sec:grbs}

\noindent
Gamma-ray bursts \citep[GRBs; for reviews,
  see][]{piran2004,Meszaros06, Gehrels09} are short ($\sim$10~s),
bright (up to 10$^{-5}$~erg~s$^{-1}$ in 10 keV -- 1 MeV), energetic
(with fluences up to more than
$10^{-4}\,\mathrm{erg}\,\mathrm{cm}^{-2}$) and frequent (about one per
day).  Observational efforts since the 1990s have provided, among
others, 1) the accurate localization and characterization of
multi-wavelength afterglows \citep{costa1997,vanparadijs1997}, 2) the
establishment of cosmological distances and supernova-like radiated
energies \citep{metzger1997}, 3) the identification of two classes of
GRBs, long and short, with a division at 2 s \citep{Kouveliotou93},
and 4) the connection of long GRBs with Type Ib/c supernovae
\citep{galama1998}.  The most important open questions in our current
understanding of GRBs involve: the physical origin of sub-classes of
GRBs (i.e., short, X-ray rich, sub-energetic and ultra-long GRBs), the
physics and geometry of the prompt emission, some unexpected early
afterglow phenomenology (e.g., plateaus and flares), the connection
between GRBs and particular kinds of supernovae, the nature of the
'central engine' and the use of GRBs as cosmological probes.

\begin{figure*}[ht!]
\centering
\includegraphics[height=0.6\textwidth]{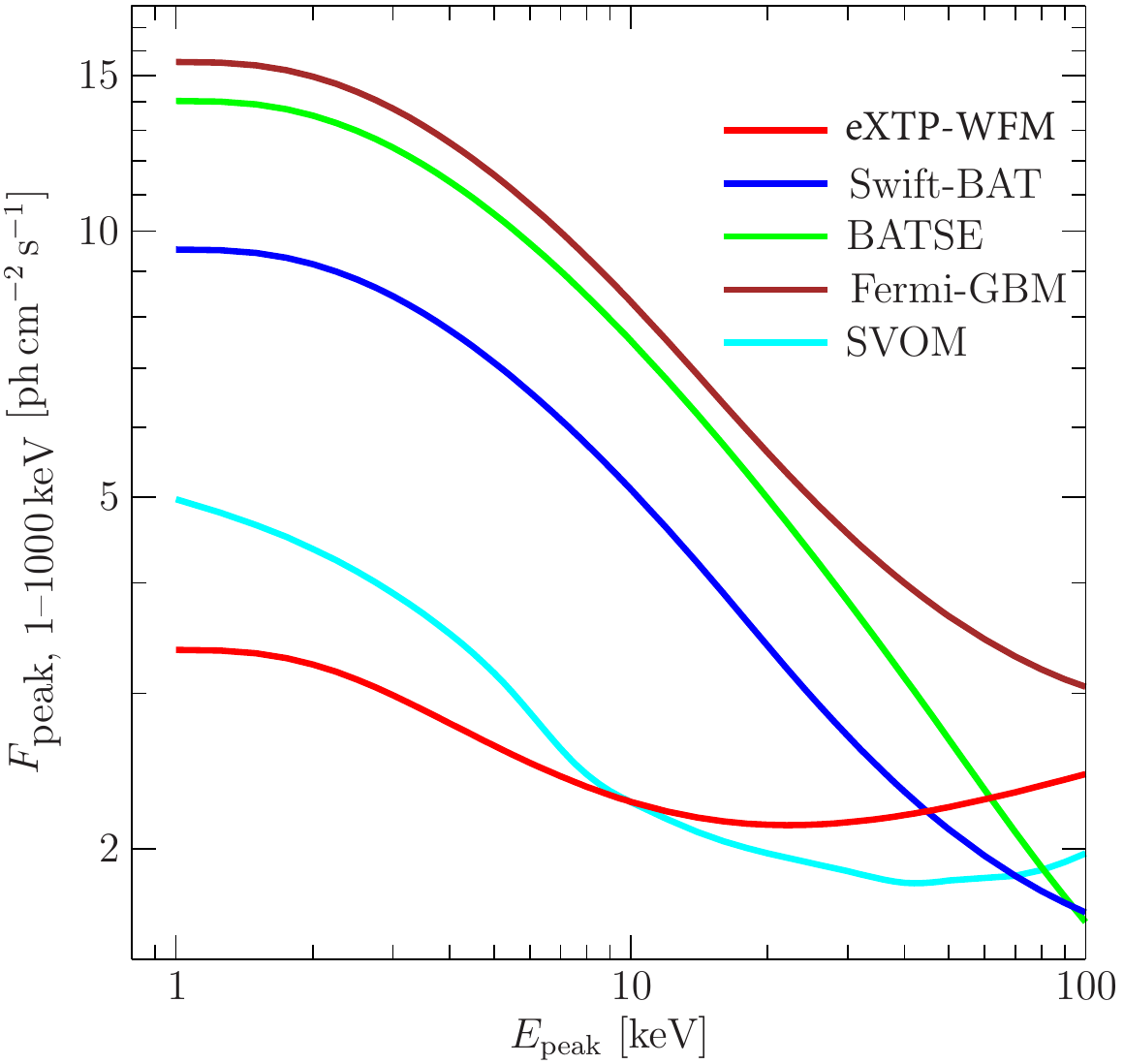}
\caption{GRB detection sensitivity in terms of peak flux sensitivity
  as a function of the spectral peak energy $E_p$ \citep{Band03} of
  the \wfm (red) for M4 configuration compared to those of
  \textsl{CGRO} BATSE (green), \textsl{Swift} BAT (blue),
  \textsl{Fermi} GBM (brown), and \textsl{SVOM} ECLAIRs
  (cyan).  \label{figgrb2}}
\end{figure*}

Short GRBs were long suspected to be connected to binary neutron-star
merger events \citep{eichler1989,kouveliotou1993}. The detection of
short GRBs associated with old stellar populations, their offsets from
their host galaxies and the identification of a candidate kilonova
explosion supported this association
\citep[e.g.][]{gehrels2005,fong2010,tanvir2013}. The suspicion
received unprecedented support through the coincident detection of the
short GRB170817A \citep{grb170817A} and the binary merger
gravitational-wave event GW170817 \citep{gw170817,grbandgw170817}. The
masses of the two binary components measured through the GWs are more
consistent with a set of dynamically measured neutron star masses than
with that of black holes. The tidal disruption of neutron stars (see
Fig.~\ref{fignsns} for stills from a simulation) is expected to yield
EM radiation because of the delayed accretion of part of the expelled
matter onto the merger product, the strong radio-active decay of
r-process elements produced during the merger, the shocks resulting
from an expanding cocoon and/or jets.

The prospect of GW detections with Advanced LIGO and Virgo, KAGRA and
LIGO-India in the 2020s have been calculated to be rather promising
(tens to hundreds per year) for distances of up to $\sim$300-500 Mpc
\citep{livingreview,yang2017}.  So far, very few \citep[4-5; see,
  e.g.,][]{gehrels06,fong15} supernova-less GRBs (including the
so-called long-short GRBs and short GRBs) have been localized within
this distance regime. This may improve with the softer response by the
\wfm and its larger field of view.  It was thought that, due to the
beaming effect in special relativity, only relativistic jets from the
mergers of two neutron stars or of a black hole and a neutron star
beamed toward the Earth can be detected as short GRBs.  The detection
of the off-axis GRB170817A \citep{grb170817A}, without any noticeable
signal of a relativistic jet in our line of sight, suggests better
prospects for the detection of off-axis X-ray emission, possibly
produced as a result of the shock propagation or by the mildly
relativistic cocoon \citep[see, e.g.,][]{salafia16,lazzati17}. This
emission has a much larger opening angle, and despite the much lower
luminosity, it might still be observable by \extp within the
LIGO-Virgo detection horizon. Moreover, independently from short GRBs,
some models predict detectable and nearly-isotropic X-ray emission
powered by the merger remnant \citep{zhang2001, metzger2008,
  rowlinson2013, zhang13, gao13, siegel16a, siegel16b}.  With its
unprecedented large field of view, \wfm may help to confirm or exclude
these models by possibly detecting EM signals from the remnant on the
needed time scale (from $\sim1$~s to longer). GRB170817A would have
been detectable with the \wfm with a signal-to-noise ratio during the
peak of up to about 10 if its spectrum would be on the 1$\sigma$ soft
side of the error margins of the spectral shape parameters measured
with Fermi-GBM \citep{grb170817A}.

\begin{figure*}[ht!]
\centering
\includegraphics[height=0.45\textwidth,angle=270]{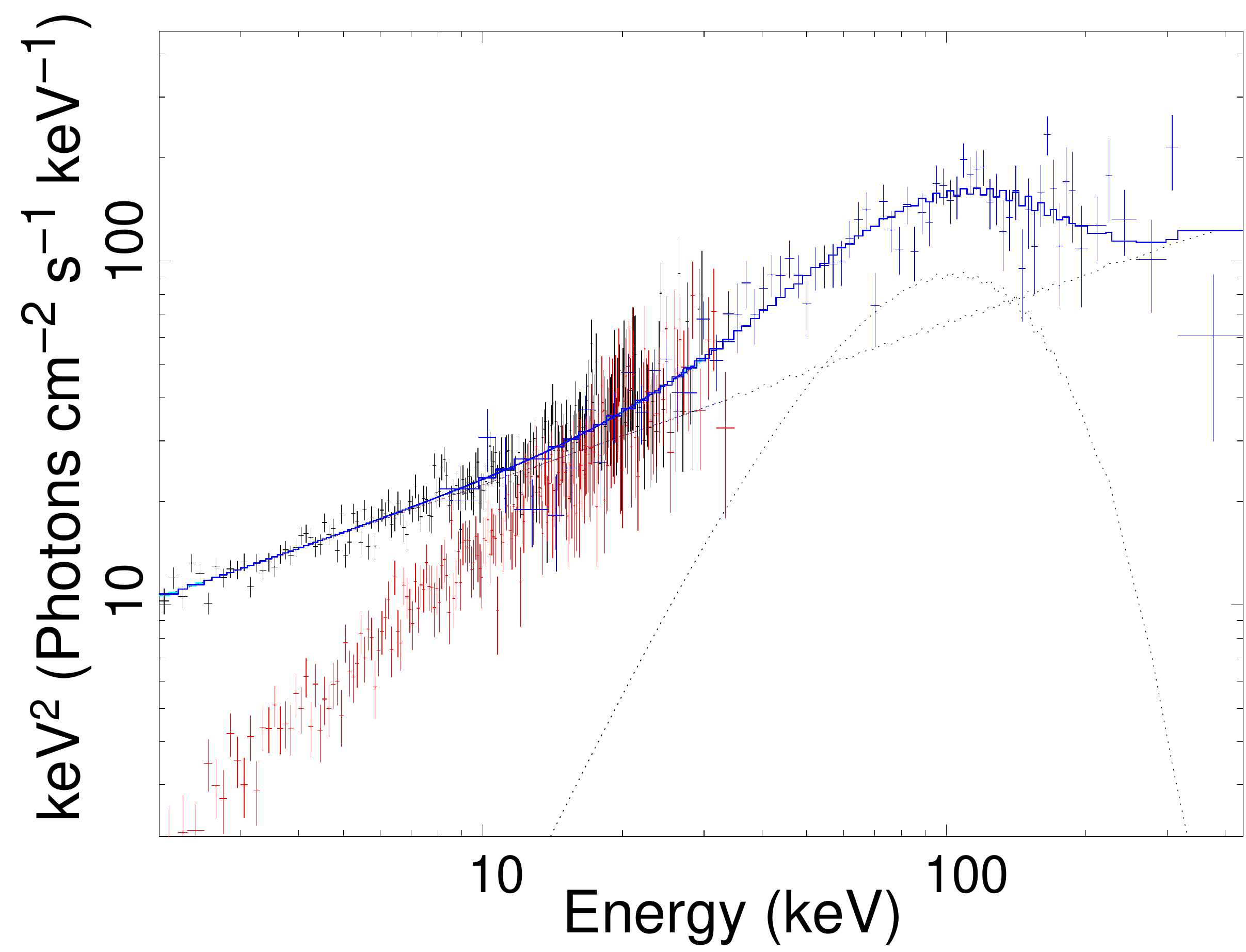}\hspace{0.4cm}
\includegraphics[height=0.48\textwidth,angle=270]{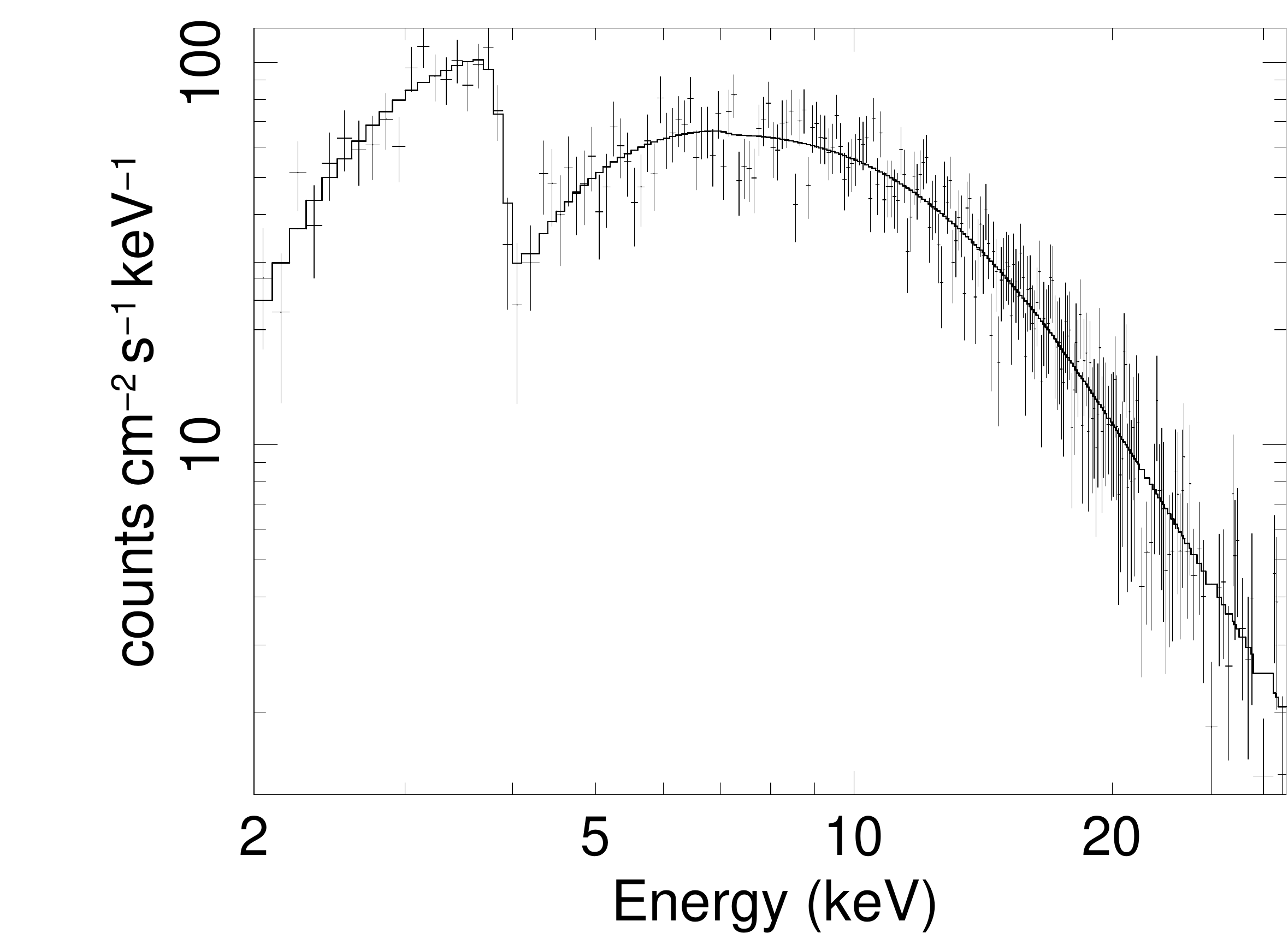} 
\caption{Illustrative plots of \wfm capabilities and of their impact
  on GRB science. \textit{Left:} Simulated \wfm spectra of the first
  $\sim$50 s of GRB\,090618 (from \citealt{Izzo12}) obtained by
  assuming either the Band-function extrapolation (black) or the
  power-law plus black-body model extrapolation (red) to the {\it
    Fermi} GBM measured spectrum (blue). The black body and power-law
  model components best-fitting the {\it Fermi} GBM spectrum are also
  shown in black dashed lines. \textit{Right:} Simulation of the
  transient absorption feature in the X-ray energy band detected by
  \textit{BeppoSAX} WFC in the first 8\,s of GRB~990705 as it would be
  measured by the \wfm \citep{amati2000a}. \label{figgrb1}}
\end{figure*}

The nature of the second supernova in tight binaries, which are
responsible for producing the binary neutron star progenitors of short
GRBs and GW sources, is an ultra-stripped supernova
\citep{tauris2013}. These supernovae produce relatively faint and
rapidly decaying light curves \citep{tauris2015,moriya2015}. A prime
candidate for such a supernova event is SN 2005ek
\citep{drout2013}. Furthermore, as a result of severe mass transfer in
the X-ray phase just prior to these supernovae (via so-called Case BB
Roche-lobe overflow), such systems are also likely to be observable
with eXTP as pulsating ultra-luminous X-ray sources (see
Sect.~\ref{sec:ulx}).

Another interesting class of soft GRBs may arrive from supernova shock
breakouts.  Wide-field soft X-ray surveys are predicted to detect each
year hundreds of supernovae in the act of exploding through the
supernova shock breakout phenomenology
\citep{horiuchi11,campana06}. Supernova shock breakout events occur in
X-rays at the very beginning of the supernova explosions, allowing for
follow-up at other wavelengths to start extremely early on in the
supernova, and possibly shedding light on the nature of the progenitor
of some types of supernovae. To date, only one supernova shock
breakout has been unambiguously detected, in {\it Swift} observations
of NGC 2770 \citep{Soderberg08}. It has a fast-rise, exponential decay
light curve with a rise time of 72 seconds, and an exponential decay
time scale of 129 seconds, and a peak X-ray luminosity of
$6.1\times10^{43}$ ergs~s$^{-1}$.

An interesting target for \extp are the recently discovered ultra-long
GRBs \citep{gendre2013,levan2015}. These events are soft gamma-ray
bursts lasting hours (instead of seconds). Only a few events are
known, and most of their properties remain to be studied. We know
however that these events present a possible faint thermal component,
and that they may be linked to progenitors very similar to population
III stars \citep[i.e. extremely massive, metal poor
  stars;][]{ioka2016}. These objects are not brighter than normal long
GRBs (albeit the total energy budget is far larger), but last for so
long that it is possible to re-point \extp and still to observe the
very bright prompt phase. This brightness, combined with the
unprecedented large collecting area of the \lad, would allow spectral
studies impossible with any other mission to date, studying in detail
the evolution of the thermal component and its still debated nature
\citep{piro2014}.

Thanks to its unique combination of field of view, broad bandpass,
spectral resolution, large effective area, and polarization
capabilities, the instruments on board \extp, particularly the \wfm
(see Fig.~\ref{figgrb2}), can address questions of fundamental
importance to understanding the above phenomena:

\begin{figure*}[ht!]
\centering
\includegraphics[height=0.48\textwidth,angle=270]{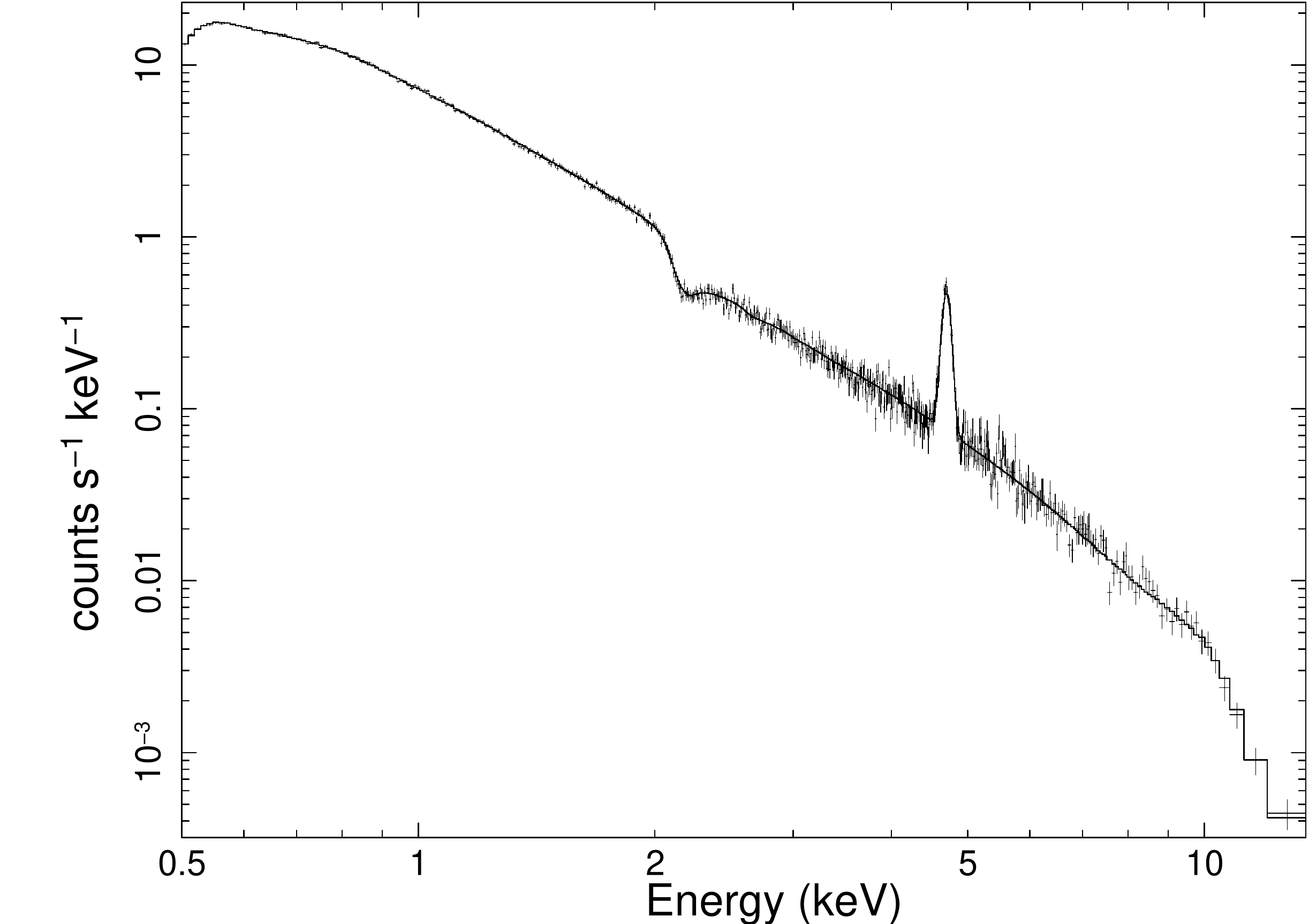}
\includegraphics[height=0.50\textwidth,angle=270]{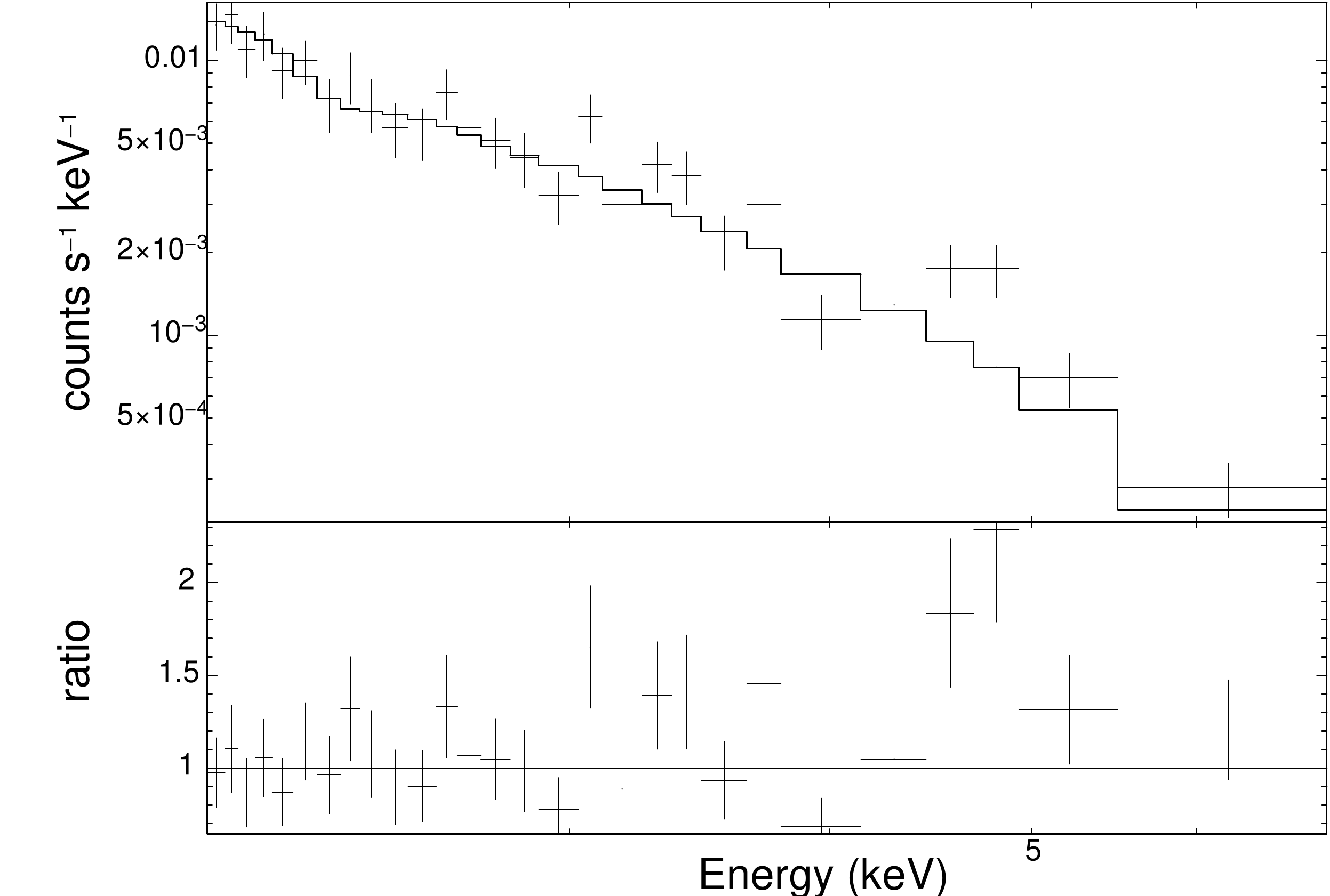}
\caption{As an example, we considered the redshifted Fe-K line that
  was measured by \textit{BeppoSAX} in the X-ray afterglow spectrum of
  GRB000210 \citep{antonelli2000}. We show in the left panel a
  simulated \lfa spectrum as obtained considering the same afterglow
  intensity and line equivalent width as detected by \textit{BeppoSAX}
  12~hr after the onset of the GRB. An exposure time of 50~ks is
  assumed (average flux of $8\times10^{-13}$~\cgsflux). The detection
  of the iron line with the \extp instrument would be outstanding.
  The right panel shows the same simulation for {\it Swift}\,/XRT. No
  line is detected. This confirms the low sensitivity of XRT to
  similar features.  These unique and unprecedented results would
  provide an important step forward in the investigation of the
  circumburst material composition and distribution, with important
  consequences for shedding light on the nature of GRB progenitors and
  emission physics.}
  \label{figgrbiron}
\end{figure*}

\begin{itemize}
\item \textit{What are X-ray Flashes?} \extp can enable a substantial
  increase (with respect to past and current missions) of the
  detection rate of X-Ray Flashes (XRFs), events commonly interpreted
  as a softer sub-class of GRBs \citep[but see also][]{Ciolfi2016}
  which likely constitutes the bulk of the GRB population but still
  lacking a good body of observational data
  \citep{heise2001,sakamoto2005,sakamoto2008}.  With respect to {\it
    HETE-2} WXM (2--25\,keV), the most efficient XRF detector flown so
  far, \wfm will increase the XRF detection rate 7-fold, with an
  expected rate of 30--40 XRF per year
  \citep[e.g.,][]{martone2017}. See also Fig.~\ref{figgrb2}.

\item \textit{What are the main radiative mechanisms at the basis of
  GRB prompt emission?}  Measurement of the GRB spectra and their
  evolution down to 2\,keV are an indispensable tool for testing
  models of GRB prompt emission and will be studied with unprecedented
  detail thanks to the \wfm's unique capabilities (see
  Fig.~\ref{figgrb1} left).

\item \textit{What is the nature of GRB progenitors?} Transient
  absorption edges and emission lines in the prompt and afterglow
  X-ray spectra of long GRBs are expected in several scenarios for
  their origin. Firm detections or deep limits on such features would
  be essential for the understanding of the properties of the
  circumburst matter (CBM) and hence the nature of GRB progenitors. In
  addition, they could enable us to determine the GRB redshift
  directly from X-ray observations. Up to now, only a few marginal
  detections were provided by \textit{BeppoSAX}, \xmm and
  \textit{Chandra}, while no significant evidence for line emission
  was found in {\it Swift} afterglow data. As illustrated in
  Figs.~\ref{figgrb1} \ref{figgrbiron} (right panels), \extp,\ thanks
  to the good energy resolution of both \wfm and \lfa, combined with a
  large field of view and a large area, respectively, can greatly
  contribute to solve this issue and possibly open a new discovery
  space in the understanding of the GRB phenomenon.

\item \textit{Can supernovae be discovered promptly?} Supernova
  explosions can be detected at their onsets thanks to the shock
  breakout effect. According to recent supernova rate estimates
  \citep{horiuchi11}, about one to two supernova breakouts per year
  should be detectable with the \wfm out to 20 Mpc.  If one takes the
  case of NGC 2770 as a template, the optimum integration time to
  detect such events out to 20 Mpc is 240 s (the \wfm sensitivity for
  the detection of impulsive events within this time interval is
  $\sim$30 mCrab).

\item \textit{What is the structure and evolution of GRB jets?}  The
  contribution to GRB science by the \lad and especially the \gpd may
  provide breakthrough information on the jet structure and on the
  nature of the afterglow flares and plateaus \citep{burrows05,
    zhang06, obrien2006, nousek06}, as different models make different
  predictions on the evolution of the polarization during these phases
  \citep{sari00, dai04}.  However, this will strongly depend on
  spacecraft re-pointing capabilities and is likely limited to the
  ~20\% brightest events.  For instance, by using the mean parameters
  of the X-ray afterglow light curves quoted by \citet{margutti2013}
  (i.e., a flux normalization at the beginning of the plateau of
  $6\times10^{-10}$~\cgsflux\, and 3.96, -0.16 and 1.59 for the decay
  indices in the three phases - steep, shallow, normal- respectively),
  we computed the average flux for an exposure of 100 ks starting at
  different epochs after the trigger. We find that sensitive
  polarimetric studies with the \gpd (i.e., average fluxes above 1
  mCrab) can be performed for a typical X-ray afterglow only if it is
  observed no later than a few minutes after the GRB trigger.
  However, when considering X-ray afterglows with a bright plateau
  phase (i.e., 10 times brighter at the start of the plateau), an
  average flux of few mCrab can be reached for a 100 ks exposure even
  starting the observations after 6 hours.

\item \textit{What is the end product of a neutron star merger?}  The
  end product of a merger of two neutron stars can either be a black
  hole or a heavy neutron star that may be hypermassive (temporarily
  supported by for example fast spins) with magnetic fields of order
  10$^{15}$~G \citep[thus a magnetar, e.g.,][]{rowlinson2013}. This
  probes a very interesting question: what is the threshold mass above
  which a neutron star cannot exist (see also Sect.~\ref{sec:lmxbs})?
  This is intimately related to the neutron star equation of state. A
  possible observational means to distinguish between both kind of
  compact objects is, apart from the GW signal \citep{WP_DM}, the
  detection of a pulsed EM signal after the merger event, for instance
  in the so-called plateau phase of GRB afterglows that is suggested
  to be due to rapidly rotating magnetar
  \citep[e.g.][]{lasky2014,rowlinson2017}. \extp would be well
  equipped to hunt for such a signal, provided the \lad is pointed
  quick enough and the prompt signal is very luminous (roughly in the
  neighborhood of its Eddington limit for a distance of 40 Mpc and a
  signal duration of 10$^4$ s, or more).

\end{itemize}

Finally we note that, as a public community service, \extp will
include a capability to localize short transients automatically
onboard using data from the \wfm. These positions, or a subset of
them, will be disseminated via a VHF network and the internet in a
similar fashion as for the SVOM mission
\citep{schanne2015,WPinstrumentation}.

\section{Objects of uncertain nature}
\label{sec:misc}

\subsection{Ultraluminous X-ray sources}
\label{sec:ulx}

Ultraluminous X-ray sources (ULXs) have observed fluxes indicating
(assuming isotropic radiation) $L_X>10^{39}$ erg~s$^{-1}$, and often
$>10^{40}$ erg$^{-1}$.  As these luminosities exceed the Eddington
limit (at which released radiation drives away in-flowing mass) for
typical stellar black hole masses, they indicate the presence of
either much larger-mass black holes, of super-Eddington accretion
rates, and/or beaming of radiation into limited angles. The recent
discovery of X-ray pulsations by XMM-Newton in three ULXs
\citep{Bachetti2014, Fuerst2016, Israel2017a, Israel2017b} proves that
some ULXs contain neutron stars, and the sinusoidal pulse shape
indicates low beaming, so these systems clearly accrete at
super-Eddington rates \citep[likely from high-mass donors,
  e.g.,][]{Rappaport2005, Motch2011}.  These systems are now being
intensively studied, to try to understand the magnetic fields of their
neutron stars, their long-term spin behavior, and the nature of the
accretion flows.

\extp can provide a substantial contribution to ULX science, by
finding and timing ULX pulsars.  \extp’s \lfa has an effective area
roughly 5.5 times that of XMM-Newton’s pn camera at 6 keV (the
relevant comparison, since ULX pulsations are stronger at high
energies), and an angular resolution of 1\arcmin, enabling high count
rates for relatively faint objects.  As pulsation detection is
generally a function of count rate and pulse fraction, the higher \lfa
effective area means that \extp should be able to detect pulsations
over twice as far away (e.g., $\sim$10 Mpc vs. 5 Mpc for $10^{40}$
erg/s ULXs with 20\% pulsed fractions), and thus within an effective
volume 8-10 times larger. A systematic \extp program to survey bright
ULXs within 10 Mpc (identified by, for example, eROSITA on
Spectrum-X-Gamma, along with Chandra and XMM-Newton surveys), and the
brightest ULXs beyond that, may discover $\sim$30 new ULX pulsars,
allowing significant statistical conclusions about the ULX
population. Repeated timing of these ULXs will measure their spin-up
rates. \extp’s high effective area will also allow studies of QPOs
\citep{Strohmayer2003}, time lags \citep{deMarco2013}, and eclipse
searches \citep[e.g.,][]{Urquhart2016}, opening up substantial
discovery space for these extreme objects.

\subsection{Fast radio bursts}

In recent years, there has been much excitement about the enigmatic
Fast Radio Bursts (FRBs; Lorimer et al., 2007). These are bright,
millisecond duration, radio bursts with large dispersion measures
pointing at extra-galactic distances and large luminosities. A few
tens have been detected since 2001 and estimates about their all-sky
rate range into the thousands per day
\citep[e.g.,][]{keane2015}. Radio telescopes are being adapted to
detect many more and aiming to distribute alerts in close to real time
\citep[e.g.,][]{petroff2017}. The origin of these bursts is
undetermined and the models range from neutron-star collapse to black
holes, neutron star or white dwarf mergers and non-cataclysmic pulsar
events \citep[for a review, see][]{rane2017}. There is one FRB source
which has repeated tens of times (Spitler et al. 2016), while all
other observed FRBs have not been observed to repeat \citep[although
  it is possible that repeats have been missed due to instrument
  sensitivity limitations and survey strategies;][]{petroff2015}. The
case of the repeater gives rise to a models where the energy is from
magnetic reconnection events from a young magnetar
\citep[e.g.,][]{katz2016} or giant pulses from a rapidly rotating
neutron star \citep[e.g.,][]{lyutikov2016}. However, there is no
consensus yet \citep{rane2017}.

To date FRBs have no counterparts outside the radio band. However,
there are models predicting FRBs associated with some GRBs and, hence,
a faint, broadband afterglow may be expected
\citep[e.g.,][]{zhang2014}. Coordinated searches of FRBs in the
repeating source revealed no detection in X-rays and gamma-rays with
0.5- 10 keV flux limits of order $10^{-7}$ erg s$^{-1}$ cm$^{-2}$
\citep[e.g.,][]{scholz2017}. This limit can be vastly improved with
\lad and \lfa to values of 10$^{-10}$ to 10$^{-11}$ erg s$^{-1}$
cm$^{-2}$. Because of the \wfm’s large field of view, the possibility
of a FRB being detected inside that field of view is
enhanced. Detection of the prompt few millisecond duration emission
will be difficult though given the flux limit of approximately
$10^{-6.5}$ erg s$^{-1}$ cm$^{-2}$. If there is a longer duration
X-ray counterpart, chances will become better.

Conversely, eXTP could instead provide candidates for targeted
searches for FRBs. Many of the progenitor theories require a young
neutron star that is rapidly rotating and/or has a high magnetic field
\citep[e.g.,][]{rane2017}. Neutron stars of this type are proposed to
be the central engines of some GRBs and superluminous supernovae (see
Section 11). Additionally, the host galaxy of the repeating FRB is
consistent with the host galaxies of long GRBs and superluminous
supernovae \citep{metzger2017}. However, for long GRBs and
superluminous supernovae progenitors, it will take a few decades to
clear the surrounding medium to allow the radio emission to escape
\citep{metzger2017}. The short GRB progenitors do not have this issue
and bursts may escape much sooner. eXTP will be able to identifying
the plateau phase \citep[the signature of the formation of a
  magnetar][]{zhang2001} following GRBs and these sources can be
targeted with rapid response and late-time FRB searches \citep[see][
  and references therein]{chu2016}.

\vspace{1cm}\noindent {\bf Acknowledgements: } We thank the external
referees for their advise prior to submission and A. Patruno for a
revision of Fig.~\ref{figMaccarone1b}. The Chinese team acknowledges
the support of the Chinese Academy of Sciences through the Strategic
Priority Program of the Chinese Academy of Sciences, Grant
No. XDA15020100. D. Altamirano acknowledges support from the Royal
Society. A. Bilous, F. Chambers and A. Watts are supported by ERC
Starting Grant 639217. Y. Cavecchi is supported by the European Union
Horizon 2020 research and innovation programme under the Marie
Sklodowska-Curie Global Fellowship (grant no. 703916). Y. Chen
acknowledges support through NSFC grant 11233001, 11773014 and
11633007 and the 973 Program grant 2015CB857100. B. Gendre
acknowledges support through NASA grant NNX13AD28A. A. Heger was
supported by an ARC Future Fellowship (FT120100363) and by NSF grant
PHY-1430152 (JINA Center for the Evolution of the
Elements). K. Iwasawa acknowledges support by the Spanish MINECO under
grant AYA2016-76012-C3-1-P and MDM-2014-0369 of ICCUB (Unidad de
Excelencia 'Mar\'ia de Maeztu'). M. Linares is supported by EU's
Horizon programme through a Marie Sklodowska-Curie Fellowship (grant
nr. 702638). A. Rozanska was supported by Polish National Science
Center grants No. 2015/17/B/ST9/03422and 2015/18/M/ST9/00541. Z. Yan
is supported by the National Natural Science Foundation of China under
grant numbers 11403074 and 11333005. A. Zdziarski has been supported
in part by the Polish National Science Center grants
2013/10/M/ST9/00729 and 2015/18/A/ST9/00746. Zhou acknowledges the
support from the NWO Veni Fellowship, grant no. 639.041.647, and NSFC
grants 11503008 and 11590781.



\vspace{0.5cm} \noindent {\bf Author contributions: } This paper is an
initiative of \extp's Science Working Group 4 on Observatory Science,
whose members are representatives of the astronomical community at
large with a scientific interest in pursuing the successful
implementation of \extp. The paper was primarily written by S. Drake
(stellar flares), Y. Chen and P. Zhou (supernova remnants),
M. Hernanz, D. de Martino (accreting white dwarfs), T. Maccarone
(binary evolution), J. in 't Zand (thermonuclear X-ray bursts),
E. Bozzo (high-mass X-ray binaries), A. de Rosa (radio-quiet active
galactic nuclei), I. Donnarumma (radio-loud active galactic nuclei),
E. Rossi (tidal disruption events) and L. Amati (gamma-ray bursts),
with major contributions by C. Heinke, G. Sala, A. Rowlinson,
H. Schatz, T. Tauris, J. Wilms, W. Yu and X. Wu. The contributions
were edited by J. in 't Zand, E. Bozzo, J. Qu and X. Li. Other co
authors provided input to refine the paper.

\end{multicols}
\end{document}